# MEIC Design Summary

## January 20, 2015

### Author List

S. Abeyratne[6], D. Barber[2], A. Bogacz[1], P. Brindza[1], Y. Cai[8], A. Camsonne[1], A. Castilla[7], P. Chevtsov[10], E. Daly[1], Y. S. Derbenev[1], D. Douglas[1], V. Dudnikov[11],R. Ent[1], B. Erdelyi[6], Y. Filatov[4], D. Gaskell[1], J. Grames[1], J. Guo[1], L. Harwood[1], A. Hutton[1], C. Hyde[7], K. Jordan[1], A. Kimber[1], G. A. Krafft[1,7](editor), A. Kondratenko[3], M. Kondratenko[3], R. Li[1], F. Lin[1], T. Mann[9], P. McIntyre[9], T. Michalski[1], V.S. Morozov[1], P. Nadel-Turonski[1], Y.M. Nosochkov[8], P. N. Ostroumov[5], K. Park[7], F. Pilat[1], M. Poelker[1], N. J. Pogue[9], R. Rimmer[1], Y. Roblin[1], T. Satogata[1], A. Sattarov[9], M. Spata[1], R. Suleiman[1], M. Sullivan[8], A. Sy[1], C. Tennant[1], H. Wang[1], M-H Wang[8], S. Wang[1], U. Wienands[8], H. Zhang[1], Y. Zhang[1], Z. Zhao[1]

### Author Institutions

[1]Thomas Jefferson National Accelerator Facility (JLab), Newport News, VA 23606 USA

[2]Deutsches Elektronen-Synchrotron (DESY), 22607 Hamburg, Germany

[3]Sci. & Tech. Laboratory Zaryad, Novosibirsk, Russia

[4]Moscow Institute of Physics and Technology, Dolgoprydny, and Joint Institute for Nuclear Research, Dubna, Russia

[5]Argonne National Laboratory, Argonne, IL 60439 USA

[6]Northern Illinois University, DeKalb, IL 50115 USA

[7]Old Dominion University, Norfolk, VA 23529 USA

[8]SLAC National Accelerator Laboratory, Menlo Park, CA 94305 USA

[9]Texas A&M University, College Station, TX 77843 USA

[10]Paul Scherrer Institute, Villigen, Switzerland

[11]Muons, Incorporated, Batavia, IL 60510 USA

# Abstract

This document summarizes the design of Jefferson Lab's electron-ion collider, MEIC, as of January 20, 2015, and describes the facility whose cost was estimated for the EIC cost review on January 26-28, 2015. In particular, each of the main technical systems within the collider is presented to the level of the best current information.

## Table of Contents

## Introduction: Collider Strategy and Recent MEIC Updates

This document summarizes the status and recent progress in the development of MEIC, the Jefferson Laboratory design for the EIC, the Electron Ion Collider. The science requirements and the conceptual design for the MEIC accelerator complex were described in a comprehensive report edited and published in 2012 [Abeyratne2012]. For a discussion of the EIC science requirements [Accardi2012] and of the MEIC physics program and for a more detailed discussion of the basis of the accelerator design we refer to the 2012 report.

The goal of this document is to recall the basic strategy and technical choices for the MEIC, and to discuss and motivate the updates that lead us to the present baseline. We present information on the baseline and the resulting luminosity and polarization performance, followed by technical descriptions of the main systems in the MEIC complex. This baseline is the basis of the present preliminary cost estimate to be reviewed by the NSAC EIC Cost Estimate Sub-Committee in January 2015. In the final section of this report we summarize the main R&D challenges and plans to validate the MEIC design, we provide a vision towards realizing the MEIC project at Jefferson Laboratory, and we discuss options for further optimizations and performance enhancements to the design.

The MEIC design, initially based on an ERL-ring option, evolved in 2006 to the present ring-ring design on the basis of luminosity performance optimization and minimization of technical risks, primarily concerning high current polarized electron sources [Merminga2003]. The overall basic strategy towards achieving high luminosity and high polarization has not changed ever since but technical design aspects have evolved. In particular the updates from the 2012 report are the results of an ongoing optimization process that focuses on performance; cost; and potential for phasing and future upgrades.

At the highest level, the main changes with respect to the design presented in 2012 are:

- The ion and electron ring circumferences have been increased to ~2.2 km.
- The design of the e-ring is now based on PEP-II High Energy Ring (HER) lattice and components (magnets, vacuum chambers, RF).
- The design of the ion-ring is based on super-ferric magnets.
- The design now features a single 8 GeV figure-8 booster based on super-ferric magnets (instead of a 3 GeV pre-booster and a 10 GeV booster co located in the main tunnel)

The motivation for the increase in circumference is the re-use of PEP-II components and in addition the increased length allowed for use of super-ferric technology in the ion ring. This opens up the possibility of upgrading the ion ring to higher energies in the future without additional tunneling and civil construction. The consolidation of 2 boosters into 1 reduces project complexity without significant impact on MEIC performance. The changes and related impacts will be discussed in more detail in the respective sections of this document.

The high-level design goals for the MEIC are:

## Energy

The center-of-mass (CM) energy of this collider should be between 15 and 65 GeV. The value of $s^2 = 4E_e E_u$ is from a few hundred to a few thousand GeV$^2$, where $E_e$ and $E_u$ are kinetic energies of electron and nucleon. The energy range of the colliding beams are:

- from 3 to 10 GeV for electrons,
- from 20 to 100 GeV for protons, and
- up to 40 GeV per nucleon for ions.

Protons or ions with energies below 20 GeV per nucleon are also potentially interesting.

## Ion species

Ion species of interest include polarized protons, deuterons, and helium-3. Other polarized light ions are also desirable. Heavy ions up to lead do not have to be polarized. All ions are fully stripped at collision.

## Multiple Detectors

The facility must be able to accommodate 2 detectors at 2 interaction points. The baseline design includes costs for one large acceptance detector. This report mainly focuses on plans developed for this detector.

## Luminosity

The luminosity goal is in the range of from low-$10^{33}$ to mid-$10^{33}$ $cm^{-2}sec^{-1}$ per interaction point over a broad energy range. The luminosity should be optimized around 45 to 50 GeV CM energy (the value of $s^2$ is around 2000 to 2500 GeV$^2$).

## Polarization

The longitudinal polarization goal for both electron and light-ion beams at the collision points is more than 70%. Transverse polarization of the ions at the collision points and spin-flip of both beams are extremely desirable. High-precision (1–2%) ion polarimetry is required.

## Positrons

Polarized positron beams colliding with ions are desirable, with a high luminosity similar to that of the electron-ion collisions. The collider design should allow positron collisions eventually; positrons are not part of the baseline design.

In addition, the MEIC accelerator design allows options for future energy upgrades: electron energies up to 12 GeV, proton energy up to 250 GeV, and ion energy up to 100 GeV per nucleon. These options will be addressed in the concluding section.

Next, we summarize the basic strategy and design principles to achieve high luminosity and polarization at the MEIC. The key to the high luminosity in MEIC is high bunch repetition rate colliding beams. Both the electron and ion beams have very short bunch length and small transverse emittances such that a strong final focusing can be applied to reduce the beam spot sizes to as small as a few $\mu$m at the collision point. This configuration combined with a high bunch repetition rate boosts the collider luminosity. An ultrahigh bunch repetition rate ensures a very small bunch charge (allowing for relatively weak collective and inter beam effects) of the colliding beams, particularly in the ion beams, while maintaining high beam current to provide high luminosity. A detailed discussion of this concept can be found in reference [Derbenev2010]. This luminosity strategy has been validated by the lepton-lepton B-factories colliders worldwide. For example, the KEK-B factory holds the present luminosity world record. [Funakoshi2004].

To illustrate this concept, first examine the standard luminosity formula as a function of a beam-beam parameter for a head-on electron-proton collision [Furman2006]

$$L = \frac{\gamma_p N_p f_c \xi_{y,p}}{2 r_p \beta^*_{y,p}} \left(1 + \frac{\sigma^*_y}{\sigma^*_x}\right)$$

where the beam-beam parameter (tune shift) for the proton beam is

$$\xi_{y,p} = \frac{r_p N_e}{\gamma_p} \frac{\beta^*_{y,p}}{2\pi\sigma^*_y(\sigma^*_x + \sigma^*_y)}$$

The other parameters in these two formulas are the collision frequency (bunch repetition rate) $f_c$, the numbers of electrons and protons per bunch $N_e$ and $N_p$, the beta function at the collision point $\beta^*_{y,p}$, and the horizontal and vertical beam spot sizes (assumed matched) at the collision point $\sigma^*_x$ and $\sigma^*_y$. Values of these parameters depend on the collider design and are usually limited by collective beam effects. The luminosity formula can also be cast using the electron beam-beam parameters.

High luminosity can be achieved by increasing both the beam current $N_p f_c$ and the beam-beam tune shift $\xi_{y,p}$, as well as decreasing the vertical $\beta^*_{y,p}$. Because the ion bunch charge is limited not only by the highly nonlinear beam-beam effects but also by additional collective beam effects such as space-charge tune shifts, increasing the bunch repetition rate is a preferable way to boost the proton beam currents. At the same time, one must watch the electron beam-beam tune shift. At the low ion charge per bunch allowed by high repetition rates, the electron beam-beam tune shift can also be kept reasonable, at already achieved levels. This is the principal advantage of high repetition rates. Present experience from years of operation of existing ring-ring colliders shows that a value of 0.035 is a practical limit for the total beam-beam tune shift for hadron beams; although recent LHC operations has indicated that such a limit may be even higher. The beam-beam tune shift is roughly a factor of four to five times larger for lepton beams thanks to their synchrotron radiation damping. Within these constraints, the luminosity of a ring-ring collider can be optimized by pushing up the bunch collision frequency and squeezing down the $\beta^*$ values.

For all existing hadron ring-ring colliders, the collision frequency, i.e., bunch repetition rate, is relatively small; therefore, there is a limited number of bunches per beam, ranging from just a handful (9 for SPS) to several dozen (36 for Tevatron). It is worth noting that RHIC now operates at up to 112 bunches (about 8.9 MHz repetition rate). The LHC has ~3000 bunches but a large circumference, so the bunch repetition rate is 34 MHz, a significant increase from all previous hadron colliders. With the large bunch charges (up to $10^{11}$ and above) necessary to maintain even a modest hadron beam current, bunch lengths are usually very long (of the order of 1 m), mainly due to limits of collective effects and scattering processes. Long bunches prevent a strong final focusing (small $\beta^*$) due to the hourglass effect, and combined with large transverse emittance without beam cooling, lead to fairly large beam spot sizes at collision points. These effects together can lead to lower luminosities.

As mentioned, the high-repetition-rate luminosity concept has been validated at today's lepton collider B-factories, which use very large bunch collision frequency (storing tens to hundreds times more bunches compared to the hadron colliders) and hundreds times smaller $\beta^*$ (and the corresponding smaller spot sizes) at the collision point through the strong final focusing enabled by short bunch lengths. The net result is several orders of magnitude increase in luminosity in a storage ring.

The MEIC luminosity concept can be summarized as follows:

- Very short bunches for both electron and ion beams
- Very small transverse emittance
- Ultrahigh collision frequency beams
- Staged electron cooling to achieve appropriate ion emittances
- Very small final focusing $\beta^*$
- Large attainable beam-beam tune shift
- Crab crossing of colliding beams

The first three items specify the design of MEIC colliding beams in terms of the phase space structure (bunch length and emittance) and time structure (bunch frequency and CW). The staged electron cooling is the essential ingredient to form low emittance high performance ion beams. The last three specify the requirements for the MEIC interaction regions in order to take advantage of high bunch repetition CW colliding beams. Crab crossing is required to recover the loss of luminosity due to the crossing angle at the interaction point (necessary to eliminate parasitic collisions and harmful long-range beam-beam effects).

The polarization requirements for both MEIC electron and ions beam are the following:

- Beam polarization above 70%
- Polarization longitudinal at the interaction points
- Sufficiently long polarization times to sustain physics stores
- Polarization of bunches can be switched at high frequency (beam flipping)

CEBAF is a fully polarized electron injector and the polarization in the electron ring can be preserved and enhanced by the Sokolov-Ternov effect. A set of energy-independent spin rotators will align the electron spins in the required longitudinal direction at the collision points and in the vertical direction in the arcs. The electron polarization lifetime in the ring will most likely not require top-off injection from CEBAF to replenish the polarization but this option is available if required.

In principle, it is a challenge to provide highly polarized ion beams in an ion ring because, unlike the electron beam, there is no synchrotron radiation and hence no self-polarizing Sokolov-Ternov effect. Consequently, polarized ion sources are necessary. In addition during acceleration the polarization must be preserved though multiple crossing of depolarizing spin resonances. The primary design choice for the MEIC polarization is to adopt a figure-8 layout for the booster and collider rings. This breakthrough concept enables energy-independent spin tune and thus effectively eliminates depolarization at all spin resonances. With Siberian snakes for proton and $^3$He or a special magnetic insert for deuterons, stable spin motions in the ion collider rings can be achieved and the desired polarization can be realized at collision points.

One important aspect of the MEIC design approach is a balance of technology innovation for maximum collider performance and overall design simplification for reduced technical challenges and required R&D. We have imposed limits on several machine or beam parameters in order to minimize accelerator R&D and to improve the robustness of the design. We have worked to stay, for the most part, within operational limits derived largely from previous lepton and hadron collider experience and the present state of the art of accelerator technology. Such a technical approach allows us to focus limited resources on the main critical accelerator R&D challenges such as high energy bunched beam electron cooling.

## Overview of the Baseline

The MEIC is designed to meet the requirements of the science program outlined in the EIC white paper [Abeyratne2012]. As already discussed, MEIC is designed to be a traditional ring-ring collider. The central part of this facility is a set of figure-8 collider rings as shown in Figure 2.1. The electron collider ring is made of normal conducting magnets reconditioned from the decommissioned PEP-II *e+e-* collider at SLAC, and will store an electron beam of 3 to 10 GeV. The ring also reuses the PEP-II vacuum chambers and RF systems. The stored electron beam current is up to 3 A, scaled down when the beam energy exceeds 7 GeV in order to satisfy the operational limit of 10 kW/m synchrotron radiation power for the PEP-II vacuum chambers. The ion collider ring is made of new super-ferric magnets, a cost-effective type of superconducting magnet with modest field strength (up to 3 T) and will store a beam with energy of 20 to 100 GeV for protons or up to 40 GeV per nucleon for heavy ions. The stored ion beam current is up to 0.5 A. The two collider rings are stacked vertically and housed in the same underground tunnel as shown in the left drawing of Figure 2.1. They have nearly identical circumferences of approximately 2.2 km, and fit the Jefferson Lab site as shown in Figure 2.2.

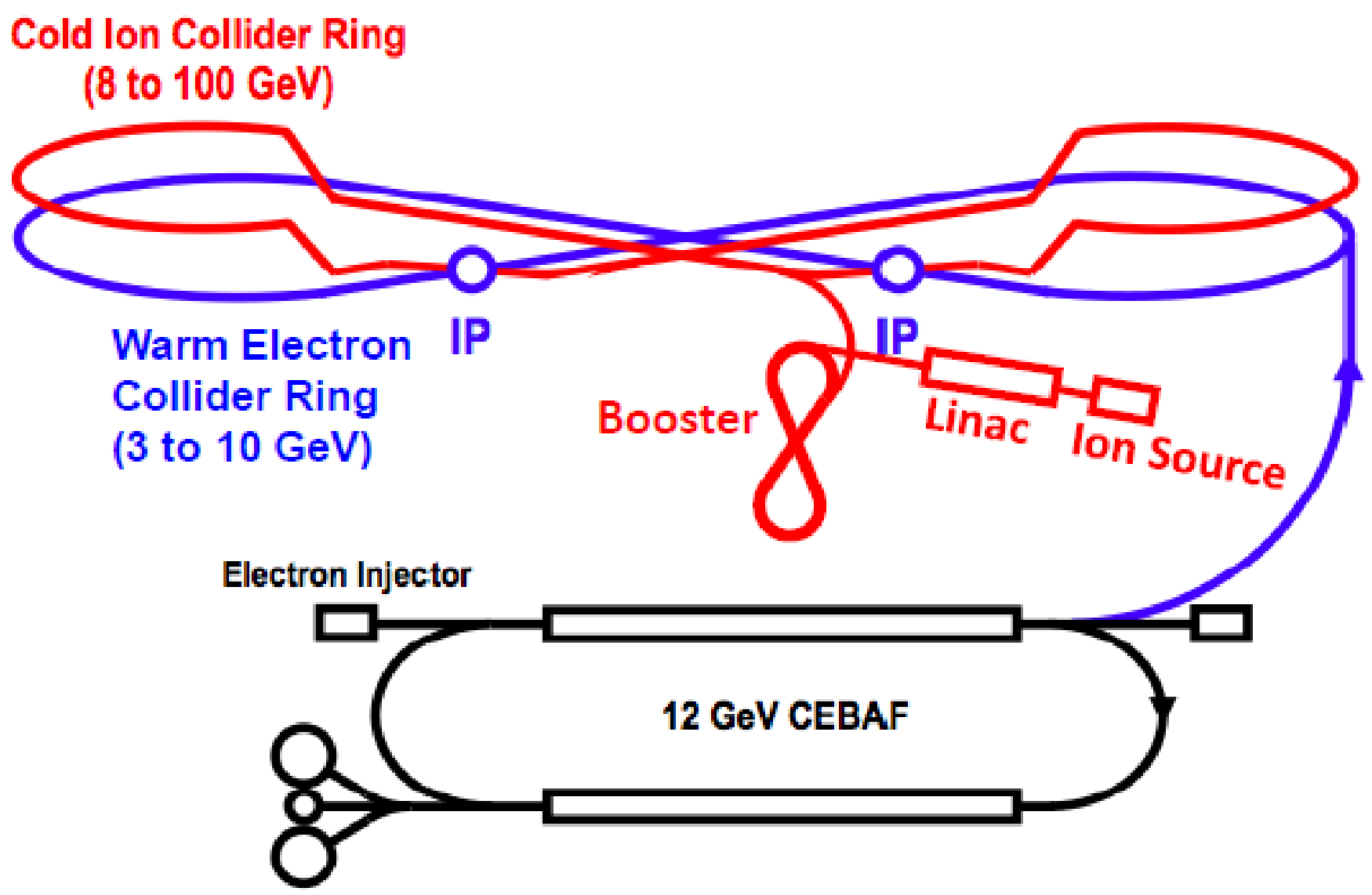

**Figure 2.1:** A schematic layout of MEIC. The ion collider ring is stacked vertically above the electron collider ring, and takes a vertical excursion to the plane of the electron ring for a horizontal crossing

The unique figure-8 shape of the MEIC collider rings has been chosen to optimally preserve the ion polarization during acceleration and store [Derbenev1996]. The crossing angle is approximately 82°, partitioning a collider ring into two arcs and two long straights. The electron and ion collider rings intersect at two symmetric points, one in each of the two long straights as shown in Figure 2.1; thus two detectors can be accommodated. The ion beam executes a vertical excursion to the plane of the electron ring to realize a horizontal crossing for electron-ion collisions. The two long straights also support other utility elements of the collider rings, among them the injection/ejection systems, the RF systems, the electron cooler, and the beam polarimeters.

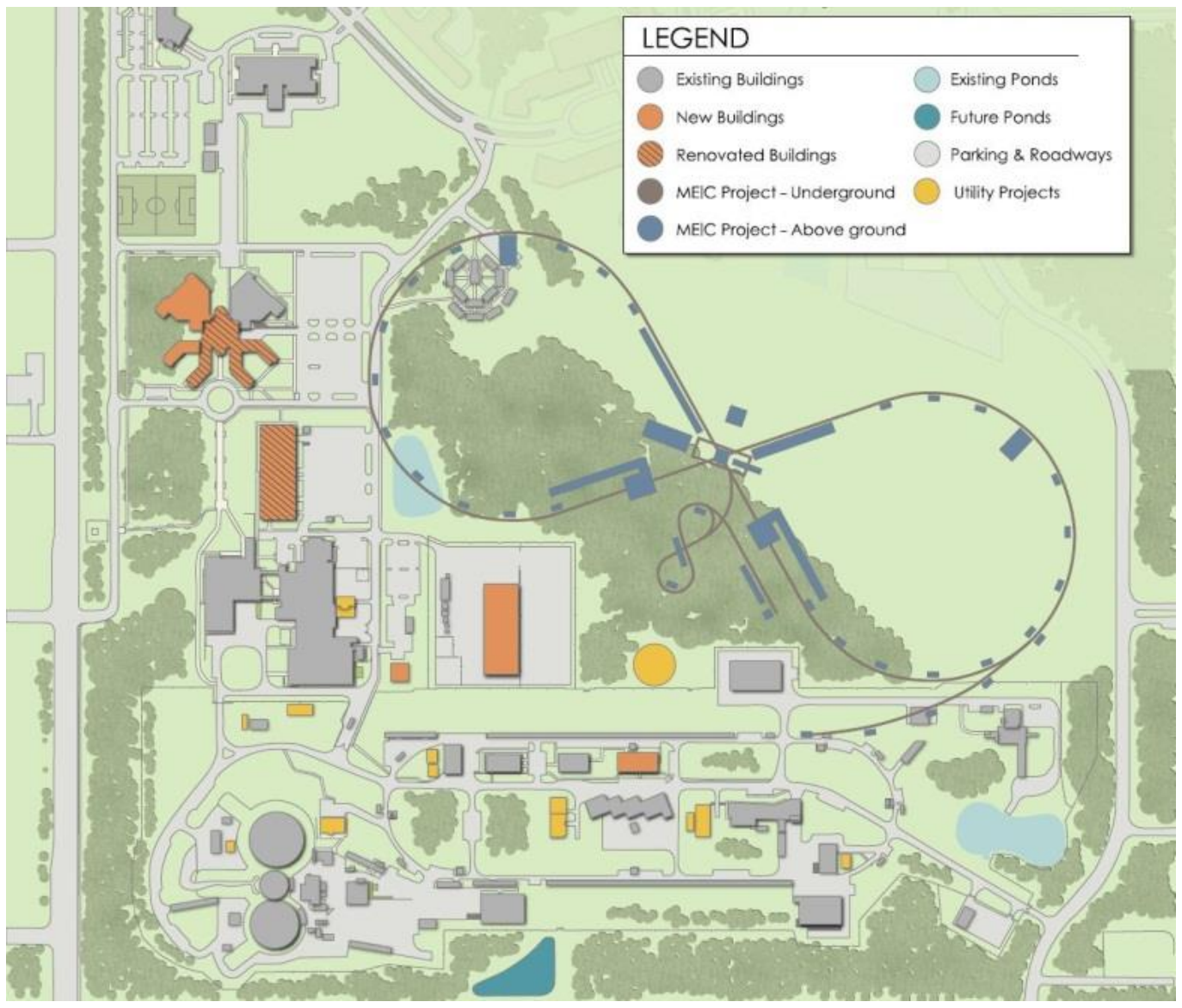

**Figure 2.2:** MEIC on the Jefferson Lab site map

The MEIC collider rings are supported by two injector complexes respectively. On the electron side, the CEBAF recirculating SRF linac will serve as a full-energy injector into the electron collider ring. The polarized linac beam will be extracted from CEBAF at the location near Experimental Hall D as illustrated in Figure 2.1. There is no need of further upgrade for beam energy, current, or polarization beyond the recently successfully commissioned 12 GeV upgrade. In the present conceptual design, the ion injector complex consists of sources for polarized light ions and non-polarized light to heavy ions, a cold ion linac, and one figure-8 shaped booster synchrotron. The booster accepts and accumulates protons or ions from the linac, accelerates protons to 8 GeV kinetic energy or lead ions to 3.2 GeV kinetic energy per nucleon, then transfers them to the ion collider ring.

As a consequence of reusing the PEP-II warm RF cavities and RF stations for the MEIC electron collider, the bunch repetition rate of the MEIC stored beams is 476 MHz, lower than the previous design value of 748.5 MHz. This choice does not affect the collider luminosity significantly. A conceptual scheme has been developed for injecting the electron bunches from the CEBAF SRF linac (which has a 1.497 GHz frequency) into the collider ring as discussed subsequently. All new SRF cavities and RF stations required for the ion collider ring will have a frequency of 952 MHz, thereby enabling cost effective future improvements in luminosity and energy.

As a critical part of the luminosity concept, MEIC selects conventional electron cooling technology for reducing the ion beam emittance. It also adopts a multi-phased cooling scheme to achieve the required high cooling efficiency. The scheme utilizes two electron coolers. One is a DC cooler in the booster synchrotron, and the other is a bunched beam cooler based on an energy recovery linac (ERL) in the collider ring. The DC cooler assists in accumulating positive ions injected from the linac and cools the ions at 2 GeV to reduce their emittance to the design values. The primary purpose of the bunched beam ERL cooler is to suppress the intra-beam scattering (IBS) and to maintain the small emittance achieved in the initial cooling stage in the booster synchrotron. Section 9 will present the MEIC beam cooling scheme in more detail.

The design of the MEIC interaction region is aimed at achieving high luminosity in an integrated full acceptance detector. The current MEIC detector design requires a magnet-free space of 7 m for the ion beam on the downstream side, and after optimization, only 3.6 m on the upstream side. For the electron beam, the first final focusing elements are permanent magnets which, due to their small transverse sizes, are placed inside the main detector and very close to the interaction point. The MEIC design adopts a finite crossing of colliding beams to avoid all parasitic collisions. Following the successful experience at KEK-B [KEKB1995], the MEIC design also utilizes a local compensation scheme based on SRF crab cavities to restore head-on collisions thus recovering the loss of luminosity caused by the crossing angle. A relatively large crossing angle (the ions cross the electrons, which are aligned with the center of the detector, with a 50 mrad angle) also enhances the detection of reacting particles. The IR design uses a combined local and global compensation scheme to control the chromatic aberrations.

A unique design feature of MEIC is a figure-8 shape of all ion rings. As pointed out above, the feature provides optimal preservation of high polarization of ion beams while accelerating and storing the polarized ion beam. The basic property is a complete cancellation of spin precession in the left and right arcs of the figure-8 ring; thus the net spin tune is zero. The spin tune can be further controlled by a weak field spin rotation device, leading to a stable spin polarization in the figure-8 ring. The figure-8 design also improves polarization of the electron beam and more importantly provides the only practical solution for accelerating and storing medium-energy polarized deuteron beams. The orientation of electron and ion beam polarization at the interaction points are achieved by spin rotators [Chevtsov2010,Kondratenko2014].

## Luminosity Performance

The MEIC nominal parameters at three representative design points in the low, medium and high CM energy regions respectively are presented in Table 3.1. The luminosity is above $10^{33}$ cm$^{-2}$sec$^{-1}$ in all these design points for the full-acceptance detector, and reaches $4.6\times10^{33}$ cm$^{-2}$sec$^{-1}$ at the medium CM design point of approximately 45 GeV. For the second detector, as an option, the interaction region design can be optimized for reaching a higher luminosity (approximately 60% increase) while still retaining a fairly large detector acceptance. But the detector space for this case must be reduced to 4.5 m so that the interaction point beta-function, $\beta^*$, may be decreased accordingly.

It is challenging to optimize the performance of MEIC over both a broad range of beam energy and a wide array of ion species. In particular, the luminosity of MEIC is affected strongly by various single bunch or multi-bunch collective beam effects. These effects limit either the bunch charges or currents of either beam, and may cause large beam emittance. Therefore, design optimization has been carried out individually in each energy region, taking into account the leading region specific performance limit. Figure 3.1 illustrates the general trends of the MEIC luminosity.

**Table 3.1:** MEIC main design parameters for a full-acceptance detector.

| CM energy | GeV | 21.9 (low) | | 44.7 (medium) | | 63.3 (high) | |
|---|---|---|---|---|---|---|---|
| | | *p* | *e* | *p* | *E* | *p* | *e* |
| Beam energy | GeV | 30 | 4 | 100 | 5 | 100 | 10 |
| Collision frequency | MHz | 476 | | 476 | | 159 | |
| Particles per bunch | $10^{10}$ | 0.66 | 3.9 | 0.66 | 3.9 | 2.0 | 2.8 |
| Beam current | A | 0.5 | 3 | 0.5 | 3 | 0.5 | 0.72 |
| Polarization | % | >70 | >70 | >70 | >70 | >70 | >70 |
| Bunch length, RMS | cm | 2.5 | 1.2 | 1.0 | 1.2 | 2.5 | 1.2 |
| Norm. emittance, vert./horz. | µm | 0.5/0.5 | 74/74 | 1/0.5 | 144/72 | 1.2/0.6 | 1152/576 |
| Horizontal and vertical $\beta^*$ | cm | 3 (1.2) | 5 (2) | 2/4 (1.6/0.8) | 2.6/1.3 (1.6/0.8) | 5/2.5 (2/1) | 2.4/1.2 (1.6/0.8) |
| Vert. beam-beam parameter | | 0.01 | 0.02 | 0.006 (0.004) | 0.014 (0.021) | 0.002 (0.001) | 0.013 (0.021) |
| Laslett tune-shift | | 0.055 | small | 0.01 | small | 0.01 | small |
| Detector space | m | 7/3.6 (4.5/4.5) | 3.2/3 (3/3) | 7/3.6 (4.5/4.5) | 3.2/3 (3/3) | 7/3.6 (4.5/4.5) | 3.2/3 (3/3) |
| Hour-glass (HG) reduction factor | | 0.89 (0.67) | | 0.89 (0.74) | | 0.73 (0.58) | |
| Lumi./IP, w/HG correction, $10^{33}$ | cm$^{-2}$s$^{1}$ | 1.9 (3.5) | | 4.6 (7.5) | | 1.0 (1.4) | |

(Values for a high-luminosity detector with a 4.5 m ion detector space are given in parentheses.)

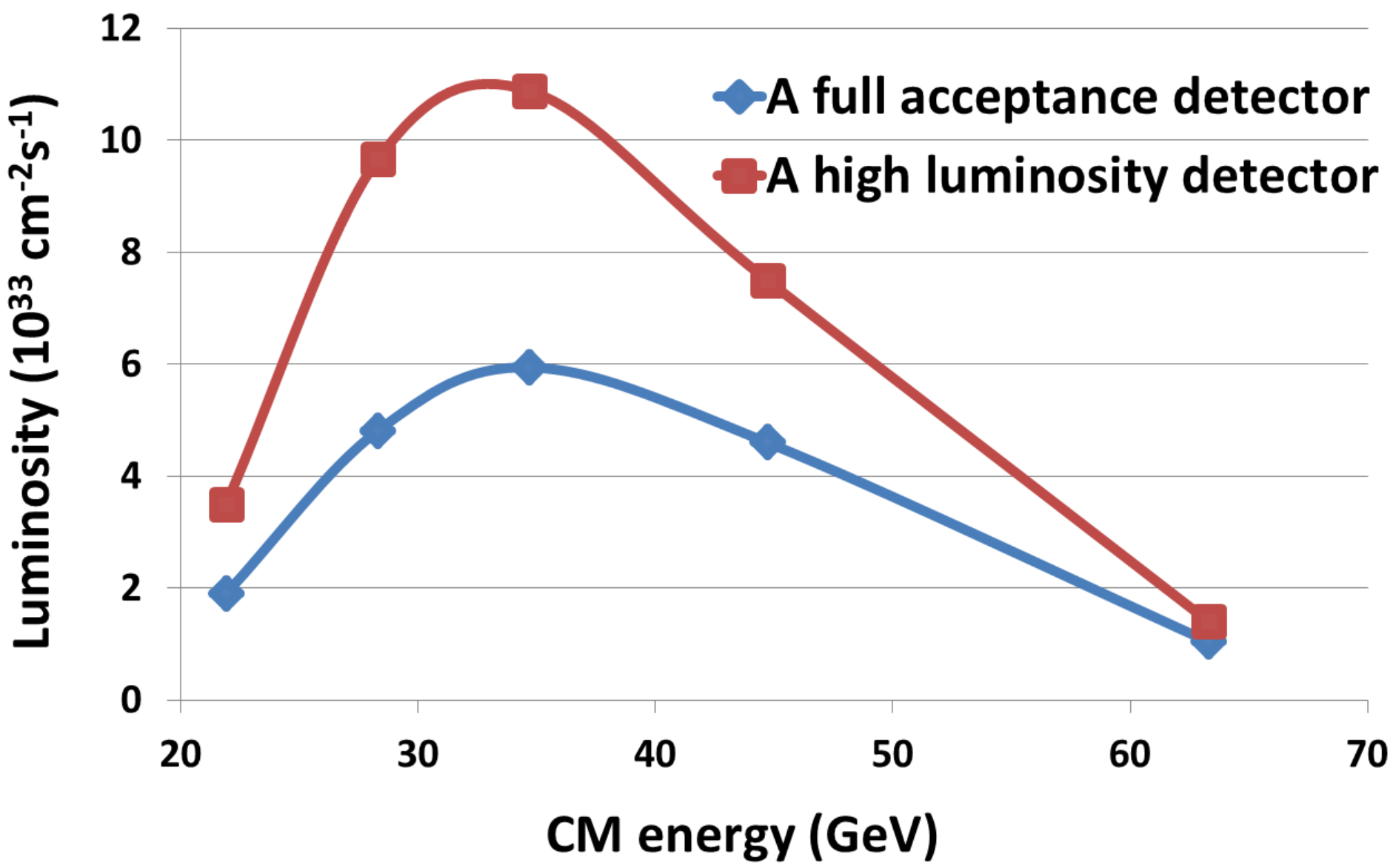

**Figure 3.1:** A luminosity plot of the MEIC *e-p* collision. The blue line is for a full acceptance detector, which is chosen as the primary one for MEIC. The red line is for an optional secondary detector optimized for achieving a higher luminosity.

At the low energy end, space charge of the low energy ion beam severely limits the bunch charge, particularly for short bunches. The design strategy is to allow a longer bunch length (2.5 cm) than at higher energy, thus accommodating the full bunch charge while remaining under the design limit for the Laslett space charge tune-shift limit of 0.06. However, because the bunch length is much larger than the $\beta^*$, there is a non-negligible (11%) loss of luminosity due to the hourglass effect. The resulting, luminosity for the low energy design point is shown in Table 3.1. We are considering advanced concepts such as the running-focusing scheme [Balakin1991] to recover the hour-glass effect induced luminosity loss. This topic will be the subject of a future R&D study.

At the high energies, synchrotron radiation of a high energy electron beam is the dominating effect. The electron beam current must be scaled down proportionally to the 4$^{th}$ power of the electron energy to reduce synchrotron radiation loading to acceptable levels. As an example, the allowable electron current is 0.72 A at 10 GeV. The design strategy is to choose a low bunch repetition rate and boost the bunch charge proportionately. For the design point of 100×10 GeV$^2$, the bunch frequency is reduced by a factor of three, leading to a factor of three increase of bunch charge. With this change, the proton bunch length must be increased to alleviate single bunch effects. Again, there is a significant luminosity loss due to the hour-glass effect; the computed luminosity includes these relevant effects. An additional concept applied is to mismatch the beam spot sizes at the collision point by taking advantage of very weak beam-beam interaction in this case. Since the electron emittance is very large, the electron beam spot size at collision is 70% larger than the proton beam spot size. In this regime the highly nonlinear beam-beam effect is a weak perturbation.

The medium-energy region of MEIC is dominated by the strong-strong beam-beam effect and so the way to achieve optimized luminosity is a combination of a high bunch repetition rate, small beam emittance, and very small $\beta^*$. This energy region delivers the highest luminosity for the MEIC. Presently, the design strategy is to relax the design values of proton or ion beam emittance because doing so does not degrade the luminosity significantly when the colliding electron and ion beam spot sizes are matched at the collision point. This design change also provides an opportunity for reducing the demands of beam cooling. This will have a great impact on the most technical R&D of MEIC. This topic will be further discussed in Section 11.

The MEIC design parameters and luminosities for several representative *e-A* collisions are summarized in Table 3.2. They are derived based on the same design principles discussed above.

**Table 3.2:** MEIC main design parameters for *e-A* collisions.

| | | Electron | Proton | Deuteron | Helium | Carbon | Calcium | Lead |
|---|---|---|---|---|---|---|---|---|
| | | *e* | *P* | *d* | $^3He^{++}$ | $^{12}C^{6+}$ | $^{40}Ca^{20+}$ | $^{208}Pb^{82+}$ |
| Beam energy | GeV | 5 | 100 | 50 | 66.7 | 50 | 50 | 39.4 |
| Particles/bunch | $10^{10}$ | 3.9 | 0.66 | 0.66 | 0.33 | 0.11 | 0.033 | 0.008 |
| Beam current | A | 3 | 0.5 | 0.5 | 0.5 | 0.5 | 0.5 | 0.5 |
| Polarization | | > 70% | > 70% | > 70% | > 70% | - | - | - |
| Bunch length, RMS | cm | 1.2 | 1 | 1 | 1 | 1 | 1 | 1 |
| Norm. emit., horz./vert. | µm | 144/72 | 1/0.5 | 0.5/0.25 | 0.7/0.35 | 0.5/0.25 | 0.5/0.25 | 0.5/0.25 |
| $\beta^*$, hori. & vert. | cm | 2. 6/1.3*1 (1.6/0.8) | 4/2 (1.6/0.8) | 4/2 (1.6/0.8) | 4/2 (1.6/0.8) | 4/2 (1.6/0.8) | 4/2 (1.6/0.8) | 5/2.5 (1.6/0.8) |
| Vert. beam-beam parameter | | 0.014*2 (0.02) | 0.006 (0.004) | 0.006 (0.004) | 0.006 (0.004) | 0.006 (0.004) | 0.006 (0.004) | 0.005 (0.004) |
| Laslett tune-shift | | | 0.01 | 0.041 | 0.022 | 0.041 | 0.041 | 0.041 |
| Detector space, up & down stream | m | 3.2 / 3 (3) | 7 / 3.6 (4.5) | | | | | |
| Hour-glass (HG) reduction factor | | | 0.89 (0.74) | 0.89 (0.74) | 0.89 (0.74) | 0.89 (0.74) | 0.89 (0.74) | 0.89 (0.74) |
| Lumi./IP/**nuclei**, w/HG correction | $10^{33}$ $cm^{-2}s^{-1}$ | | 4.6 (7.5) | 4.6 (9.2) | 2.2 (3.7) | 0.77 (1.37) | 0.23 (0.38) | 0.04 (0.08) |
| Lumi./IP/**nucleon**, w/HG correction, | $10^{33}$ $cm^{-2}s^{-1}$ | | 4.6 (7.5) | 9.2 (15.1) | 6.6 (11.1) | 9.2 (15.1) | 9.2 (15.1) | 7.8 (17.3) |

(Values for a high-luminosity detector with a 4.5 m ion detector space are given in parentheses.)

*1 For a full acceptance detector, the horizontal and vertical electron $\beta^*$ are 2.7 and 1.35 cm respectively for *e-d* collisions. For *e*-Pb collisions, the values are 4 and 2 cm for the electron beam.

*2 For *e*-Pb collisions, the electron vertical beam-beam parameter is 0.014 for a full-acceptance detector and 0.021 for a high luminosity detector.

To derive the sets of MEIC parameters in Tables 3.1 and 3.2, limits were imposed on several key machine or beam parameters in order to reduce accelerator R&D challenges and to improve the robustness of the design. These limits are based largely on previous lepton and hadron collider experience, particularly, that of the components of the PEP-II electron ring that are being reused, and on the present state-of-the-art in accelerator technology:

- The stored beam currents are up to 0.5 A for protons or ions and 3 A for electrons.

- Electron synchrotron radiation power density should not exceed 10 kW/m.
- Maximum bending field of super-ferric magnets is 3 T.
- Maximum betatron function at a beam extension area near an IP is 2.5 km.
- The direct space-charge tune-shift of ion beams is limited to 0.06.
- The hadron beam-beam tune-shift is limited to 0.03 per interaction point.
- The electron beam-beam tune-shift is limited to 0.15 per interaction point.

## Detector Region

MEIC's primary detector [Abeyratne2012] is designed to provide essentially full acceptance to all fragments produced in collisions. It features a relatively large 50 mrad beam crossing angle that allows for a quick separation of the two colliding beams near the interaction point (IP) to make sufficient space for placement of interaction region magnets and to avoid parasitic collisions of shortly-spaced 476 MHz electron and ion bunches. It also moves the spot of poor resolution along the solenoid axis into the periphery and minimizes the shadow of the electron final focusing quadrupoles. A local deflective crab-crossing scheme is used to restore effective head-on collisions of the incident bunches and preserve the luminosity.

A schematic layout of the full-acceptance detector [Abeyratne2012] is shown in Fig. 4.1. The central detector is based on a 5 m long solenoid offset by 50 cm from the IP for kinematic considerations. The forward hadron detection is done in three stages: (1) fragments with scattering angles down to a few degrees are detected in a 2 m long end-cap, (2) fragments up to a few degrees are detected after passing through a 1 m long 2 Tm spectrometer dipole in front of the final focusing quads (FFQs), and (3) fragments up to about one degree pass through the apertures of the FFQs and are detected in a 4 m space before and a 16 m space after a second 4 m long 20 Tm spectrometer dipole. On the forward electron side, the large-angle reaction products are detected in the second end-cap. Electrons scattered at small angles are detected in a low-$Q^2$ tagger consisting of large-aperture electron FFQs and a spectrometer dipole with a few meters of instrumented space on either side.

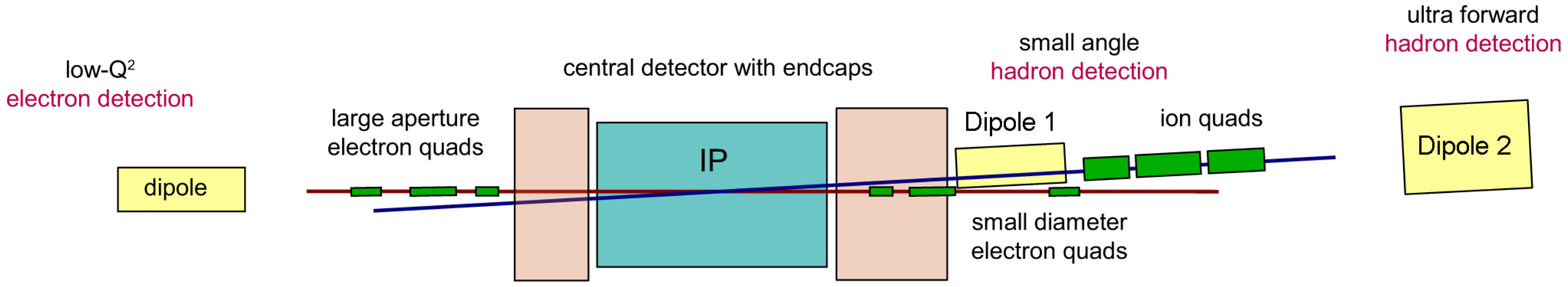

**Figure 4.1:** Schematic layout of the full-acceptance detector.

The interaction region optics is optimized to meet the detection requirements [Morozov2012,Lin2013,Morozov2014] as shown in Figs. 4.2(a) and 4.2(b) for ions and electrons, respectively. In particular, the detector space is made asymmetric by leaving a large 7 m distance from the IP to the first ion FFQ in the downstream ion direction where the reaction products tend to go, while having the upstream ion FFQs placed closer to the IP at 3.5 m to minimize their chromatic contribution. A weak spectrometer dipole is placed in front of the downstream ion FFQs. Electron FFQs nearest to the IP are made of permanent magnets. They have small size and are placed 3 m from the IP on each side. Change of their focusing strength with energy is compensated by adjusting the nearby regular electro-magnetic quads.

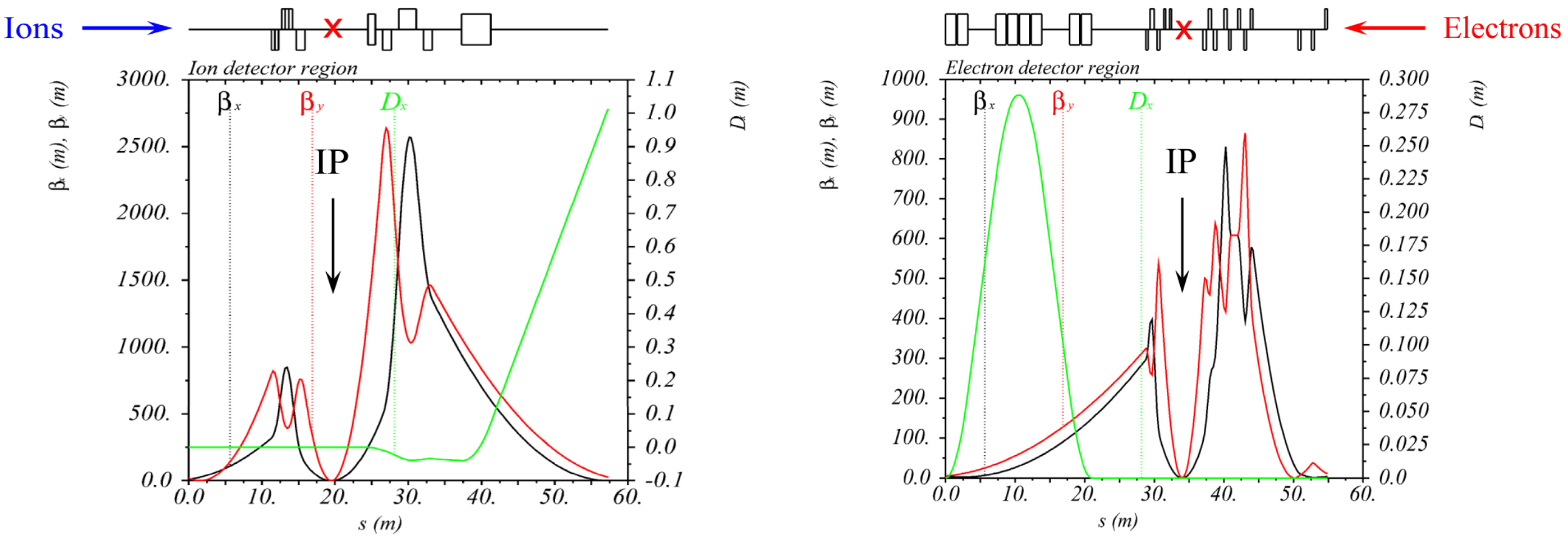

(a) (b)

**Figure 4.2:** Optics of the ion (a) and electron (b) detector regions.

The downstream ion and electron final focusing quads are designed with large apertures for forward detection and are followed by spectrometer dipoles. Additionally, as shown in Figs. 4.2(a) and 4.2(b), both the ion and electron beams are focused again towards the ends of the element-free spaces downstream of the respective spectrometer dipoles to allow closer placement of the detectors at those locations, which, in combination with the relatively large dispersion values there, enhances momentum resolution of the forward detector. The dispersion generated by the spectrometer dipoles is suppressed on the ion side by a specially designed section, which also controls the beam line geometry, while on the electron side the dispersion suppression is done by a simple dipole chicane whose parameters are chosen to avoid a significant impact on the electron equilibrium emittances.

Special attention is paid to sizes and positions of the detector region elements to avoid them interfering with each other and with the detector functionality. Layout of the detector region magnets in the forward ion direction is shown in Fig. 4.3.

The detector performance has been optimized using particle tracking [Morozov2012]. Figure 4.4 shows a complete 3D model of the whole forward detector region developed using GEANT4-based G4beamline [G4beamline] and GEMC [GEMC]. Figure 4.5(a) shows an example of the detector acceptance to small-angle forward-scattered hadrons in terms of the particle's initial horizontal angle $\theta_x$ and momentum offset

Δp/p (or equivalently Δ(m/q)). Figures 4.5(b) and 4.5(c) demonstrate examples of the detector resolution properties at a point 16 m downstream of Dipole 2. The simulations have shown that the far-forward detection system is capable of accepting (1) neutrals in a cone with a full angle of about 25 mrad down to zero degrees, (2) recoil baryons with up to 99.5% of the beam energy for all scattering angles, and (3) recoil baryons with scattering angles down to 2-3 mrad for all energies.

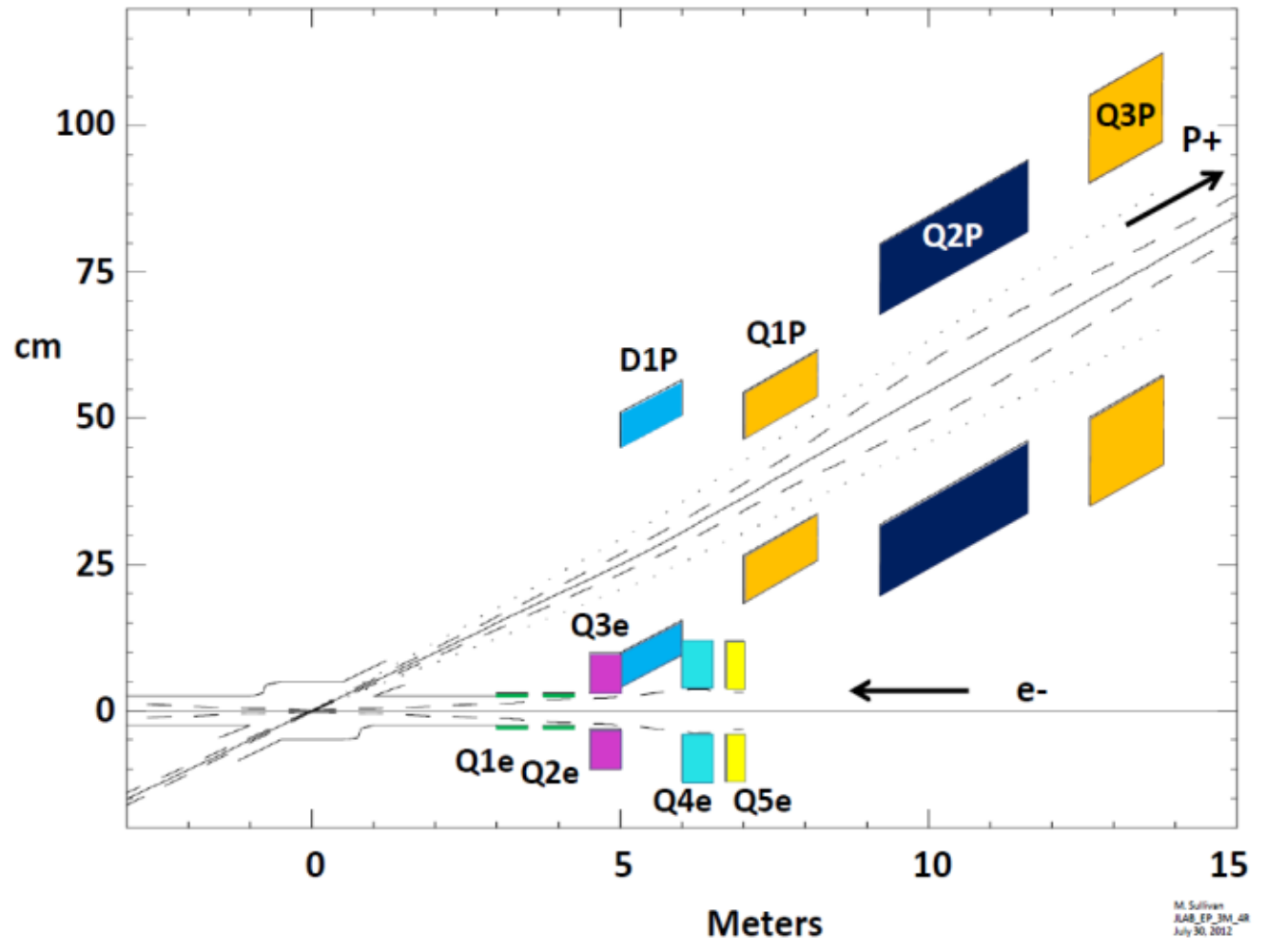

**Figure 4.3:** Layout of the detector region magnets in the forward ion direction.

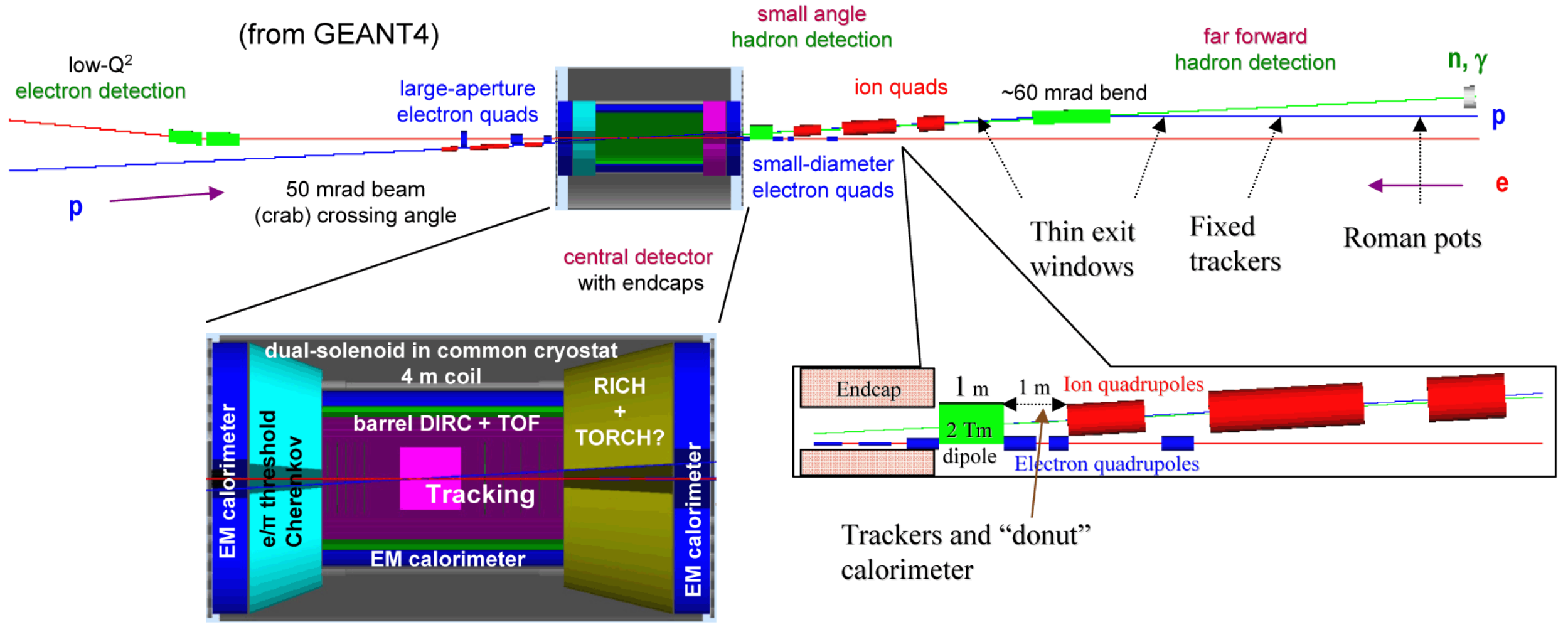

**Figure 4.4:** 3D model of the detector region in GEANT4-based G4beamline and GEMC.

The full-acceptance detector region has been integrated into the electron and ion collider ring lattices with necessary optical and geometric matching as illustrated in Fig. 4.6. The detector is placed far from the electron arc exit to minimize the synchrotron radiation background and close to the ion arc exit to minimize the hadronic background due to the ion beam scattering on the residual gas. The present design includes a single detector in one of the figure-8 straights. The capability of adding a similar or different second detector in the other straight is retained.

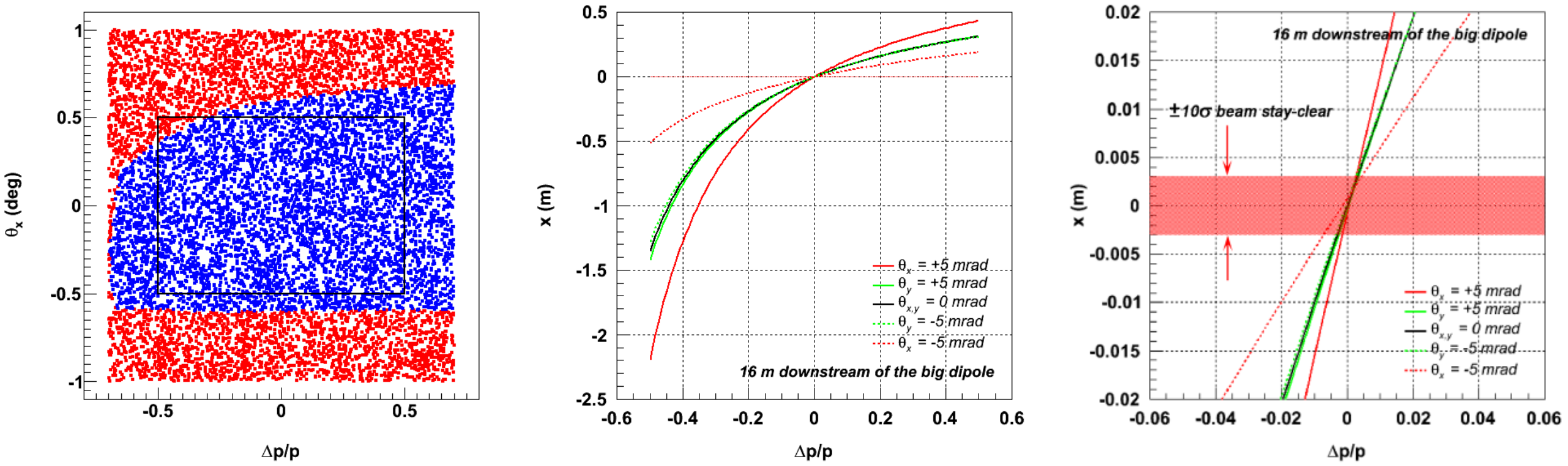

(a) (b) (c)

**Figure 4.5:** (a) Detector's far-forward acceptance (blue), (b) detector's far-forward momentum resolution for a characteristic set of initial angles, (c) expanded version of (b).

Figure 4.7 shows an expanded view of the detector region layout. The end section of the ion arc upstream of the IP is shaped to produce a 50 mrad horizontal crossing angle between the ion and electron beams while the ion beam line segment downstream of the IP is designed to produce a 1.5 m transverse separation between the ion and electron beams. The electron detector region has no net bend or shift. This makes the collider ring geometry somewhat independent from the detector region design and simplifies its optimization [Lin2013,Morozov2014].

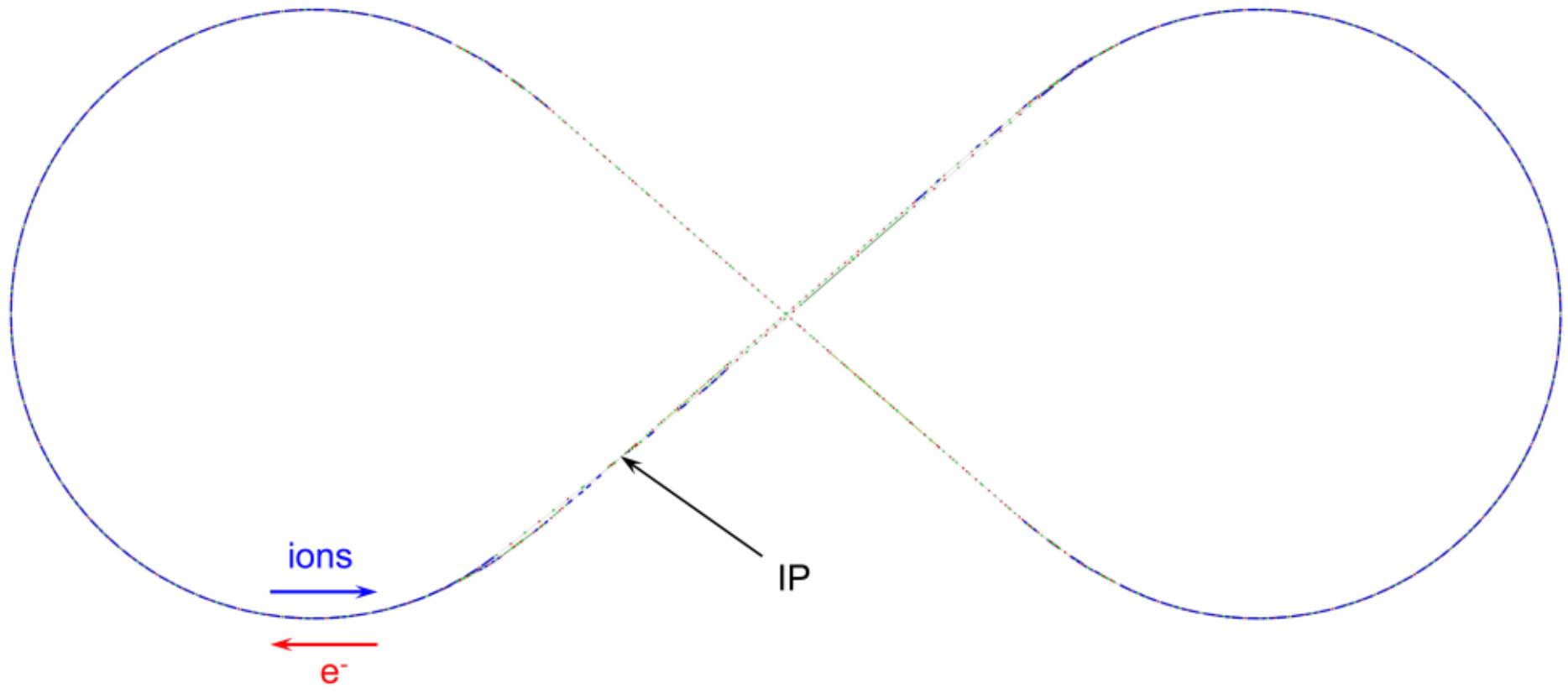

**Figure 4.6:** Collider rings' layout.

Due to the strong beam focusing at the IP, the chromatic effect of the FFBs in both the ion and electron collider rings is very significant and requires proper compensation but is manageable. Chromatic compensation is done using properly arranged sextupole fields in the rings' dispersive regions. Both in the electron and ion rings, there is a sextupole next to each of the arc quadrupoles. Additionally, a dedicated chromaticity compensation block with zero net bend is placed in the electron ring's detector straight for local compensation of the chromaticity induced by the electron FFQs. This provides a substantial flexibility for exploration of various chromaticity compensation schemes. These studies are currently underway with encouraging results [Lin2012,Morozov2013].

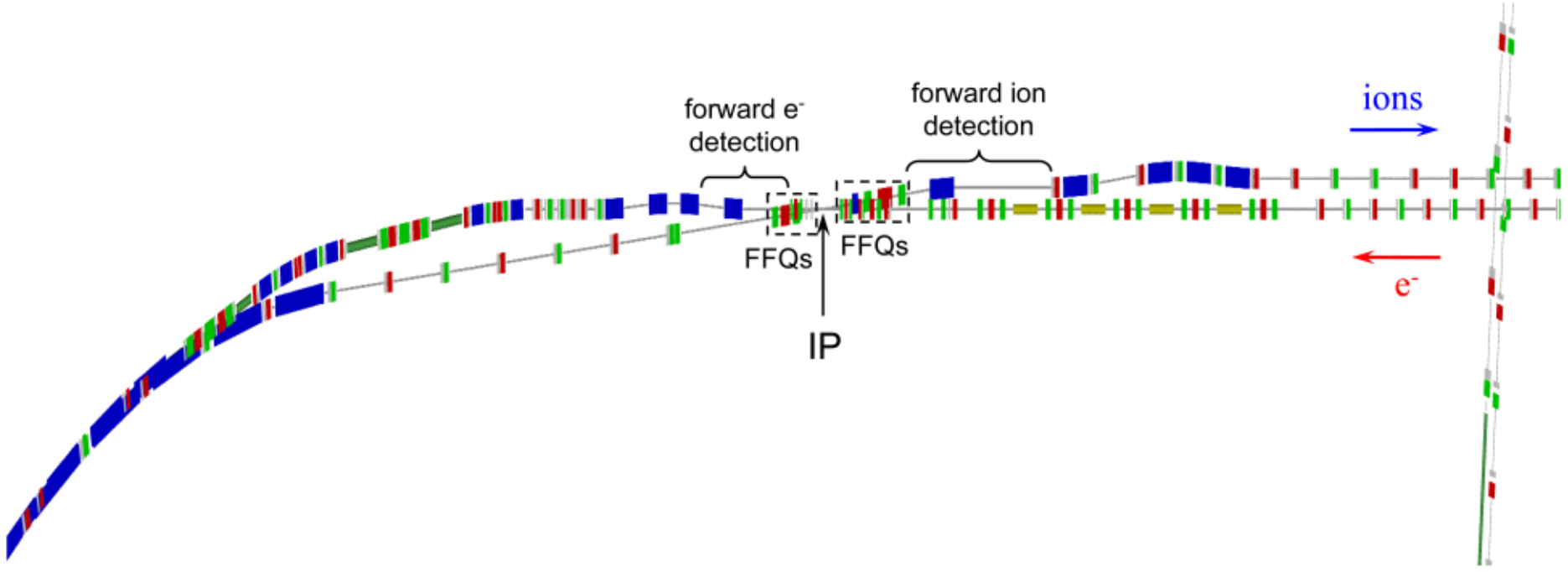

**Figure 4.7:** Detector region layout.

# Ion Injector Complex and Transfer Lines

The ion injector complex is a new facility, consisting of an ion source, a linac, and a booster along with injection/extraction transfer lines. The complex is aimed at accumulating and accelerating low current beams from the sources to reach full design current and energy for polarized proton, deuteron, and $^3$He beams (polarization of the light ion beams exceeds 70%), as well as un-polarized ion beams up to lead. The source is followed by an ion linac, configured with both normal and superconducting RF cavities to provide fast acceleration of ion beams, to minimize space-charge effects and to improve current and emittance of the beam injected into the booster. The booster configured as a figure-8 synchrotron, provides the next stage of acceleration. The ring was carefully designed, so that crossing of transition energy will not be allowed for any ion species during acceleration, so that particle loss will be minimized.  In order to achieve the designed beam parameters, the ions need to go through multiple stages of acceleration and accumulation. The evolution of the beam parameters throughout the ion complex is shown in Table 5.1 for protons and Table 5.2 for lead ions.

**Table 5.1:** Evolution of polarized proton beam in the ion complex

| | (units) | *ABPIS Source | Linac | Booster | |
|---|---|---|---|---|---|
| | | | Entrance | Injection | Extraction |
| Charge status | | $H^-$ | $H^-$ | $H^- \rightarrow H^+$ | $H^+$ |
| Kinetic energy | MeV | ~0 | 13.2 | 285 | 7062 |
| $\gamma$ | | | | 1.3 | 8.52 |
| $\beta$ | | | | 0.64 | 0.993 |
| Pulse current | mA | 2 | 2 | 2 | |
| Pulse length | ms | 0.5 | 0.5 | 0.22 | |
| Charge per pulse | μC | 1 | 1 | 0.44 | |
| Protons per pulse | $10^{12}$ | 3.05 | 3.05 | 2.75 | |
| Pulses | | | | 1 | |

**Table 5.2:** Evolution of un-polarized Pb beam in the ion complex

| | (units) | *EBIS Source | Linac | Booster | |
|---|---|---|---|---|---|
| | | | After stripper | Injection | After acceleration |
| Charge status | | $^{208}Pb^{30+}$ | $^{208}Pb^{67+}$ | $^{208}Pb^{67+}$ | $^{208}Pb^{67+}$ |
| Kinetic energy | MeV/u | ~0 | 13.2 | 100 | 670 |
| $\gamma$ | | | | 1.11 | 1.71 |
| $\beta$ | | | | 0.43 | 0.83 |
| Pulse current | mA | 1.3 | 0.1 | | |
| Pulse length | ms | 0.01 | 0.01 | | |
| Charge per pulse | μC | 0.075 | 0.015 | | |
| Ions per pulse | $10^{10}$ | 1.0 | 0.2 | | |
| Pulses | | | | 28 | |

For the remainder of this paragraph we will describe all three components of the ion injection complex.

## Ion Sources

All required ion species will be generated by two sources, namely, an Atomic Beam Polarized Ion Source (ABPIS) [Clegg1995] for polarized or non-polarized light ions and an Electron Beam Ion Source (EBIS) [Alessi2006] for un-polarized heavy ions up to lead. A third type, an Electron Cyclotron Resonance ion source (ECR) [Geller1990], could also be a candidate for heavy ion production. All ions will be extracted in pulses of many micro bunches from the sources. Detailed parameters and time structure of the proton and lead ion beam options are summarized in Tables 5.1 and 5.2.

## Ion Linac

The ion pulses from the source will be accelerated (to 285 MeV for protons and 112 MeV/u for lead ions) by a linac consisting of both warm and cold RF cavities. Depending on the ion species, a certain amount of electrons will be stripped out in the linac. A complete technical design of a SRF ion linac had been developed at Argonne National Laboratory as a heavy ion accelerator for FRIB [Mustapha2007]. The linac is very effective in accelerating a wide variety of ions from proton $H^-$ (280 MeV) to lead ion $^{208}Pb^{67+}$ (100 MeV/u). Economic acceleration of lead ions (to 100 MeV/u) requires a stripper in the linac with an optimum stripping energy of 13 MeV/u.

**Table 5.3:** Ion linac parameters

| **Ion species: p to Pb** | |
|---|---|
| Ion species for the reference design | $^{208}Pb$ |
| Kinetic energy (p, Pb) | 285 MeV<br>100 MeV/u |
| Maximum pulse current: Light ions (A/Q<3)<br>Heavy ions (A/Q>3) | 2 mA<br>0.5 mA |
| Pulse repetition rate | up to 10 Hz |
| Pulse length: Light ions (A/Q<3)<br>Heavy ions (A/Q>3) | 0.50 ms<br>0.25 ms |
| Maximum beam pulsed power | 680 kW |
| Fundamental frequency | 115 MHz |
| Total length | 121 m |

The ion linac includes a room temperature RFQ and an inter-digital IH structure operating at a fixed velocity profile, similar to the CERN lead-ion linac [Haseroth1996] and to the BNL pulsed heavy-ion injector [Alessi2006]. The initial linac section provides 4.8 MeV/u energy for all ion species, and is considered highly effective, especially for pulsed machines. The ion beams will be subsequently accelerated by the SRF section of the linac, which consists of two types of accelerating cavities to cover velocity range from 0.1 to 0.5 of the speed of light [Shepard2003]. The quarter wave resonator (QWR) and half wave resonator (HWR) have been developed for the next generation heavy ion driver linacs [Mustapha2007]. The basic linac parameters are listed in Table 5.3.

## Booster Ring

The linac pulses of 285 MeV $H^-$ will be injected into the booster, utilizing multi-turn injection with combined longitudinal and transverse painting and the charge exchange mechanism (for $H^-$ and $D^-$ ions). The resulting coasting beam is captured with the first harmonic RF, and then accelerated to 8 GeV (for protons; the corresponding energies for ions are scaled by the mass-to-charge ratio to preserve magnetic rigidity). The proposed design of the injection line matched to the booster is illustrated in Figure 5.1.

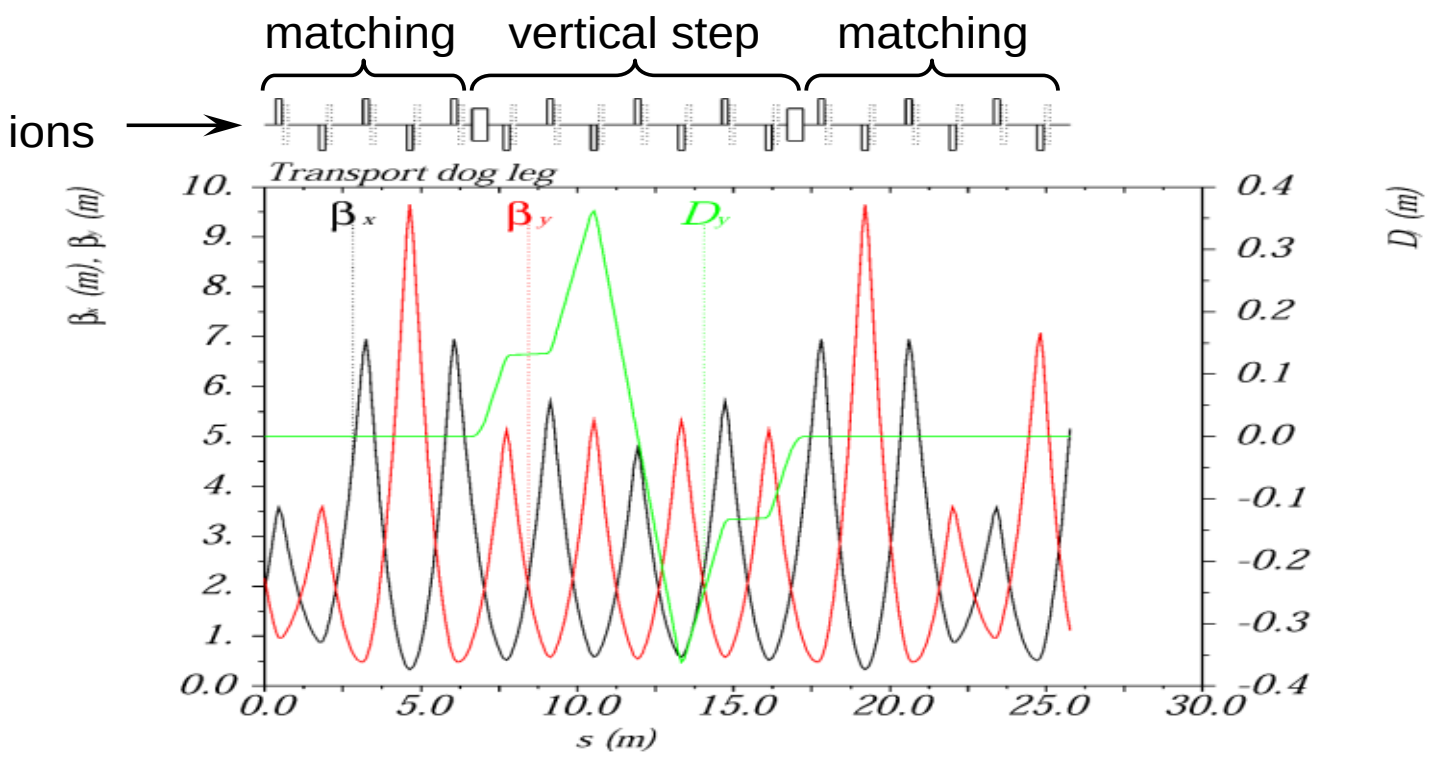

**Figure 5.1:** Lattice for the linac-to-booster transfer line, featuring a vertical step achromat.

The figure-8 ring features two $255^0$ arcs connected by two dispersion free straights (total circumference of 273 m), as illustrated in Figure 5.2.

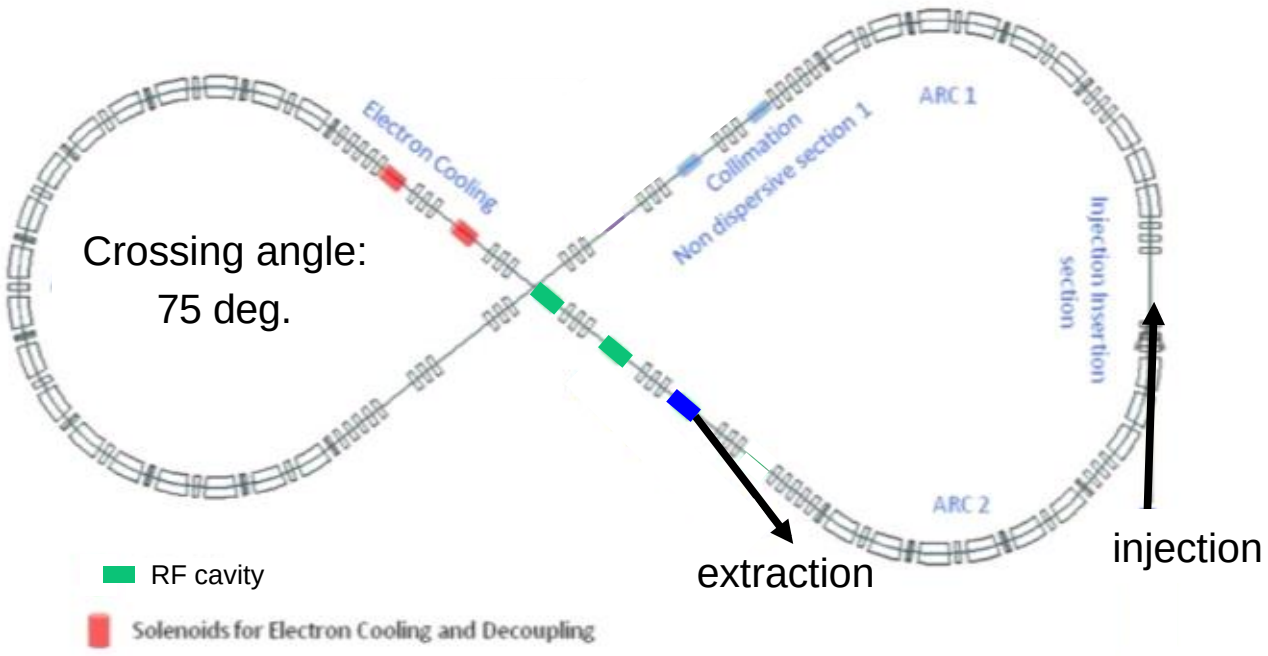

**Figure 5.2:** Layout of the booster, indicating injection and extraction.

Both achromatic arcs were carefully designed, to avoid transition crossing for all ion species during the course of acceleration. The arc optics, as shown in Figure 5.3, is based on a perturbed $120^0$ FODO, where the horizontal dispersion is driven partly negative for the inward bending arc to minimize momentum compaction (the net momentum compaction of 273 cm yields the transition gamma of 10).

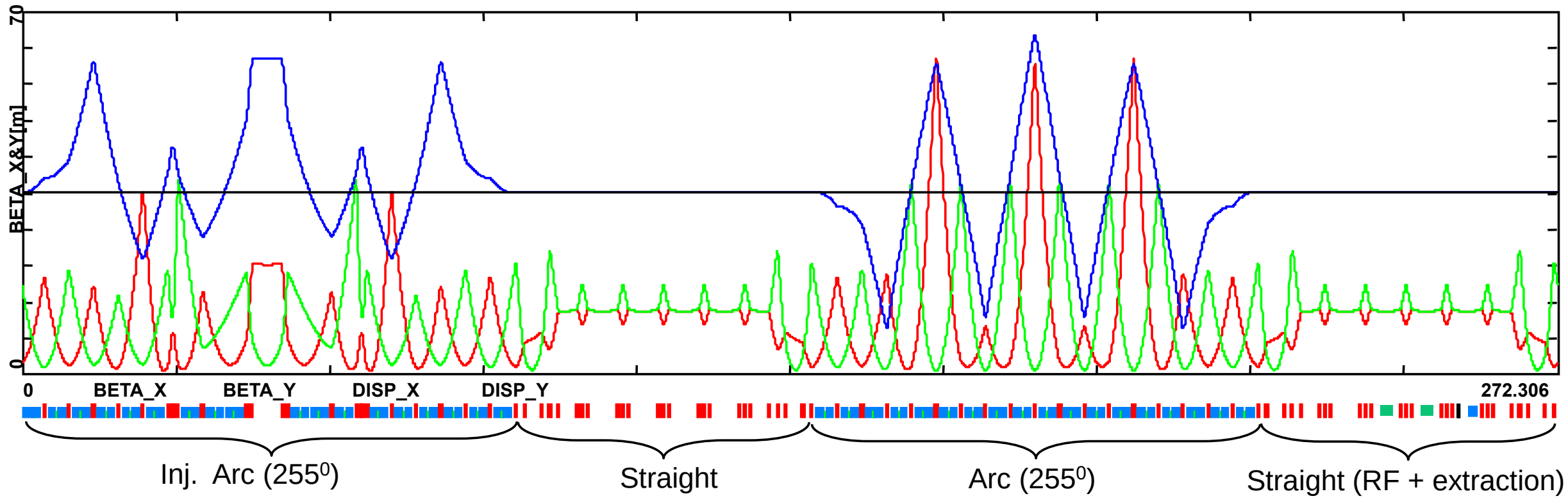

**Figure 5.3:** Complete booster lattice based on low momentum compaction optics and high dispersion injection insert. The circumference of the booster is one eighth of the ion ring.

Both straights are configured with the triplet lattice featuring long drifts (5 m) to accommodate RF cavities, extraction and electron cooling inserts. Extraction from the booster is achieved by a single magnetic kicker and an outward bending septum, followed by a long mirror-symmetric arc. The booster-to-ion ring transfer line ends with an identical septum-kicker pair, as illustrated in Figure 5.4.

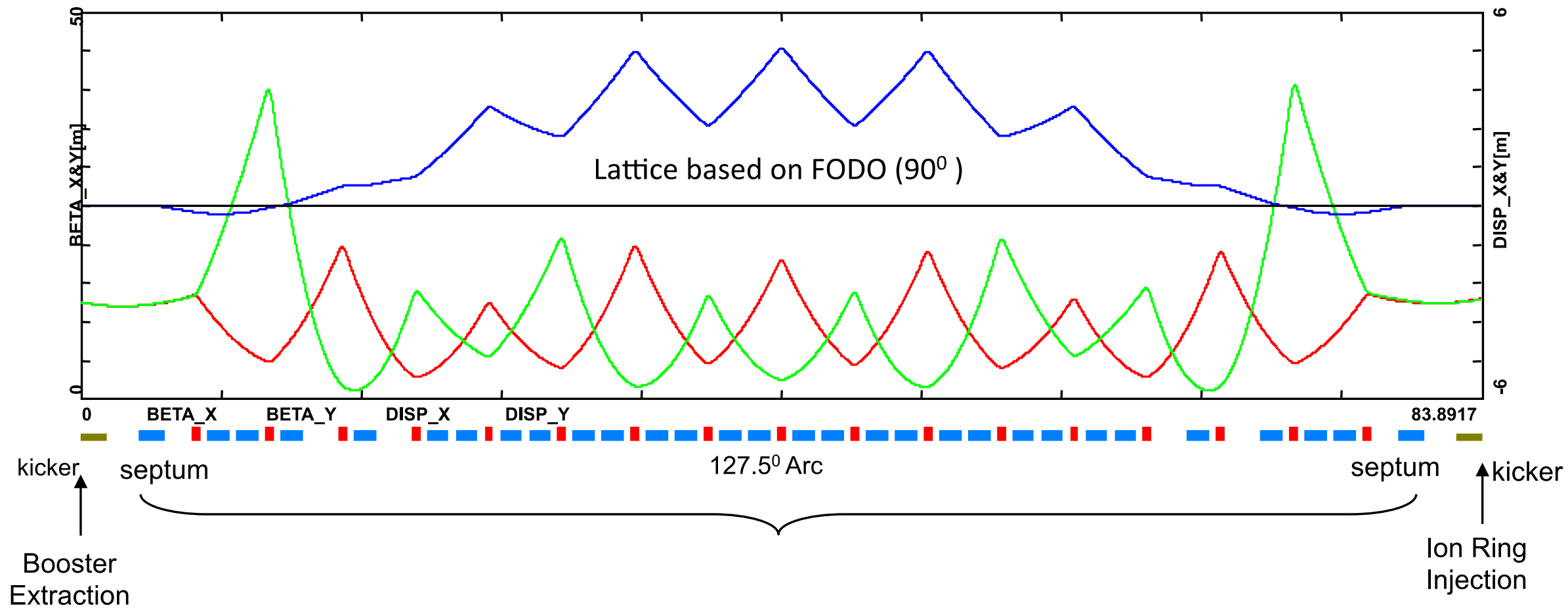

**Figure 5.4:** Lattice for the booster-to-ion ring transfer line, featuring a mirror symmetric arc achromat, symmetric extraction, and injection kicker-septum pairs.

## CEBAF Injector, Transfer Line, and Electron Collider Ring

The electron complex of MEIC is designed to accumulate and store a high-current polarized electron beam for collisions with an ion beam. Detailed design requirements [Abeyratne 2012] are:

- The electron collider ring should use warm magnets.
- The electron collider ring should accommodate electrons in an energy range of 3 to 10 GeV.
- The stored beam current can reach 3 A up to an energy above 6 GeV.
- The stored beam should have a short bunch length (~1 cm) and small transverse emittances over a wide energy range to support the luminosity requirement.
- The maximum linear density of synchrotron radiation power should not exceed 10 kW/m and the total power should be less than 10 MW.
- The stored beam polarization should be 70% or above with a reasonably long lifetime.
- The polarization should be longitudinal at collision points.

In order to reduce the cost and engineering effort needed to bring the project to fruition, we will reuse PEP-II components, such as magnets, vacuum chambers, power supplies, RF system, etc. The following three sections describe each subsystem of the electron complex: CEBAF as a full energy injector, the transfer line, and the electron collider ring. PEP-II hardware parameters were adopted and achieve the design goal.

## CEBAF Injector

The MEIC electron collider ring utilizes the CEBAF 5.5 pass SRF recirculating linac as a full energy injector with an energy range of 3 to beyond 10 GeV. No further upgrade is needed beyond the CEBAF 12 GeV upgrade in terms of beam current and polarization (>85%). In the baseline design, the PEP-II 476 MHz RF system will be reused in the electron collider ring as discussed in the Section 10. To synchronize the RF buckets between CEBAF whose RF frequency is 1497 MHz and the electron collider, we note that 7/22 of the CEBAF linac frequency is 476.3 MHz. This frequency is well within the operational range of the PEP-II 476 MHz cavities and klystrons.

Considering the electron polarization design with two polarization states coexisting in the electron collider, two long, oppositely polarized bunch trains are injected into the collider with gaps in between for beam abort, ion cleaning, turning on/off the injection kicker, and detector response to different polarizations. To synchronize the two bunch trains between the CEBAF and the electron collider, the harmonic number of the ring is chosen to be an even number and a multiple of 7. Figure 6.1 shows the time structure of the 476.3 MHz polarized electron beam injection at 6 GeV. The harmonic number of the ring is 3416 and the revolution time is 7.17 $\mu$s. The length of each polarization bunch train is 3.233 $\mu$s, or 1540 RF buckets. The length of each gap is 353 ns, which is about 5% of the ring circumference or 168 times the 476 MHz ring RF buckets.

The polarized electron source is a photoelectron gun operating at 68.05 MHz repetition rate (1/7 of the e-ring and 1/22 of the CEBAF SRF frequencies), which synchronizes with both CEBAF and the electron collider ring and allows an injection to one of every 7 ring RF buckets each time. The first bunch train with an up-polarization is injected to the 1st of every 7 ring buckets in the first half ring, which is about 3.2 $\mu$s and contains 220 bunches. A same second bunch train but with a down-polarization is injected to the 1st of every 7 ring buckets in the second half ring. The time interval between the two bunch trains is 72 $\mu$s, which leaves enough time for the source to change the laser helicity and flip the electron polarization. The injection idles for 12-700 msec, about two damping times depending on the beam energy, after these two bunch train injections to allow the beam to damp to the closed orbit. Injection is then resumed in the next mid-cycle for the 2nd of every 7 e-ring buckets. The synchronization can be achieved by slightly reducing the CEBAF RF frequency and the gun drive clock so that the timing in the ring will be one bucket ahead after one mid-cycle. The change of frequency is very small with a range of 3×10-9 to 2×10-7 at various energies. After 7 mid-cycles, the gun trigger needs to jump one 68.05MHz cycle ahead to start with the 1st of every 7 ring buckets again. The injection cycles repeat until the ring is filled to the required beam current.

Considering a future MEIC upgrade, such an injection scheme also works for a 952.6 MHz RF system by doubling the repetition rate of the source gun.

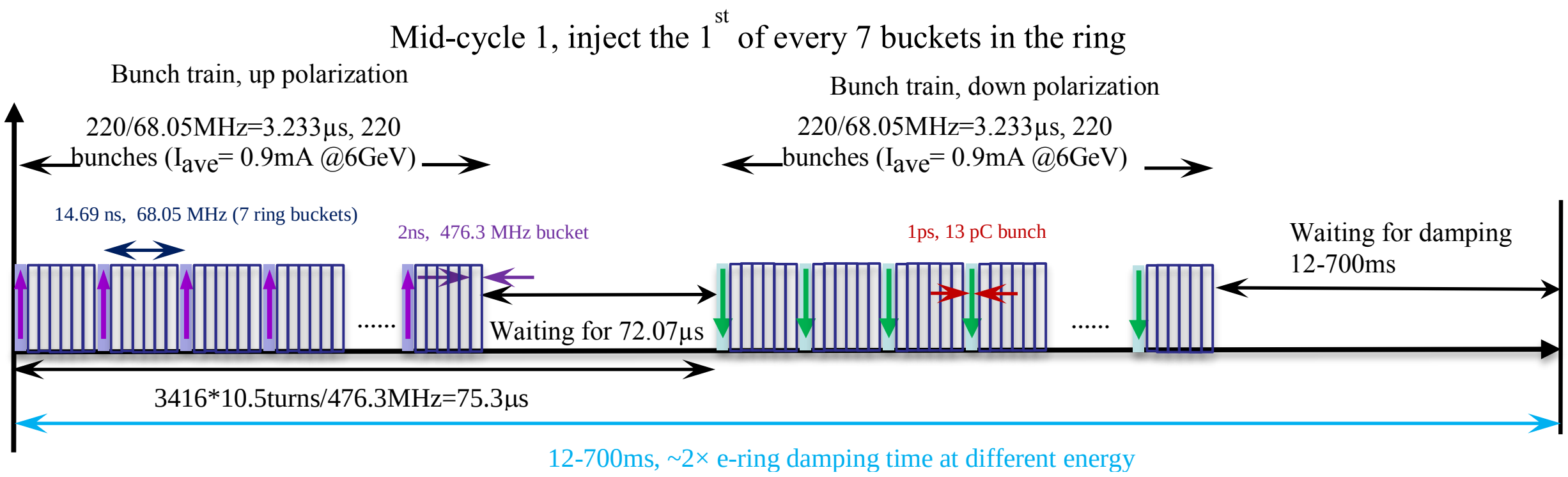

**Figure 6.1:** Time structure of the polarized electron beam injection at 6 GeV with the PEP-II 476.3 MHz RF system.

Given a stored beam current in the electron ring, the injection time depends on the CEBAF beam current, which is limited to a CW beam power of 1 MW. This beam power is ultimately determined by the power rating of the existing beam dump which is approximately matched to the linac's RF power capacity. In the proposed injection scheme, the heads and tails of the bunch trains do not overlap in the CEBAF linac. Therefore, the beam power within the bunch train can gain 1 MW during each of the 5.5 beam passes, yielding an extracted beam power of 5.5 MW total. The resulting peak current is about 0.55 mA at 10 GeV, 0.9 mA at 6 GeV, and 1.8 mA at 3 GeV. The charge per bunch with a 68.05 MHz gun repetition rate is about 8 pC at 10 GeV, 13 pC at 6 GeV, and 26 pC at 3 GeV, which are 1-2 orders of magnitude higher than the capacity of the current CEBAF gun, but not inhibitive.

The transient beam loading for such a bunch train with a peak current close to the CEBAF CW beam will cause a significant gradient droop when the linac's klystrons run at a constant power output. However, with a feedforward system, we can lower the klystrons' output during the time periods that the cavities do not see the beam to mitigate the gradient droop to an acceptable level.

The total injection time can be roughly estimated as

$$T_{inj} \approx 2\tau_d \frac{I_r E}{5.5MW}$$

where $\tau_d$ is the transverse damping time of the e-ring, $I_r$ is the designed e-ring beam current, and $E$ is the injection energy. Figure 6.2 shows the estimated injection time and designed beam current at various beam energies.

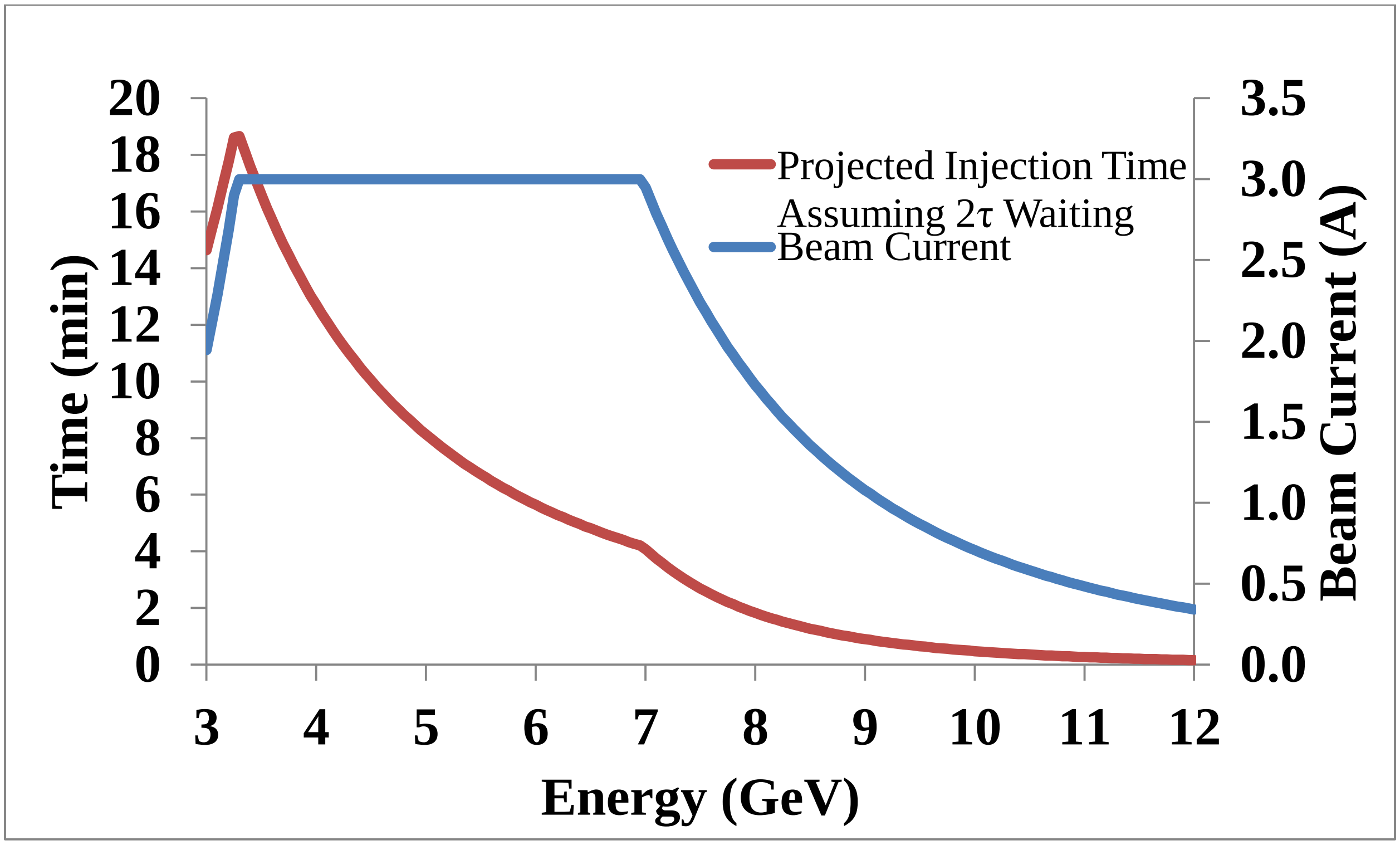

**Figure 6.2:** Estimated electron ring injection time and designed beam current at various energies.

## Transfer Line

After the last recirculation in the CEBAF accelerator, the beam is sent to a transfer line located upstream of the ramp to the experimental hall after CEBAF's north linac. The transfer line between the linac and the MEIC electron collider ring is made up of FODO cells with 120 degrees of phase advance. There are fifteen such cells, each of length 20.93 meters. We reused the PEP-II Low Energy Ring (LER) quadrupoles and dipoles to construct the transfer line. We were able to use the LER components (designed for an energy of 3.5 GeV/c) by doubling up the quadrupoles, changing the FODO structure and utilizing dipoles in groups of six. We use a total of 156 dipoles and 68 quadrupoles, which is below the total available count in the PEP-II. Dipoles are combined in two groups of six in each FODO cell. Each dipole is powered at 0.57 Tesla at 10 GeV and bends 0.44 degrees, well within the original specifications of these dipoles. The quadrupoles are used in pairs yielding a maximum individual gradient of 2.8 T/m, again well within their specifications [Henderson1995].

We chose a phase advance of 120 degrees for a number of reasons. First, it keeps the quadrupole gradients low enough that we can reuse the PEP-II quads. Secondly, it is compatible with a missing dipole scheme for dispersion suppression. Lastly, this choice for phase advance insures that there is no significant emittance growth due to synchrotron radiation during the transport from CEBAF to the MEIC electron collider ring.

The whole beamline is rendered achromatic by altering the first two and last two cells to have 8 and 4 missing dipoles, respectively. The optics of the transfer line is shown in Figure 6.3.

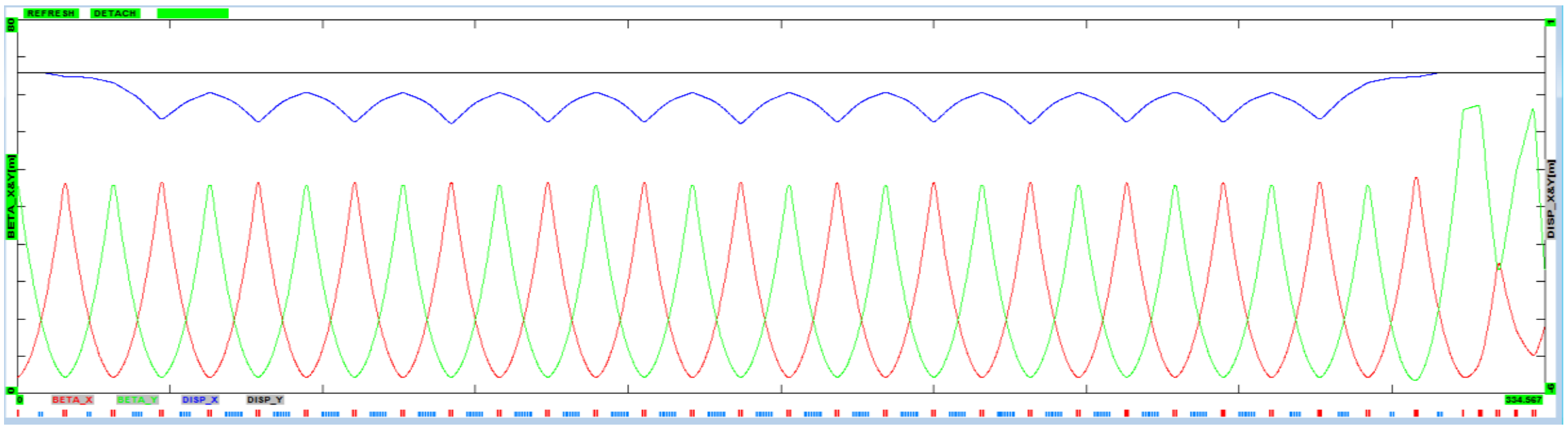

**Figure 6.3:** Transfer beam line optics between the CEBAF and the MEIC electron collider ring: beta functions (x in red, y in green) and horizontal dispersion (in blue).

The injection into the electron ring occurs at the 90 degrees point from the straight and the optical matching is accomplished by a straight section comprised of five quadrupoles. The length of the entire beam line is about 333 meters.

New vacuum chambers are considered for the dipoles in order to minimize the number of flanges. Each group of six dipoles will be strung on a single curved stainless steel vacuum chamber terminated by flanges. Pumping ports will be installed after each such section.

The LER powering scheme will be reused, which consists of utilizing a single magnet bending string with the dipole magnets connected in series by water cooled aluminum cables and powered by two 500 V, 640A power supplies regulated at 0.01%. The quadrupoles, also utilizing aluminum cables, can also be connected in two separate strings (for the focusing and defocusing quadrupoles) and powered likewise. Individual power supplies will be required for the last five quadrupoles which are part of the matching section.

A separate water cooling system for the transfer line is considered because both quads and dipoles are using hollow core aluminum conductors. We expect to be able to reuse the existing LER design (but not the components) which calls for about 1 gpm at 130 PSI.

## Electron Collider Ring

The electron and ion collider rings of the MEIC are designed to follow the same footprint and share the same underground tunnel. The electron collider determines the overall geometry, which the ion collider follows with the exception of the region around the IP due to the 50 mrad crossing angle. The dimension of the electron ring should be able to accommodate all machine components. The electron collider ring will reuse the PEP-II High Energy Ring (HER) components, such as magnets, vacuum chamber, power supplies, RF system, etc. The ring optics design will be presented broadly in the following.

The electron ring is designed as a FODO lattice in both arcs and straights, using the majority of PEP-II arc and straight dipoles and quadrupoles within the limit of their strengths. The magnet inventory of PEP-II HER is given in Table 6.1 from the SuperB CDR [Bona2007]. Particular machine blocks, such as spin rotators, interaction regions and RF sections etc., are designed as modules using new dipoles and quadrupoles, inserted and matched into the base line lattice.

Each arc has 34 normal FODO cells and 8 matching cells (4 in each end of the arc). The normal FODO cell is 15.2 m long and filled with two 5.4 m-long PEP-II dipoles and two 0.56 m-long PEP-II arc quadrupoles. Each quadrupole is followed by a PEP-II sextupole for chromatic compensation in the arcs. Though the dipole field strength of 0.3 T at 10 GeV in the MEIC is higher than the 0.27 T shown in Table 6.1, PEP-II dipoles can reach 0.363 T because they were originally designed for the PEP 18 GeV electron beam. For the MEIC application, each dipole has a 3.3 cm sagitta based on a bending angle of 2.8°. This sagitta is 1.1 cm larger than the PEP-II dipole sagitta of 2.2 cm, but it is still within the dipole good field region of 5 cm. The phase advances in the horizontal and vertical planes are chosen to be 108° and 90° so that the field gradients of quadrupoles are within their maximum specification of 16.96 T/m at 10 GeV and the third-order geometrical aberrations generated by sextupoles are cancelled within an achromat, which consists of 5 cells in the horizontal plane and 2 in the vertical plane. The matching cell has the same length and magnets of the normal FODO cell, except that the quadrupoles are adjusted to match the optics between the normal FODO cell and the Universal Spin Rotator (USR). The spin rotator is designed with interleaved solenoids and dipoles, and quadrupoles in between for the optics, to rotate the electron polarization between the vertical (in arcs) and longitudinal (at IP) direction from 3 to 12 GeV [Chevtsov2010]. Note that the spin rotator does not change the design orbit over the entire range of electron beam energies. The transverse orbital coupling induced by the longitudinal fields in the solenoids is neutralized by placing quadrupoles between half solenoids. In total, four such spin rotators are located in both ends of two arcs with a bending angle of 13.2° for each of them. Each spin rotator has two 2.5 m-long solenoids, two 5 m-long solenoids, six 2 m-long new dipoles and 21 new quadrupoles (11 families) in order to meet the optics requirements and save space. The optics of normal arc FODO and matching and USR section is plotted in Figure 6.4.

**Table 6.1:** Magnet inventory of the PEP-II HER.

**Table 3-38.** *PEP-II High Energy Ring magnets.*

| Dipoles (Location) | Length (m) | Aperture (mm) | Field (T) | Int. Strength (Tm) | Current (A) | Quantity |
|---|---|---|---|---|---|---|
| Arc | 5.4 | 60 | 0.27 | 1.45 | 950 | 194 |
| IR Soft bends | 2 | 150 × 100 | 0.092 | 0.184 | 170 | 6 |

| Quadrupoles (Location) | Length (m) | Aperture (mm) | Gradient (T/m) | Int. Strength (T) | Current (A) | Quantity |
|---|---|---|---|---|---|---|
| Arc | 0.56 | R 50 | 16.96 | 9.5 | 350 | 202 |
| Inj. sect. | 0.45 | R 50 | 11.11 | 5 | 200 | 4 |
| Straight sect. | 0.73 | R 50 | 17.53 | 12.8 | 350 | 81 |
| IR | 1.5 | | 6.67 | 10 | 650 | 2 |
| IR | 1.5 | | 10 | 15 | 1150 | 2 |
| Global skew | 0.3 | R 90 | 2.33 | 0.7 | 250 | 4 |
| IR skew | 0.2 | R 50 | 0.32 | 0.064 | 50 | 4 |
| IR skew | 0.3 | R 50 | 1.33 | 0.4 | 12 | 4 |

| Sextupoles (Location) | Length (m) | Aperture (mm) | Strength ($T/m^2$) | Int. Strength (T/m) | Current (A) | Quantity |
|---|---|---|---|---|---|---|
| Arc | 0.3 | R 60 | 210 | 63 | 400 | 104 |

| Correctors (Location) | Length (m) | Aperture (mm) | Field (T) | Int. Strength (Tm) | Current (A) | Quantity |
|---|---|---|---|---|---|---|
| Arc X | 0.3 | 90 × 50 | 0.018 | 0.0054 | 12 | 96 |
| Arc Y | 0.3 | 90 × 50 | 0.01 | 0.003 | 12 | 96 |
| Straight | 0.3 | R 50 | 0.012 | 0.0036 | 12 | 91 |

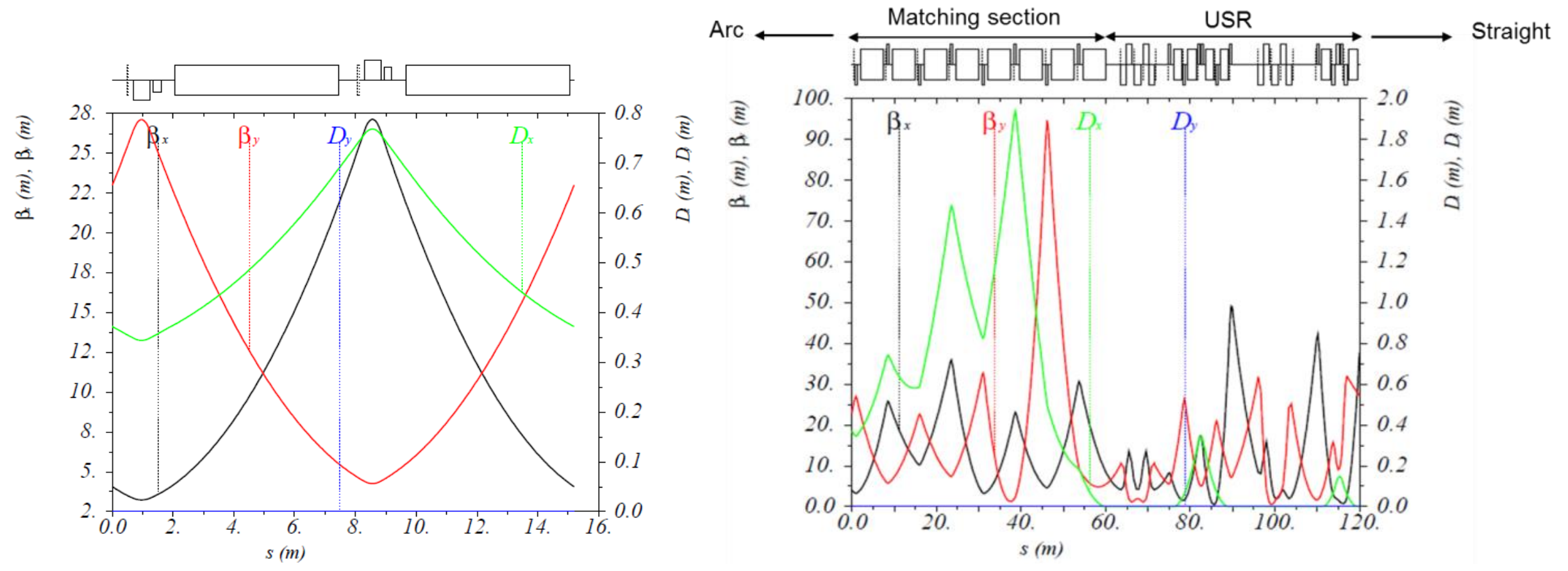

**Figure 6.4:** Optics of normal arc FODO cell (left) and matching and USR section (right).

The straight FODO cells with a 60° phase advance and matching sections between various machine blocks are designed using the 0.73 m-long PEP-II straight quadrupoles. The optics is adjusted to have most of the quads within their maximum field gradient of 17.53 T/m at 10 GeV. For those quadrupoles with higher field gradient than the maximum specification, they can be tuned down by further optimization of the optics or built as new ones.

A dedicated Chromaticity Compensation Block (CCB) has been developed to compensate the chromatic effects induced by the extra strong focusing at the IP region [Morozov2013]. In this design approach, the number of aberration conditions at the IP is greatly reduced by requiring certain symmetries of the beam orbital motion and of the dispersion and by utilizing a symmetric arrangement of quadrupoles, sextupoles and/or octupoles with respect to the center of the CCB. Such a scheme should allow simultaneous compensation of the 1st-order chromaticities and chromatic beam smear at the IP without introducing significant 2nd-order aberrations. The lattice functions are plotted in Figure 6.5. Such a CCB has two 5 m-long dipoles and four 2 m-long dipoles with a maximum field of 0.58 T at 10 GeV, 13 quadrupoles in 7 families with a maximum field gradient of 25 T/m at 10 GeV, and 4 sextupoles in 2 families for the chromaticity compensation. Except for the sextupoles from the PEP-II, all dipoles and quadrupoles are new magnets.

There are two RF sections in two straights, where the beta functions are designed to be a relatively small 5 m in order to improve the coupled beam instability thresholds. There are 15 quadrupoles in 2 families needed in each section with maximum field strength of 25 T/m at 10 GeV. The optics is plotted in Figure 6.6.

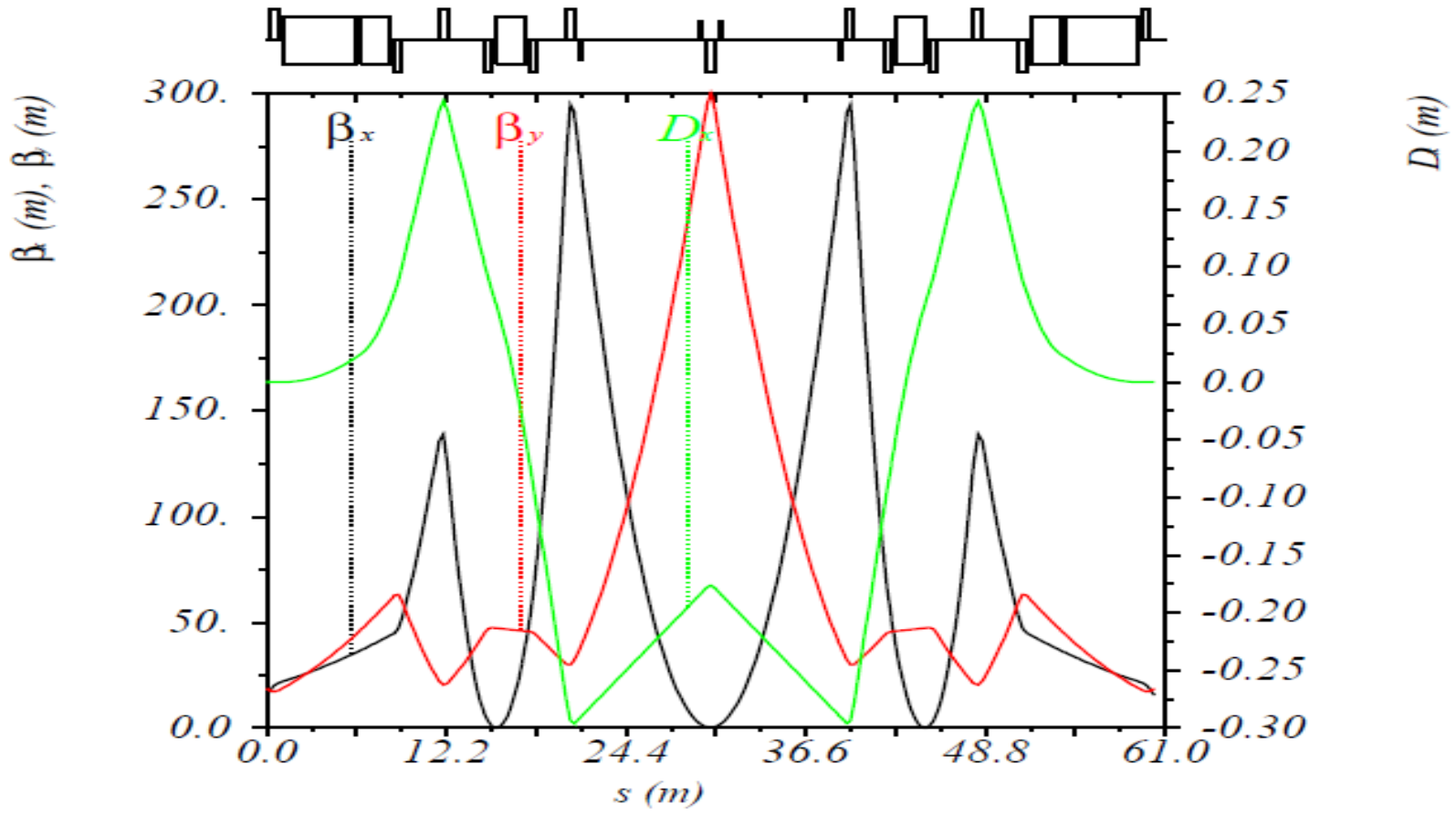

**Figure 6.5:** Optics of Chromaticity Compensation Block (CCB).

Figure 6.7 and Figure 6.8 show the optics and layout of the detector region. Details of the detector region design, including the integrated Compton polarimetry design, are discussed in the sections covering the Detector Region and Electron Polarization. Note that the electron collider ring adopts a local deflective crab-crossing scheme to restore effective head-on collisions of the two beams and preserve the luminosity.

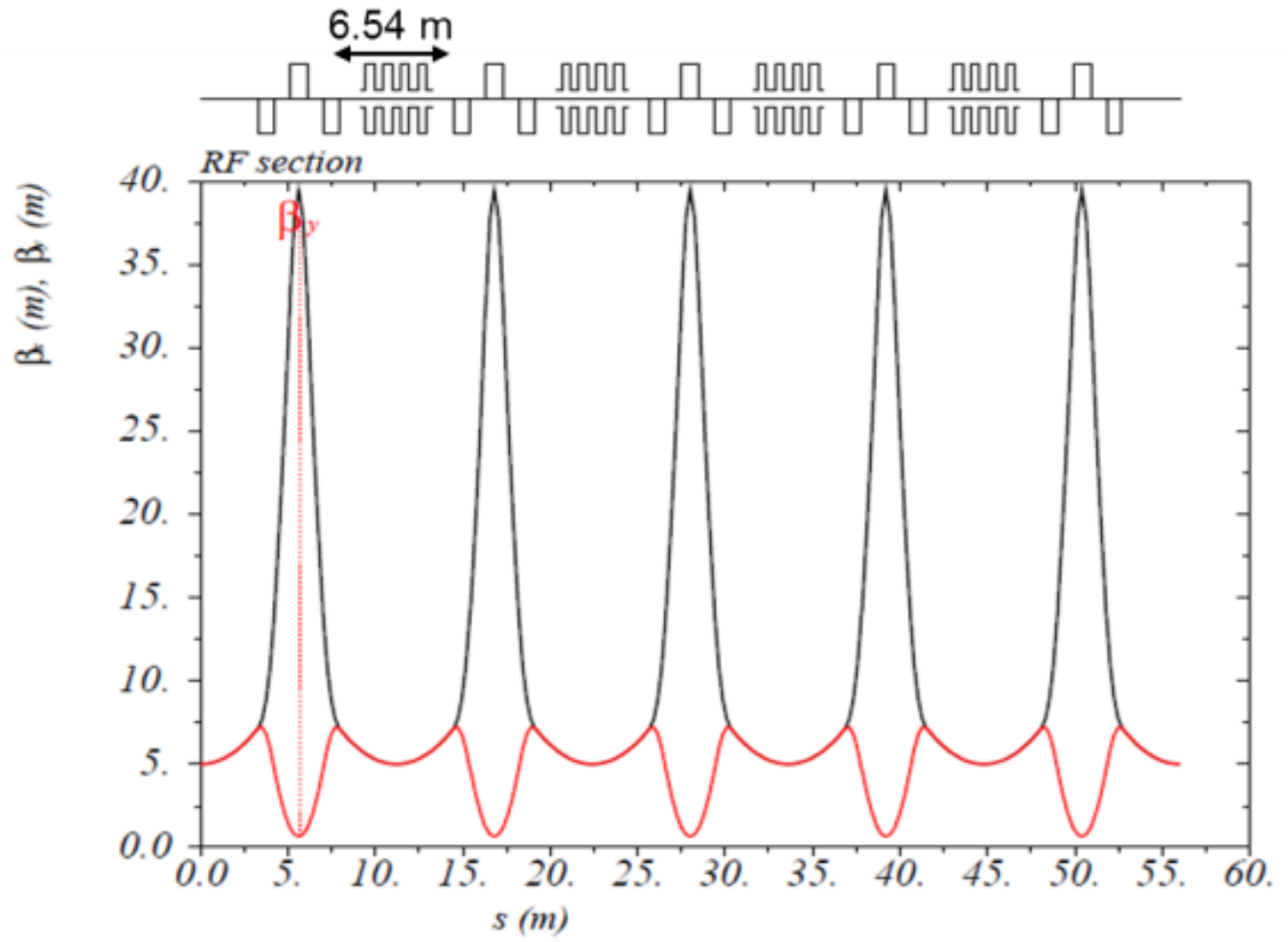

**Figure 6.6:** Optics of RF section.

The optics and geometric layout of the complete figure-8 electron collider ring with one IP are illustrated in Figure 6.9 and Figure 6.10. The circumference of the electron collider is 2154.28 m with two identical 754.84 m-long arcs and two straights of 322.31 and 322.29 m, respectively. The Figure-8 crossing angle is 81.7°. Some high level optics and synchrotron radiation parameters are listed in Table 6.2 and Table 6.3. The electron currents at high energies are determined by the total synchrotron radiation power of 10 MW within the PEP-II vacuum chamber specification.

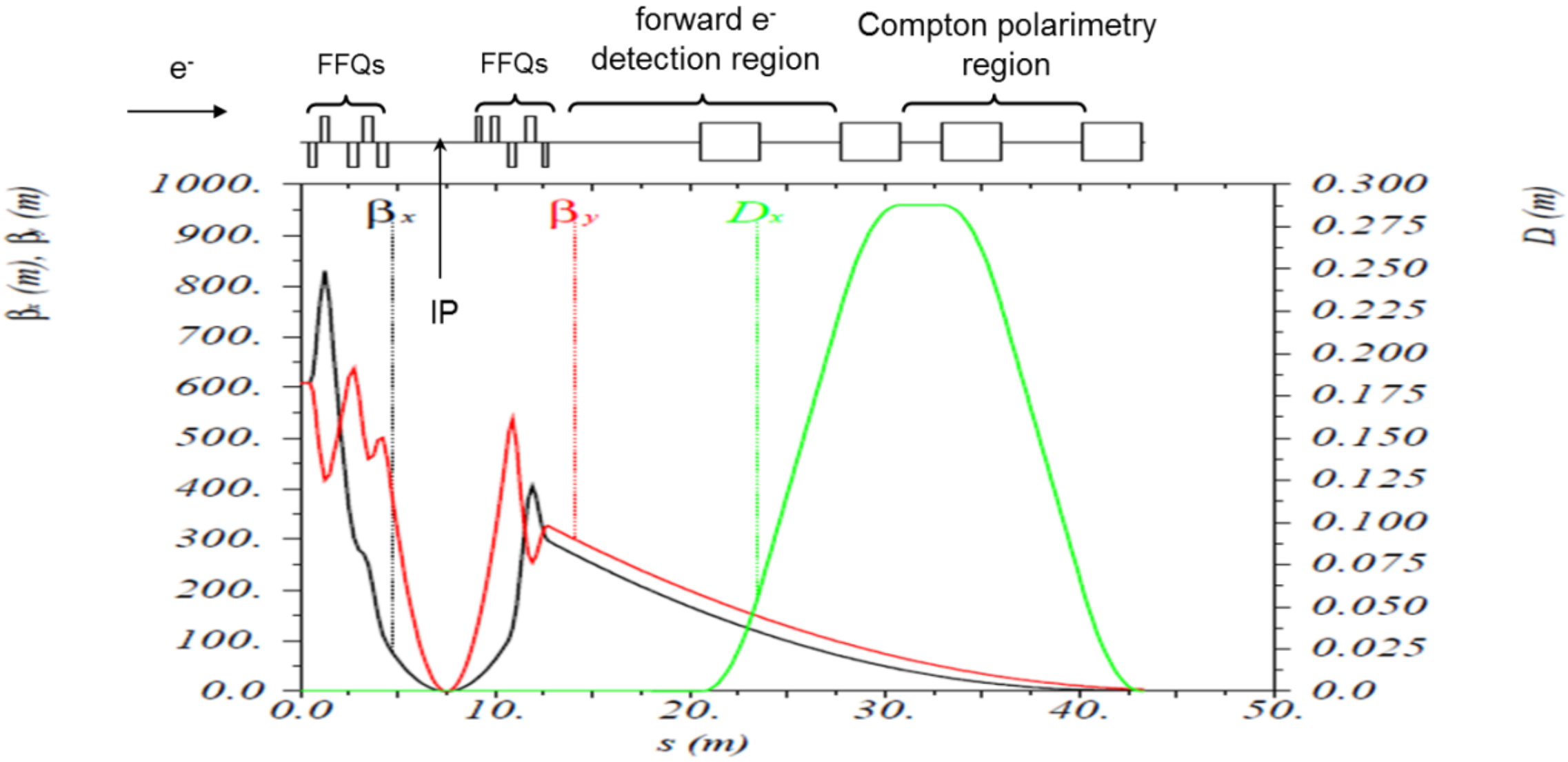

**Figure 6.7:** Optics of electron collider ring detector region.

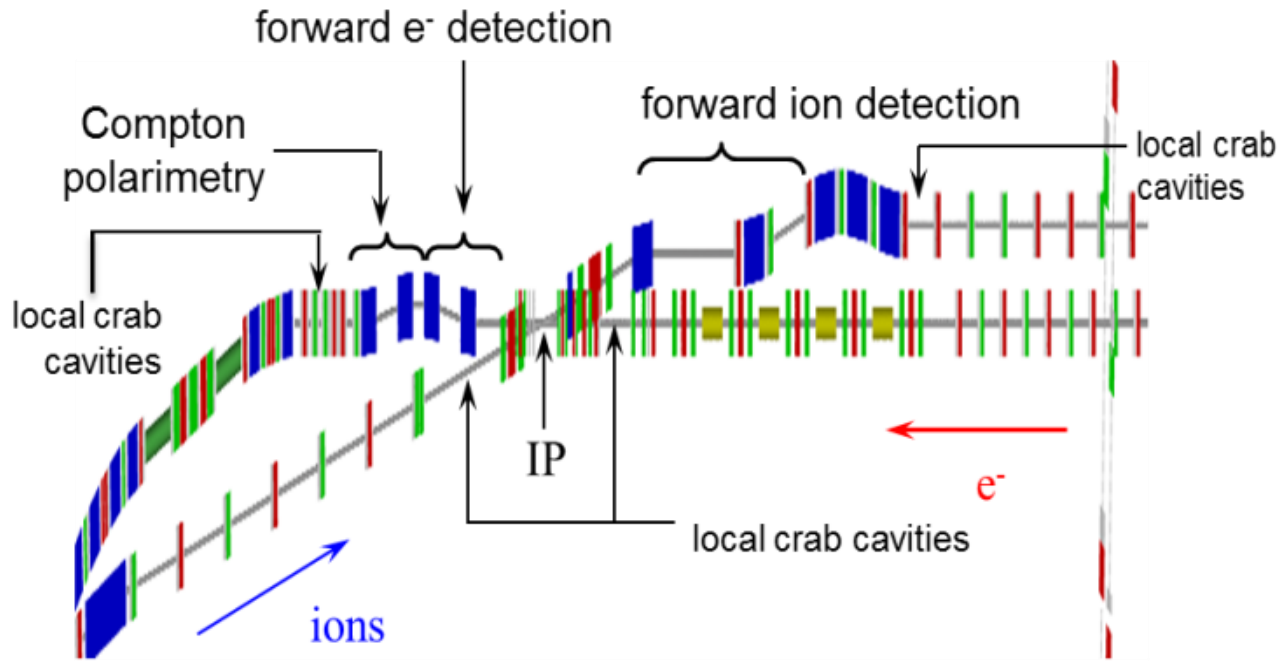

**Figure 6.8:** Detector region layout with both ion and electron beam lines.

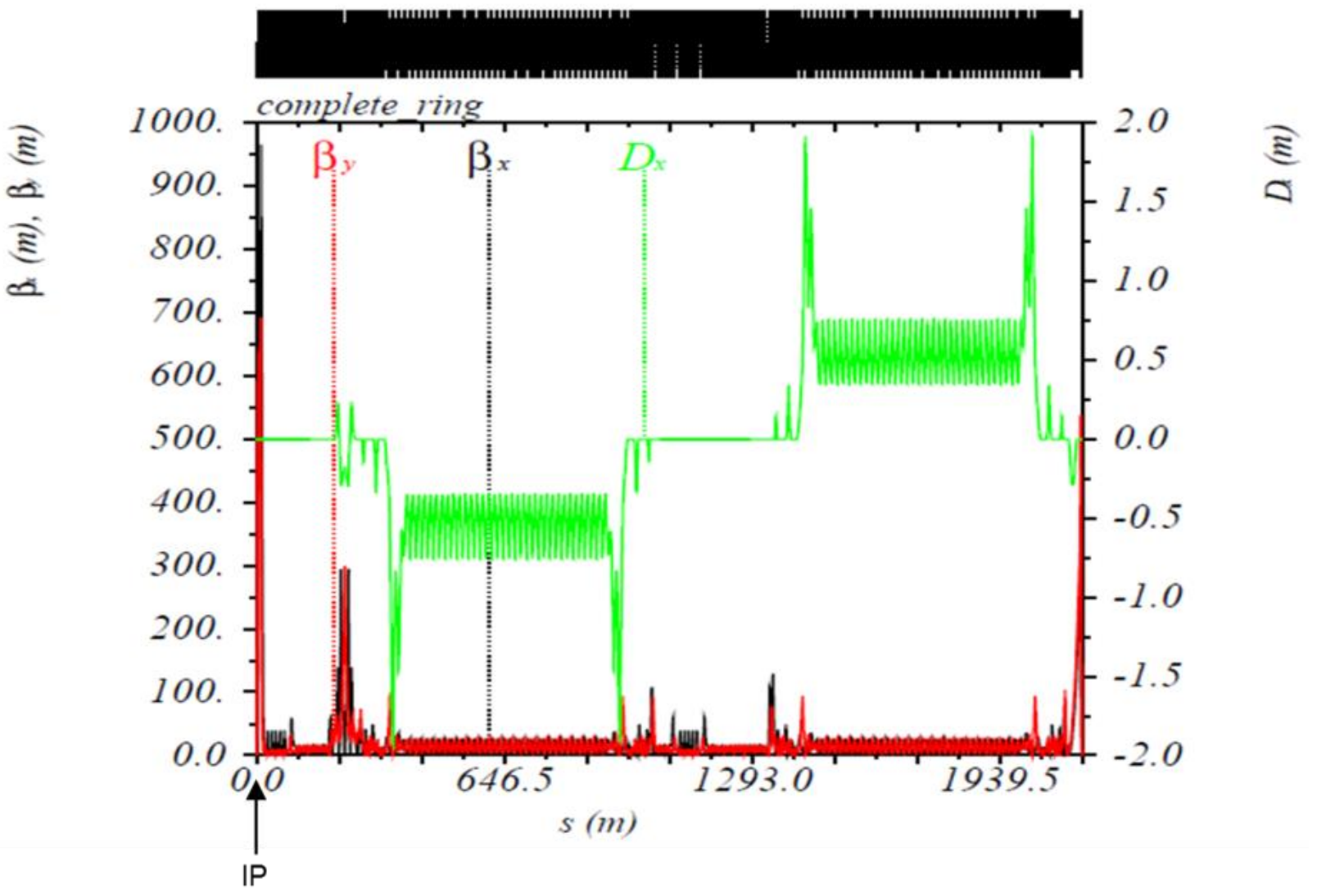

**Figure 6.9:** Optics of the complete figure-8 electron collider ring with one IP.

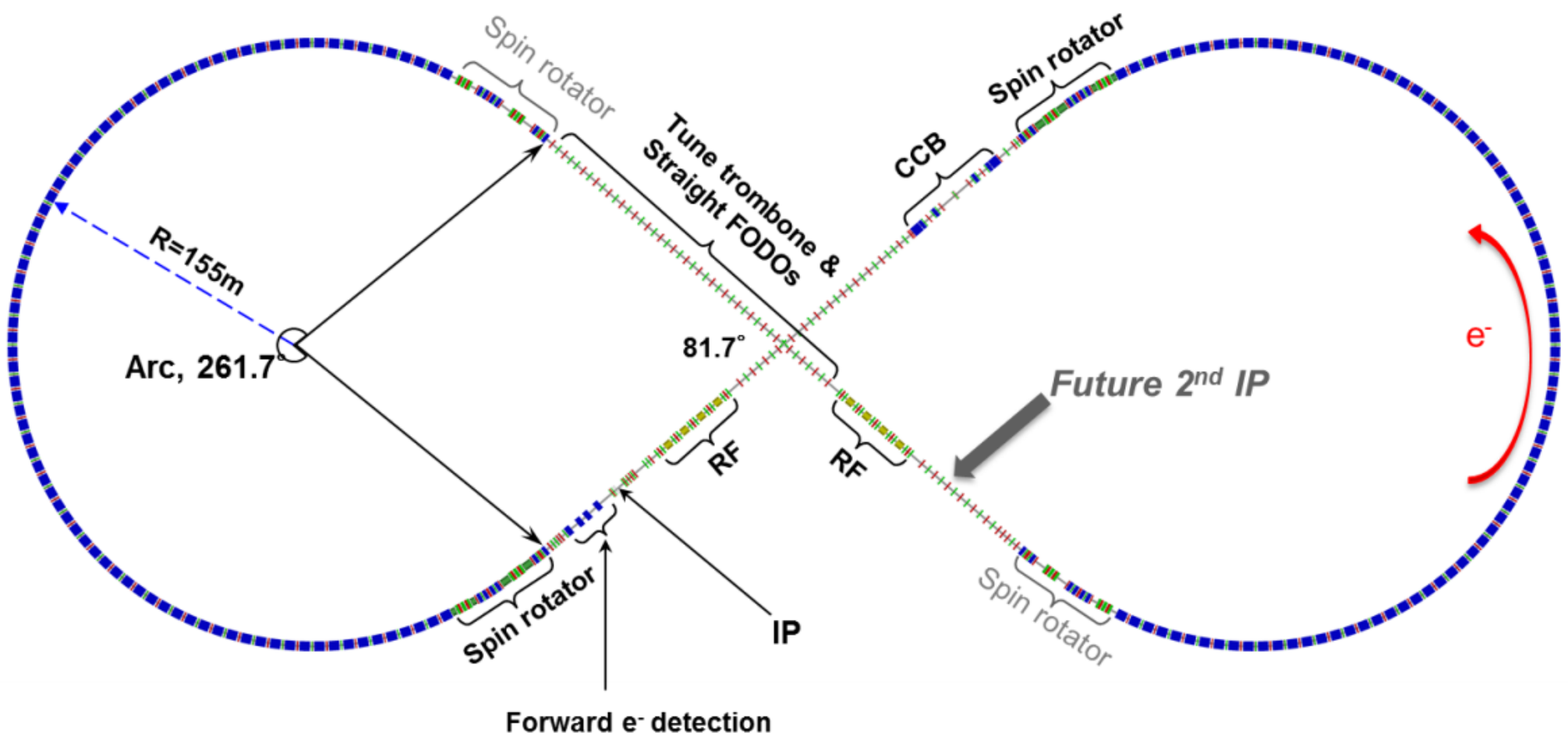

**Figure 6.10:** Geometric layout of the figure-8 electron collider ring with one IP and major machine sections.

**Table 6.2:** Electron ring optics parameters.

| | | |
|---|---|---|
| Electron beam momentum | GeV/c | 10 |
| Circumference | m | 2154.28 |
| Figure-8 crossing angle | degrees | 81.7 |
| Maximum horizontal/vertical $\beta$ functions | m | 949/692 |
| Maximum horizontal/vertical dispersion Dx,y | m | 1.9/0 |
| Horizontal/vertical betatron tunes $\nu$x,y | | 45.(89)/43.(61) |
| Horizontal/vertical chromaticities $\xi$x,y | | -149/-123 |
| Momentum compaction factor $\alpha$ | $10^{-3}$ | 2.2 |
| Transition energy $\gamma$t | | 21.6 |
| Horizontal/vertical normalized emittance $\varepsilon$x,y | $\mu$m-rad | 1093/378 |
| Maximum horizontal/vertical rms beam size $\sigma$x,y | mm | 7.3/2.1 |

**Table 6.3:** Synchrotron radiation parameters of electron beam at various energies.

| | | | | | | |
|---|---|---|---|---|---|---|
| Beam energy | GeV | 3 | 5 | 6.95 | 9.3 | 10 |
| Beam current | A | 1.4 | 3 | 3 | 0.95 | 0.71 |
| Total SR power | MW | 0.16 | 2.65 | 10 | 10 | 10 |
| Linear SR power density (arcs) | kW/m | 0.16 | 2.63 | 9.9 | 9.9 | 9.9 |
| Energy loss per turn | MeV | 0.11 | 0.88 | 3.3 | 10.6 | 14.1 |
| Energy spread | $10^{-3}$ | 0.27 | 0.46 | 0.66 | 0.82 | 0.91 |
| Transverse damping time | ms | 376 | 81 | 26 | 14 | 10 |
| Longitudinal damping time | ms | 188 | 41 | 13 | 7 | 5 |
| Normalized emittance | $\mu$m-rad | 30 | 137 | 425 | 797 | 1093 |

Table 6.4 lists the magnet inventory of the proposed MEIC electron collider ring optics design with both PEP-II and new magnets. Only the maximum strength of each magnet category is presented. As one can see, the majority of the PEP-II magnets can be reused in the electron collider ring within their original specifications. BPMs and correctors are also allocated along the electron collider ring for orbit measurement and correction.

**Table 6.4:** Magnet inventory of the MEIC electron collider ring.

| Magnet category | PEP-II magnet | | New magnet | |
|---|---|---|---|---|
| | Number | Max. Strength | Number | Max. Strength |
| Dipole | 168 | 0.3 T | 34 | 0.64 T |
| Quadrupole | 263 | 17 T/m | 151 | 25 T/m |
| Sextupole | 104 | 600 T/m2 | 32 | 600 T/m2 |
| Skew quadrupole | 12 | 2.33 T/m | | |
| BPM | | | 331 | |
| Corrector | 283 | 0.02 T | 48 | 0.02 T |

The complete MEIC electron complex layout is shown in Figure 6.11.

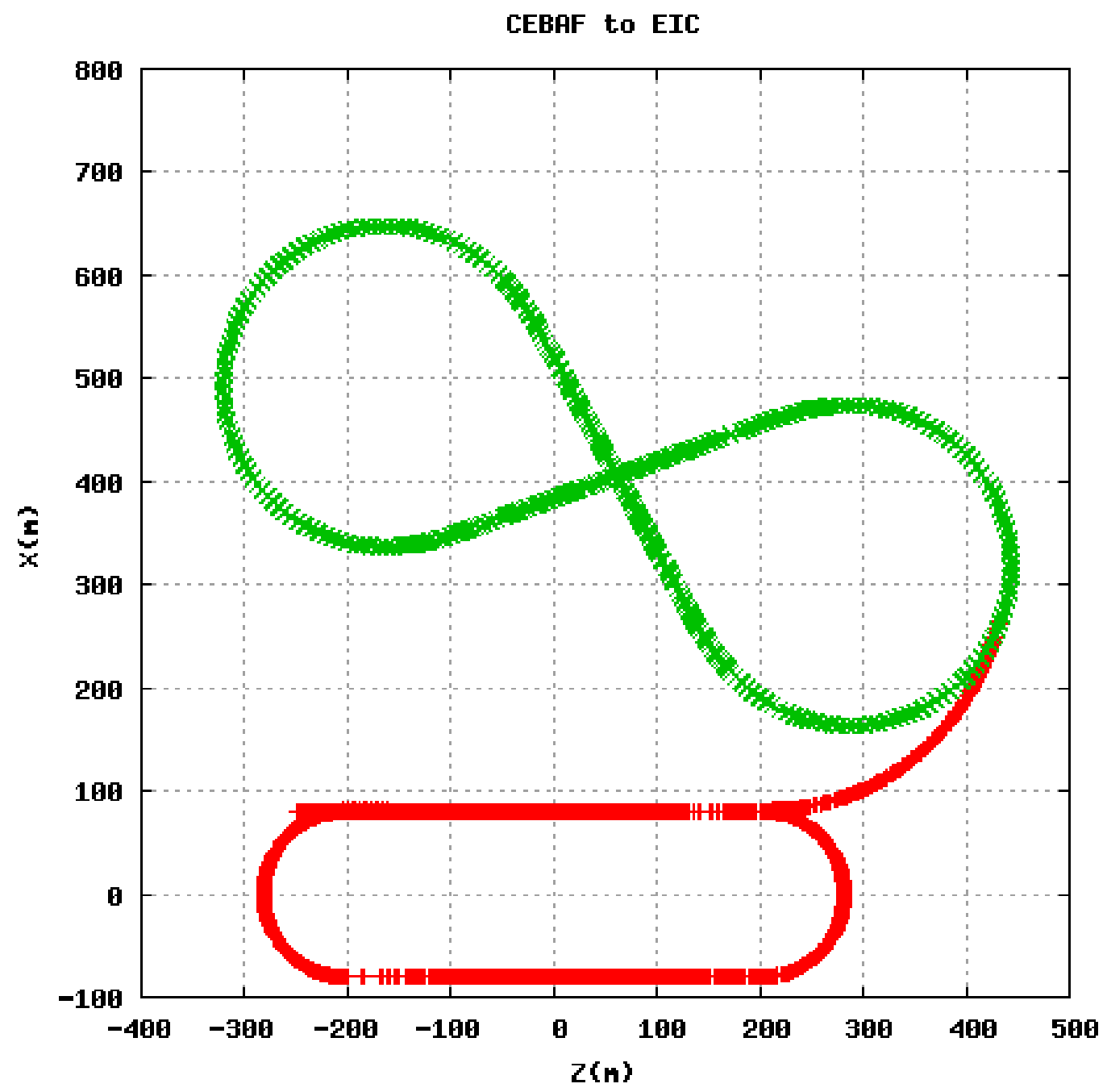

**Figure 6.11:** MEIC electron complex layout: CEBAF, transfer line, and electron collider ring.

## Ion Collider Ring

The ion collider ring accelerates protons from 9 to up to 100 GeV/c or ions in the equivalent momentum range and is designed to provide luminosity above $10^{33}$ $cm^{-2}s^{-1}$ in the momentum range from 20 to 100 GeV/c. The overall layout of the ion collider ring indicating the main components is shown in Figure 7.1. The ring consists of two 261.7° arcs connected by two straight sections intersecting at an 81.7° angle. The ion collider ring's geometry is determined by the electron collider ring [Lin2013]. The ion arcs are composed mainly of FODO cells. The last few dipoles at either end of each arc are arranged to match the geometry of the ion collider ring. One of the straights houses an interaction region and is shaped to make a 50 mrad crossing angle with the electron beam at the interaction point. The second straight is mostly filled with FODO cells, while retaining the capability of inserting a second interaction region. The overall ion ring circumference is 2153.89 m. The main building blocks of the ring are described below.

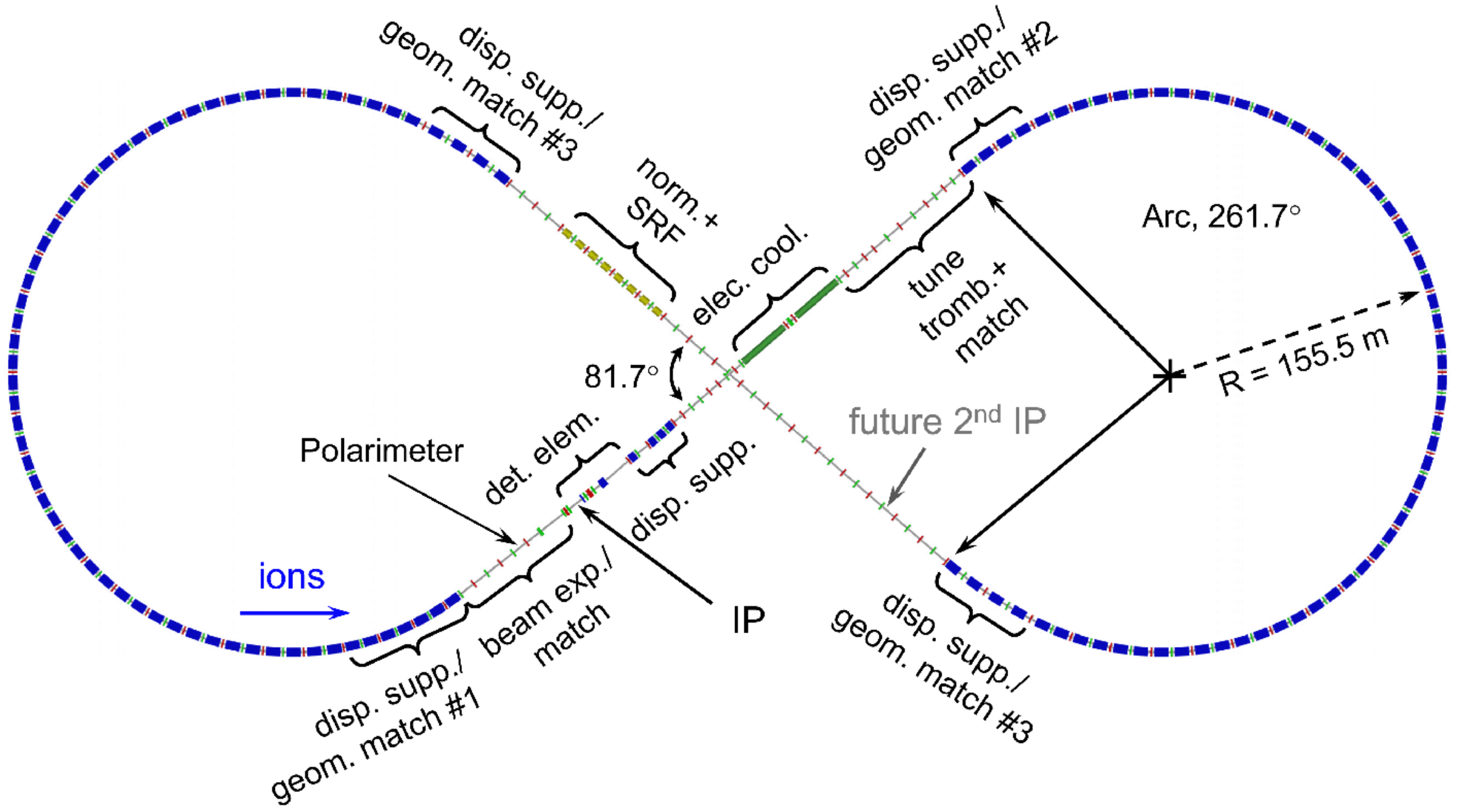

**Figure 7.1:** Layout and main components of the ion collider ring.

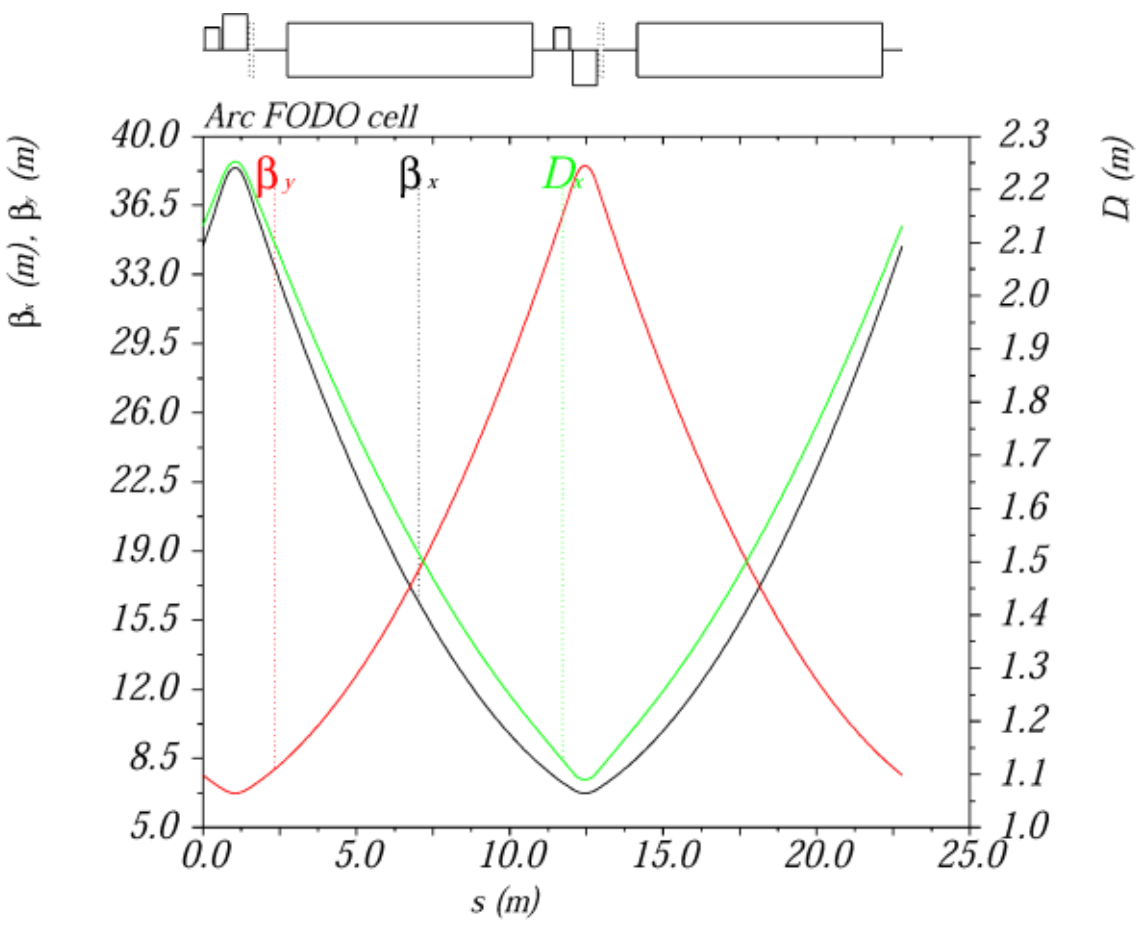

**Figure 7.2:** Ion arc FODO optics.

The main building block of the ion arcs is a FODO cell shown in Figure 7.2. It has been designed considering a balance of geometric, engineering and beam dynamical aspects. It has the same average bending radius as the electron arc. The ion FODO cell length is chosen at 22.8 m to be 1.5 times that of the electron FODO cell. Such a size allows for use of super-ferric magnets [Husen1985] up to a proton momentum of about 100 GeV/c. Each 8 m long dipole has a maximum field of about 3.06 T and bends the beam by about 4.2° with a bending radius of about 109.1 m.

The required magnet apertures are determined using a sum of a ±10 rms beam size at injection (including betatron and dispersive contributions), a ±1 cm closed orbit allowance, and, in case of dipoles, plus or minus half of the orbit sagitta. To make the dipole horizontal aperture size more manageable, each dipole is implemented as two 4 m long straight pieces reducing the sagitta to about 18 mm. The resulting required dipole good-field region is ±5 and ±3 cm in the horizontal and vertical planes, respectively. The required good-field region for all other magnets is circular with a radius of 4 cm. Sufficient space has been reserved in the lattice for magnet coil extensions: 14 cm at each end for dipoles and 5 cm at each end for most other magnets.

The maximum gradient of the 0.8 m long FODO cell quadrupoles is about 53 T/m for a 90° betatron phase advance in both planes, which is straightforward to achieve with the super-ferric technology. We place a 0.5 m long corrector package on one side and a 15 cm long BPM on the other side of each quadrupole. Each corrector package includes horizontal and vertical orbit correctors, a skew quadrupole, and higher-order multipoles. The sextupoles in the corrector packages located in the dispersive regions are used for chromaticity compensation. With the above FODO cell parameters, the dispersive and betatron components of the beam size are comparable, which provides for an efficient use of the sextupole strength for chromaticity compensation. Nominally, each focusing sextupole of a FODO cell with a 3 T field at 4 cm radius adds 34.8/-7.1 units of horizontal/vertical chromaticity. Similarly, each defocusing sextupole adds -3.7/18.1 units of horizontal/vertical chromaticity per FODO cell. This leaves a sufficient margin for exploration of various chromaticity compensation schemes [Morozov2013, Fartoukh2013]. Two arc sextupole families could be used to generate horizontal and vertical chromatic beta waves to compensate the chromatic kicks of the final focusing quadrupoles. Two additional sextupole families could be used to compensate the residual chromaticities.

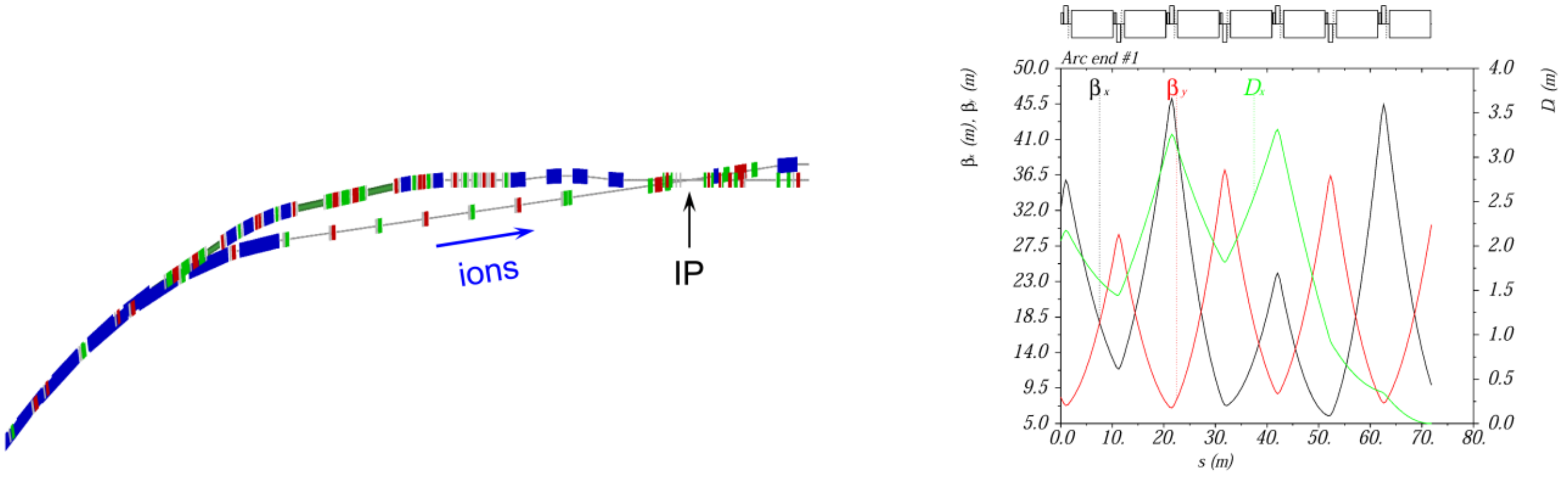

(a) (b)

**Figure 7.3:** Geometry (a) and optics (b) of the arc end upstream of the IP.

The bending angles and the spacing of the seven dipoles at the end of the arc upstream of the IP are adjusted to match the electron ring geometry and form a 50 mrad crossing angle at the IP as shown in Figure 7.3a. The quadrupole strengths in this section are adjusted to suppress the dispersion while keeping the beta functions under control as shown in Figure 7.3b. A four-dipole section at the entrance to the arc downstream of the IP is adjusted to suppress the dispersion and provide a 1.5 m separation between the ion and electron beams. Similarly, four-dipole arc-end sections near the ends of the other straight are used to suppress the dispersion and match the shape of the electron ring.

The concept of the detector region design is described in Section 4 [Lin2013,Morozov2014]. Figure 7.4 shows the detector region optics. It starts at the end of one of the arcs and consists of a matching/ beam expansion section, upstream and downstream triplet Final Focusing Blocks (FFB), a spectrometer section, a geometric match/dispersion suppression section, and a matching/ beam compression section. The matching sections contain a sufficient number of quadrupoles to control both the beta functions at the IP and the betatron phase advance for beta squeeze and chromaticity compensation. The upstream FFB is closer to the IP than the downstream FFB to minimize the total chromatic contribution while satisfying the detector requirements as discussed earlier in this document. The nominal horizontal and vertical $\beta^*$ values at the IP are 10 and 2 cm, respectively. After the downstream FFB, there is an approximately 56 mrad spectrometer dipole followed by a machine-element-free 16 m space instrumented with detector elements. The subsequent four-dipole section suppresses the dispersion generated by the spectrometer dipole as well as adjusts the ion beam to be parallel to, and separated by 1.5 m from, the electron beam. This separation matches that at the arc end downstream of the IP formed by shaping the ion arc end as described above. Such a design makes the interaction region somewhat modular and decouples from the ring geometry providing for ease of integration into the ring lattice [Lin2013].

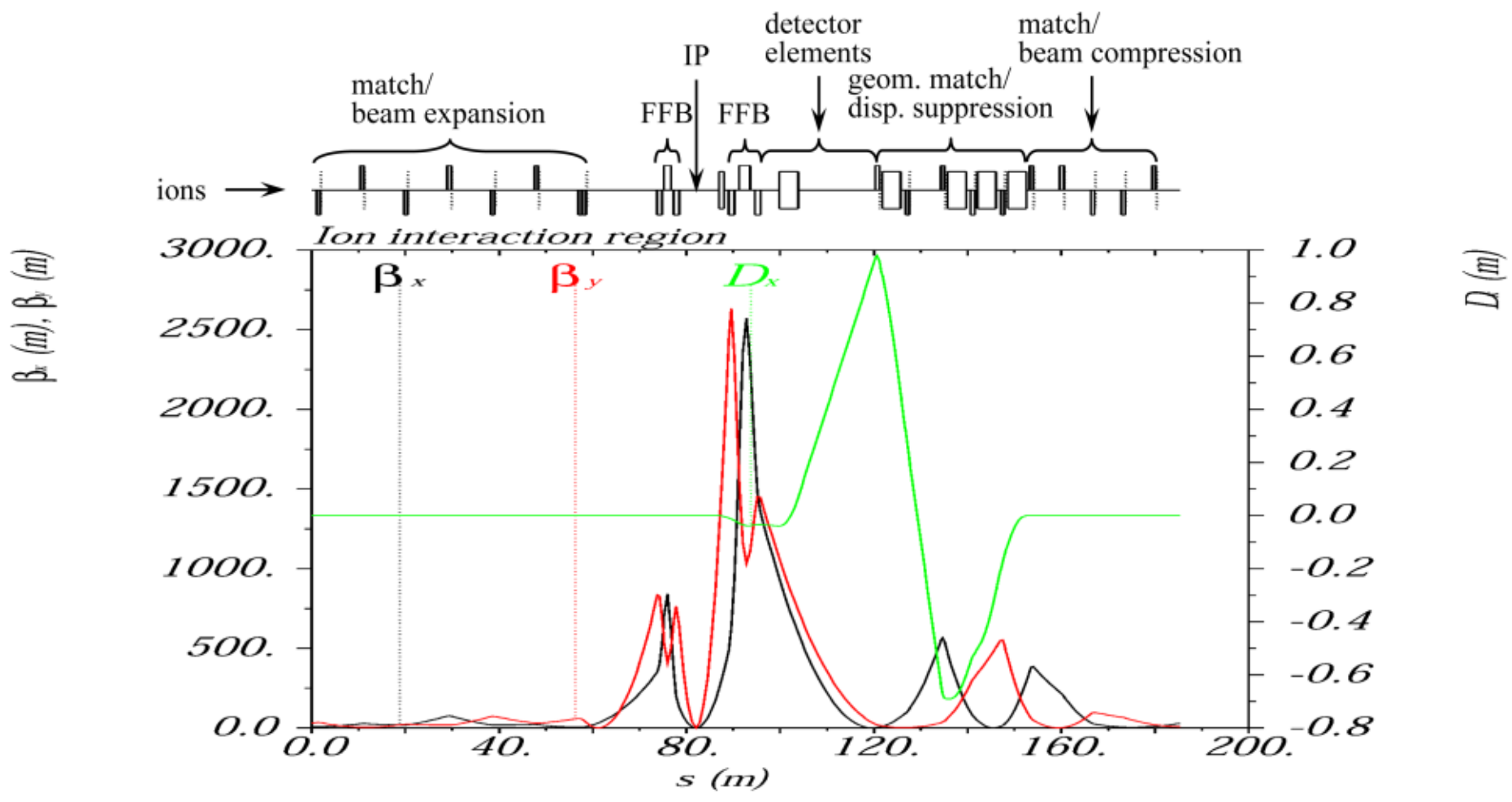

**Figure 7.4:** Ion interaction region optics.

Another major component housed in the same straight as the interaction region is the electron cooling section [Zhang2013] shown in Figure 7.5. Two 30 m long drifts have been reserved for electron cooler solenoids. We plan to alternate the solenoid fields so that their net longitudinal field integral is zero to compensate their coupling and their effect on the ion spin. Optics based on triplet focusing is used to provide such long drifts. There is a matching segment at each end of the cooling section connecting it the interaction region on one side and a straight FODO of a tune trombone on the other side.

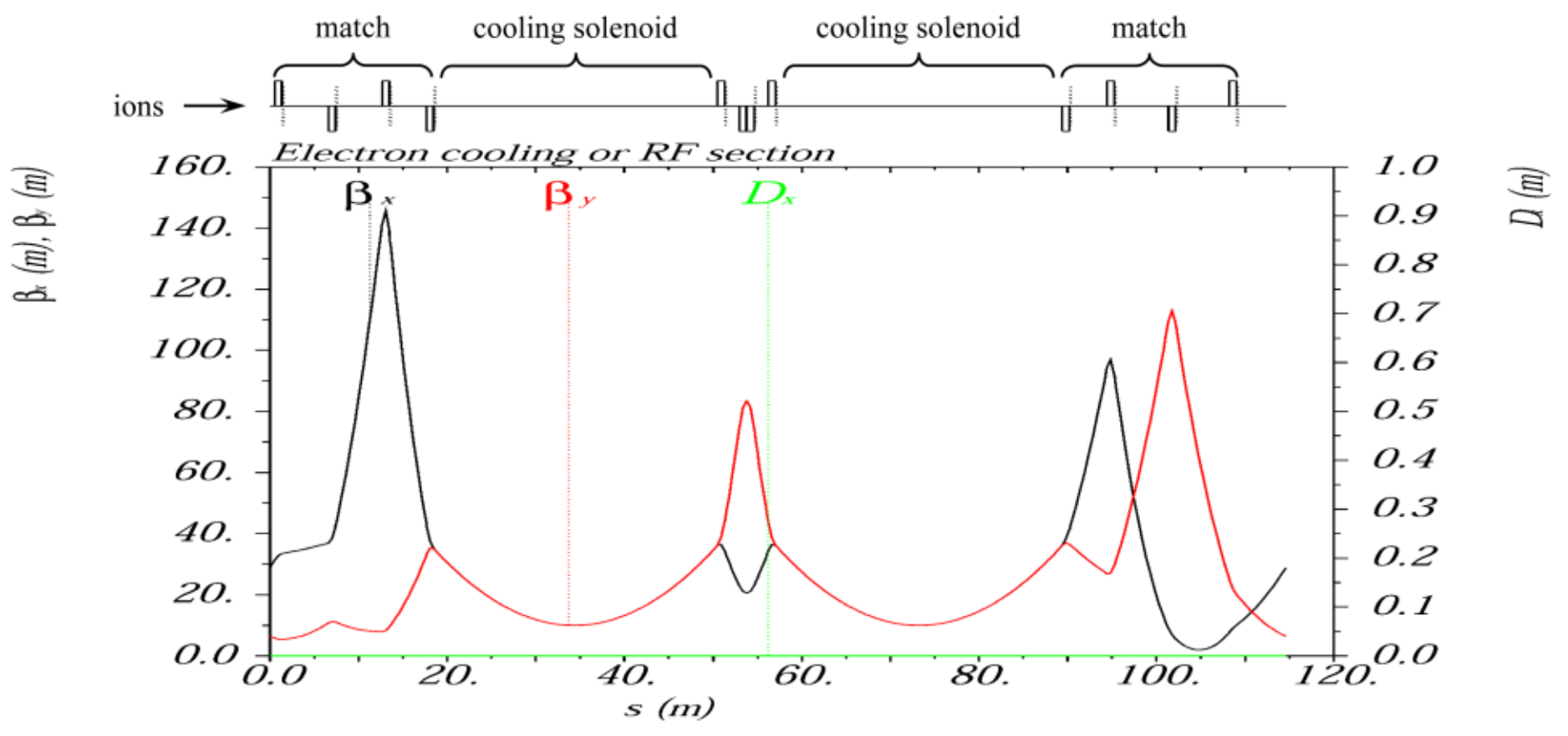

**Figure 7.5:** Electron cooling section.

The rest of the IP straight is occupied by a tune trombone for betatron tune adjustment. It consists of two FODO cells surrounded by matching sections. One of the matching sections is shared with the electron cooling section. The other matches the FODO to the adjacent arc.

The second straight reserves space for a second IP and is presently filled with a FODO lattice connected to the arcs by matching sections as shown in Figure 7.6. Both accelerating and bunching RF cavities [Wang2013] are placed in this straight between the quadrupoles of the FODO lattice. The complete ion collider ring optics consisting of the components discussed above is shown in Figure 7.7. Some of the ring's global parameters are summarized in Table 7.1. Note that crossing of the transition energy occurs during acceleration. However, existing experience shows that it can be handled efficiently using standard techniques [Kewisch2003]. A number of special elements has not been incorporated into the ring lattice at this stage but has been accounted for in the cost estimate; these include the injection kicker, beam abort system, electron cooler solenoids, spin control elements [Kondratenko2014], polarimeter, crab cavities [Ahmed2011], and collimators.

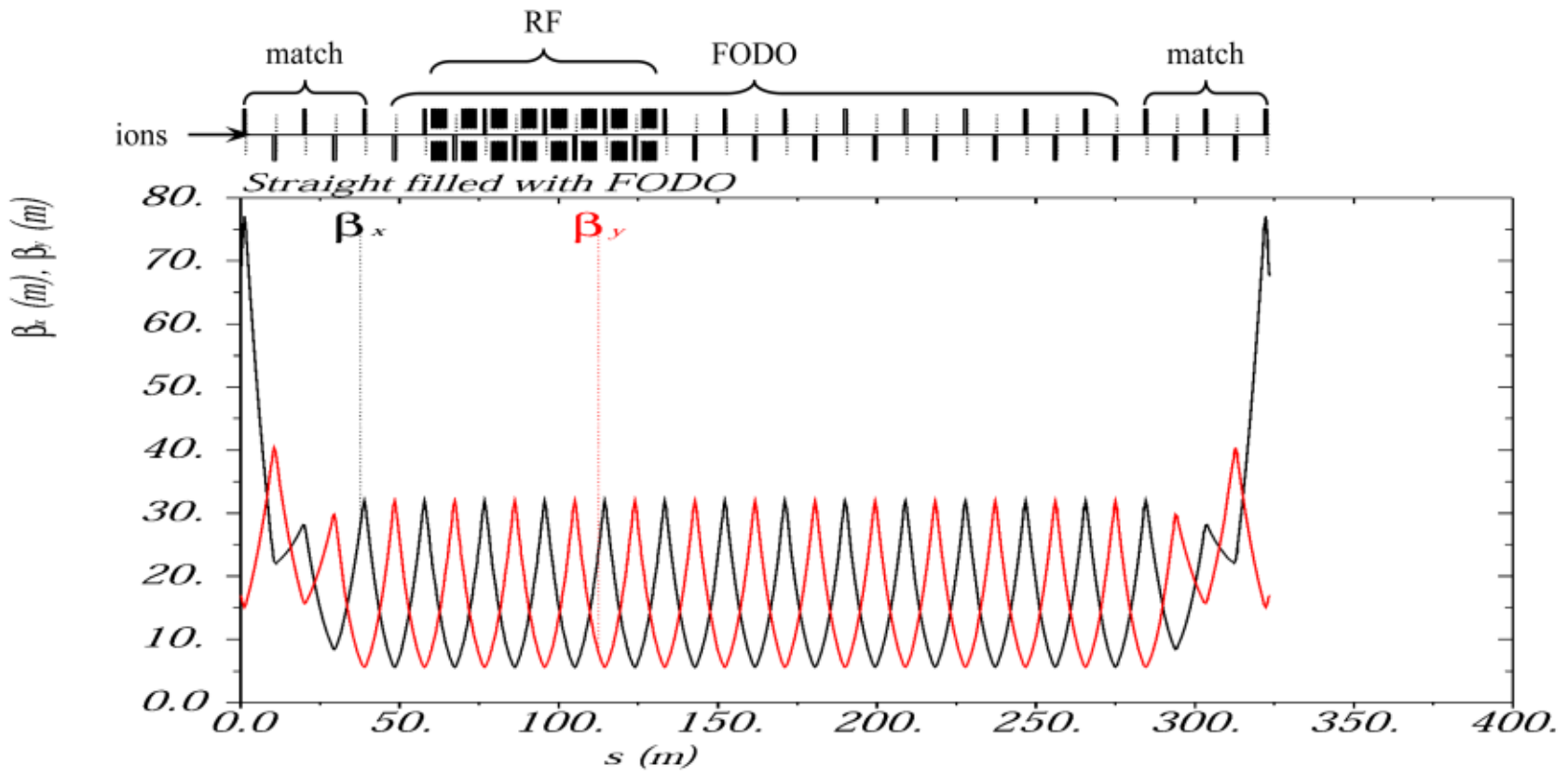

**Figure 7.6:** Optics of the ion ring's second straight.

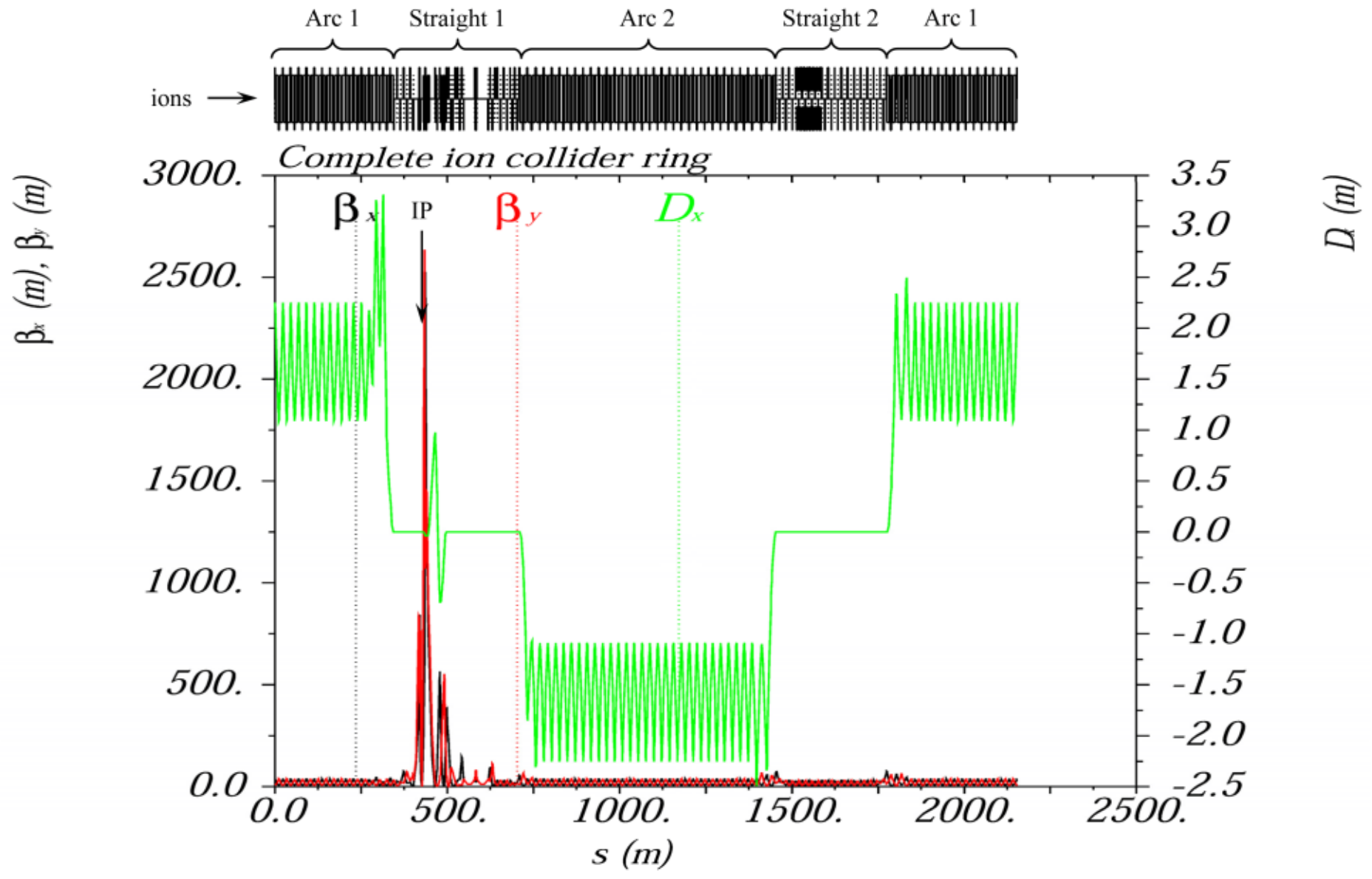

**Figure 7.7:** Complete ion collider ring optics.

**Table 7.1:** Some of the global parameters of the ion collider ring.

| Parameter | Unit | Value |
|---|---|---|
| Proton momentum range | GeV/c | 9-100 |
| Circumference | m | 2153.89 |
| Arc bending angle | degrees | 261.7 |
| Straight sections crossing angle | degrees | 81.7 |
| Maximum horizontal / vertical beta functions | m | 2301 / 2450 |
| Maximum horizontal dispersion | m | 3.28 |
| Horizontal / vertical betatron tunes | | 24(.38) / 24 (.28) |
| Horizontal / vertical natural chromaticities | | -101 / -112 |
| Momentum compaction factor | | $6.45 \cdot 10^{-3}$ |
| Transition gamma | | 12.46 |
| Horizontal / vertical normalized emittances | $\mu$m·rad | 0.35 / 0.07 |
| Maximum horizontal/ vertical rms beam size | mm | 2.8 / 1.3 |

# Polarization

## Electron Polarization

The MEIC electron ring at Jefferson Lab is designed to preserve and manipulate a highly polarized electron beam as required by the nuclear physics program: i) polarization of 70% or above, ii) longitudinal polarization at collision points, iii) polarization flipping at the required frequencies [Abeyratne2012]. To achieve these goals, various strategies have been carefully considered from different points of view and investigated as follows.

A highly vertically-polarized electron beam (>85%) is injected from CEBAF into the MEIC electron collider ring at a full energy from 3 to 10 GeV. Such an injection with vertical, instead of longitudinal, polarization has three advantages. First, it avoids spin decoherence caused by the energy variation during the acceleration in CEBAF. Second, it simplifies polarization transport between CEBAF and the MEIC collider ring. Third, electron beams with vertical polarization are injected into the MEIC electron collider ring in the arc section. Injection here can significantly reduce the background in the detector because dipole magnets in the arc sweep out incompletely injected electrons so that they have less chance to propagate through the chamber and hit the detectors.

The polarization in the MEIC electron collider ring is designed to be vertical in the arcs to minimize spin depolarization, and longitudinal at the collision points for experiments, as shown in Figure 8.1. Proper orientation of the spin is accomplished using four universal spin rotators [Chevtsov2010] located at each end of two arcs. These spin rotators, composed of interleaved solenoid and dipole fields as shown schematically in Figure 8.2, are designed to rotate electron polarization in the whole energy range from 3 to 12 GeV. The transverse orbital coupling induced by the longitudinal fields in the solenoids is neutralized by placing quadrupoles between half solenoids [Zholentz1984, Sayed2010].

The polarization configuration in the MEIC electron collider ring is determined by the solenoid field directions in the pair of spin rotators in the same long straight. These were chosen to have opposite solenoid polarities [Lin2013], as shown in Figure 8.3. Then the polarization is anti-parallel to the vertical guiding field in one arc and parallel to the guiding field in the other one, regardless of the choice of two possible opposite longitudinal polarizations at the IPs (purple solid and dashed arrows in Figure 8.3). Therefore, the Sokolov-Ternov self-polarization [Sokolov1964] process has a net depolarization effect integrated over the whole collider ring, and both polarization states from the polarized source will be equally affected.

In addition, with opposing longitudinal solenoid fields in the pair of spin rotators in the same long straight, the net field integral is zero. As a result, the 1$^{st}$ order spin perturbation in the solenoids for off-momentum particles vanishes. This significantly extends the polarization lifetime and reduces the burden on the spin matching and ring-optics design. Though this polarization configuration has zero equilibrium polarization, with highly polarized injected beams, the estimated polarization lifetimes at various energies (shown in Table 8.1) are adequate for detectors to collect data.

The electron polarization configuration, combined with a figure-8 geometry of the collider ring, produces a net zero spin precession. Hence the spin tune on the design orbit is zero and independent of beam energy. This significantly reduces the synchrotron sideband resonances. In addition, since there is no preferred direction of the polarization, the polarization can be easily controlled  and stabilized by using relatively small magnetic fields, for example spin tuning solenoids in the straights where the polarization is longitudinal.

The desired spin flipping in the MEIC electron ring is likely to be implemented by alternating the helicity of the photo-injector drive laser at the source to provide oppositely polarized bunch trains. Here, two polarization states coexist in the collider ring and have similar polarization degradation in the aforementioned MEIC polarization configuration. Two long oppositely polarized bunch trains have been considered in the collider ring, as shown in Figure 8.4. This simplifies the Compton polarimeters for polarization measurements in the collider ring.

The design of a Compton polarimeter has been integrated into the interaction region design to provide rapid and high precision measurements of the electron beam polarization in the MEIC. The design of the low-$Q^2$ tagging system in the electron collider ring is based on a four dipole chicane located a few meters after the downstream electron final focusing block. In addition to serving as spectrometers providing resolution down to very small momentum offsets, the center of the chicane is a convenient place for a Compton polarimeter as shown in Figure 8.5. Because there is no net spin rotation between the polarization measurement point and the interaction point (IP), the polarimeter allows for continuous non-invasive monitoring of the electron polarization at the IP. The polarimeter is shifted from the electron beam axis at the IP eliminating the background due to the synchrotron radiation coming from the IP.

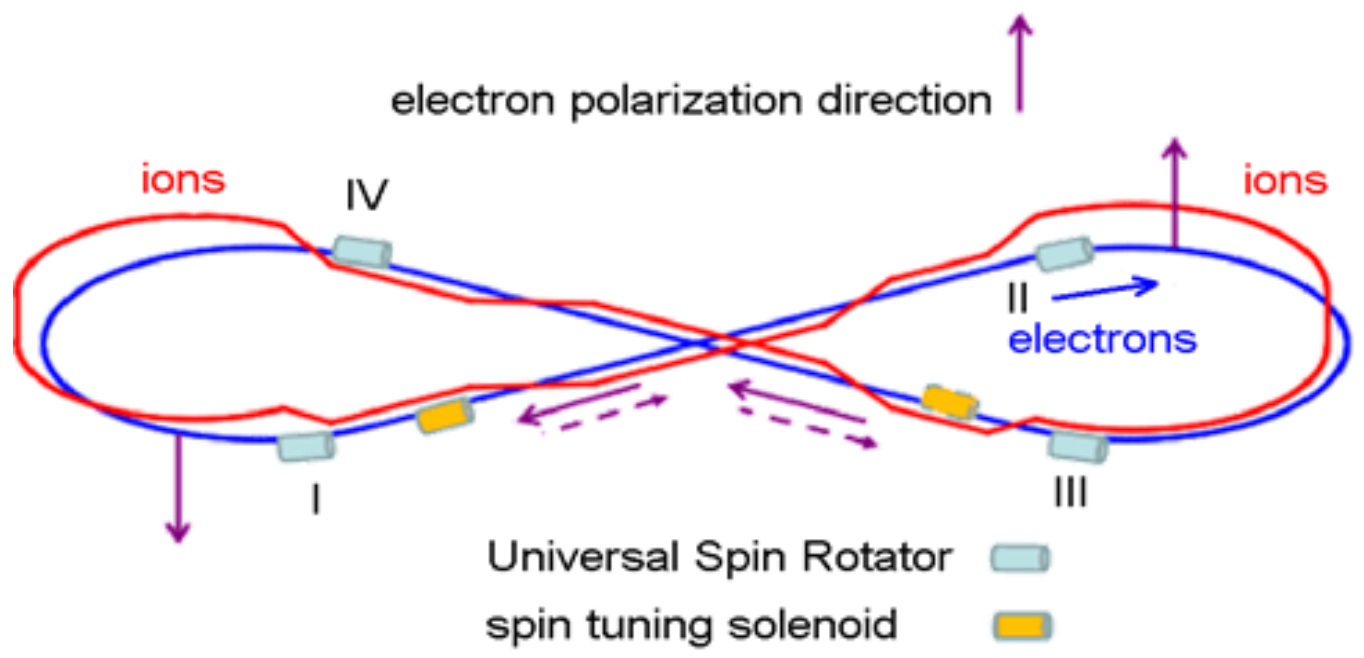

**Figure 8.1:** Electron polarization design in the MEIC electron collider ring.

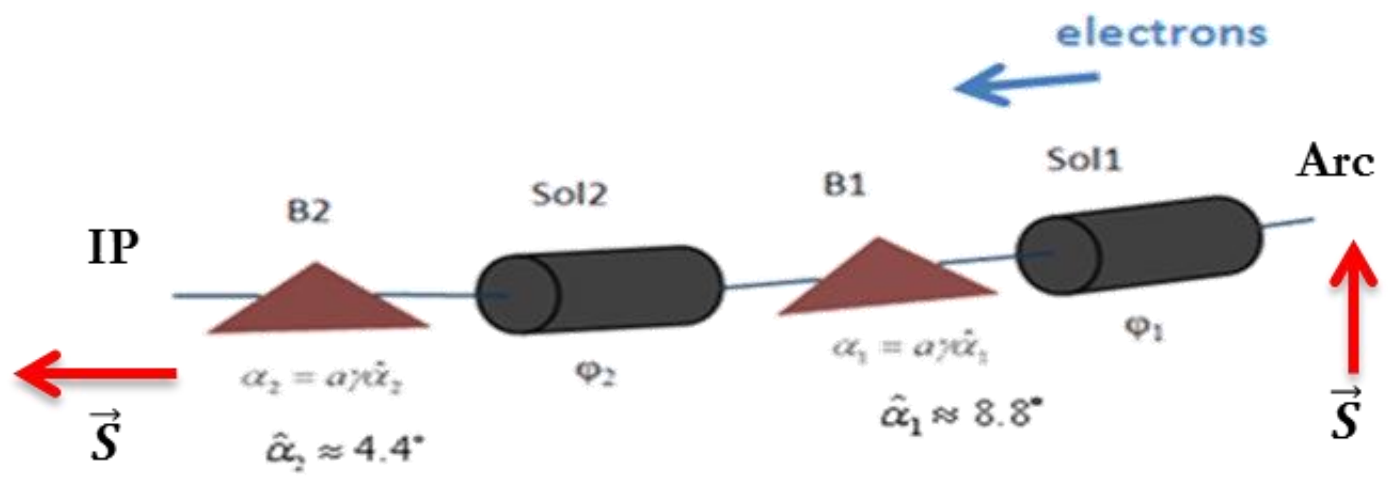

**Figure 8.2:** A schematic drawing of a spin rotator.

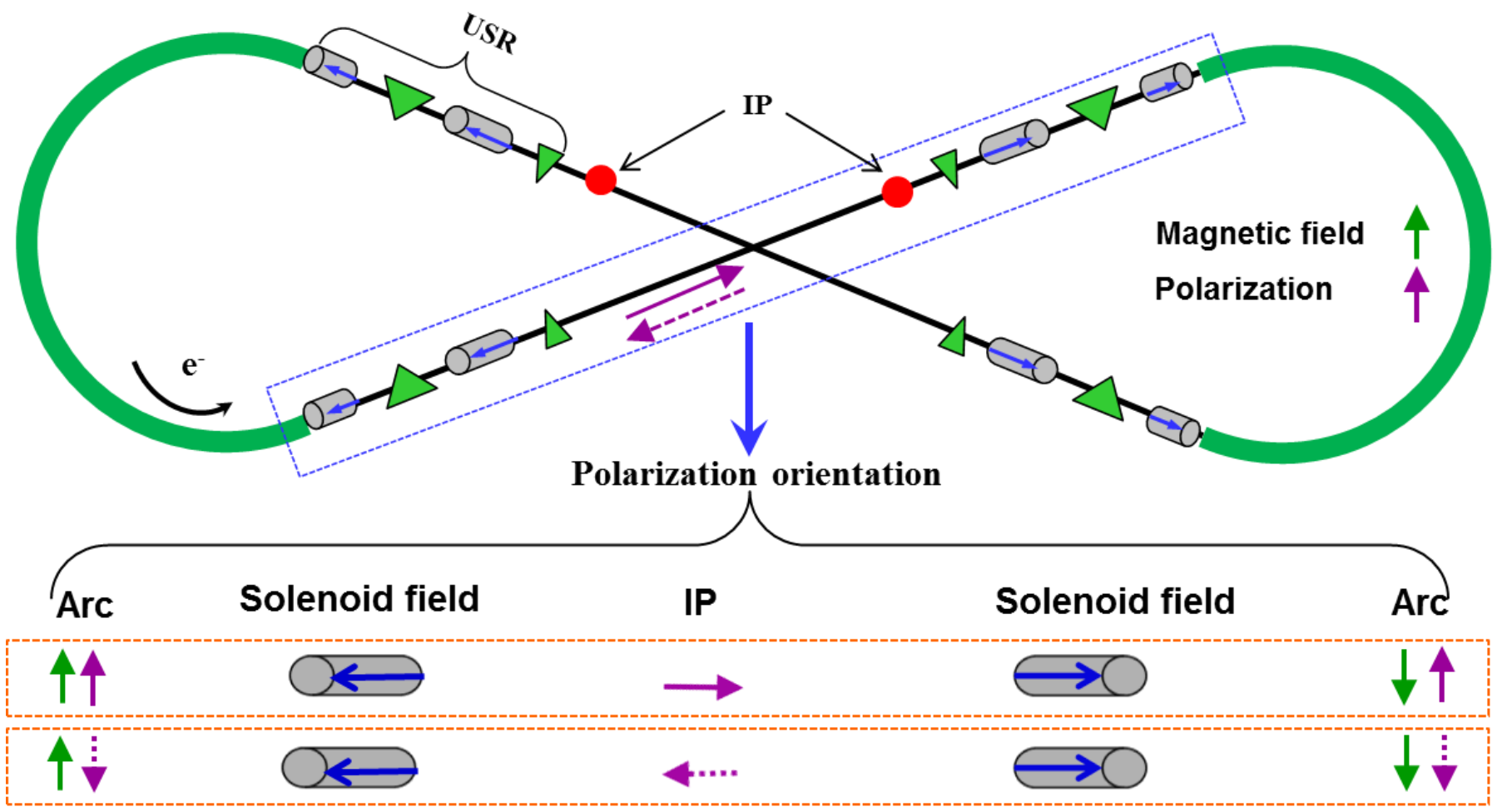

**Figure 8.3:** Polarization ($\vec{P}$, purple arrow) directions remain the same in the two arcs by having opposite longitudinal solenoid field directions in the same long straight. The blue arrow in the solenoid represents the field direction. The polarization orientation in one of the two long straights (shown in the blue dash-line box) is exhibited under the half brace with two different polarization states at the IP.

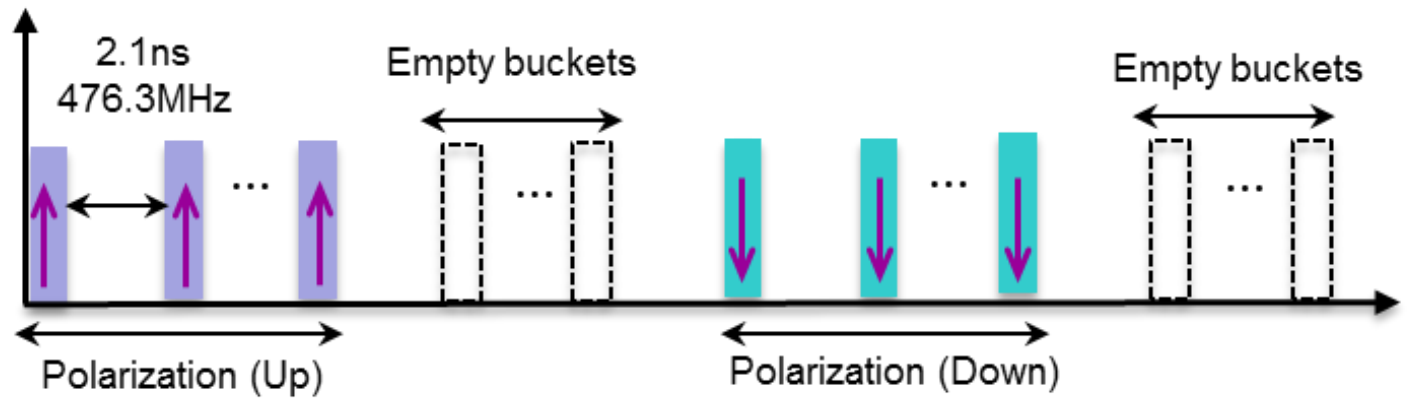

**Figure 8.4:** Bunch train and polarization pattern in the MEIC electron collider ring.

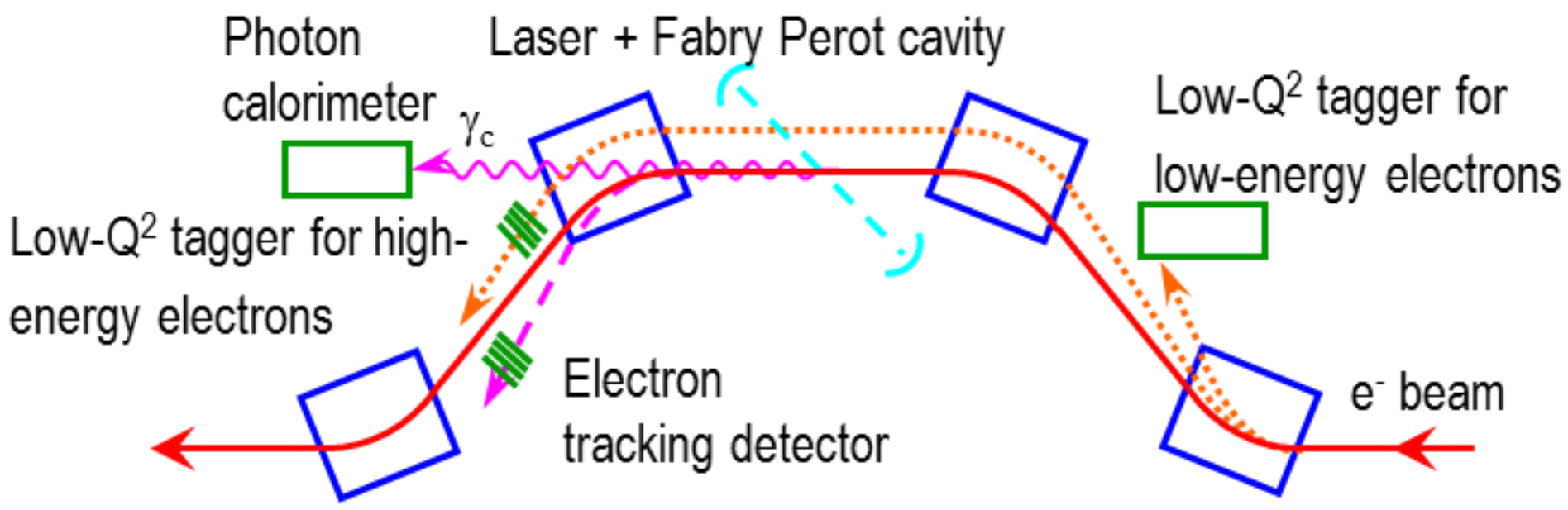

**Figure 8.5:** Schematic of the low-$Q^2$ tagger and electron polarimeter setup.

**Table 8.1:** Estimated electron polarization lifetime at various energies.

| Energy (GeV) | Estimated Pol. Lifetime (hours) |
|---|---|
| 3 | 66 |
| 5 | 5.2 |
| 7 | 2.2 |
| 9 | 1.3 |
| 10 | 0.86 |

## Ion Polarization

The main challenge of accelerating polarized ion beams to high energies in synchrotrons with vertical bending fields is the necessity to overcome numerous spin depolarizing resonances. Use of solenoidal and helical (dipole) Siberian snakes solves the problem of polarization preservation in the regions of low and high energies, respectively. However, usage of snakes may become problematic in the medium energy range from a few GeV to a few tens of GeV due to the longitudinal-field snakes already being too weak while the transverse-field snakes still causing large orbit excursions. Using figure-8 ring geometry [Derbenev1996] is an elegant way to preserve and control the polarization of a beam of any particle species during its acceleration and storage at any energy.

In a figure-8 collider, the spin first rotates about the vertical field in one arc. This rotation is then undone by the opposite field in the other arc. The resulting effect of the “strong” arc dipoles on the spin dynamics reduces to zero over one particle turn and the whole ring becomes “transparent” for the spin. Any spin orientation at any orbital location repeats every turn. In other words, in a figure-8 accelerator, the spin tune is zero, and there is no preferred polarization orientation because the particle is in the zero spin resonance region. To stabilize the beam polarization direction at the interaction point, it is sufficient to use compact insertions for polarization control, which utilize already existing collider magnets and solenoids with small field integrals (“weak” solenoids) [Kondratenko2013, Kondratenko2014a-c]. The spin tune and polarization direction are determined not by the "strong" fields of the collider's main lattice but by the weak solenoids introduced. Weak solenoids do not affect the closed orbit at all and essentially do not change the beam orbital parameters. This property is universal and does not depend on the particle type. Figure-8 colliders provide a real opportunity for obtaining intense polarized deuteron beams with energies greater than a few tens of GeV.

Figure 8.6 shows the scheme for polarization preservation in the booster of MEIC [Kondratenko2013, Kondratenko2014a-c]. A weak solenoid stabilizes the polarization in the longitudinal direction in the straight where it is placed. There is no problem with ramping the field of such a solenoid during the acceleration cycle. The required solenoid field integral does not exceed 0.7 T·m at the top energy of the booster for both protons and deuterons. It provides sufficient spin tune shifts of 0.01 and 0.003 from zero for protons and deuterons, respectively.

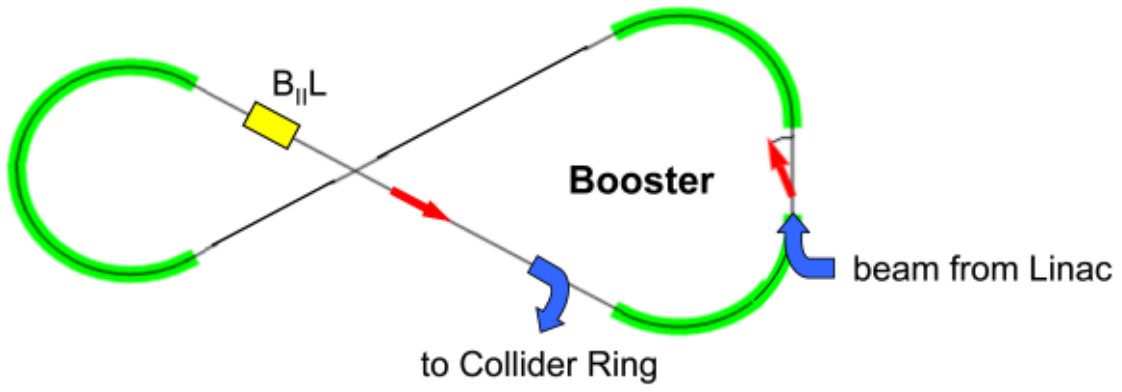

**Figure 8.6:** Schematic of the polarization control in the booster.

The beam polarization of any particle (p, d, $^3$He, ...) is controlled in the ion collider ring using universal 3D spin rotators [Kondratenko2014a-c]. The 3D rotators are designed using weak solenoids and allow performance of the following tasks: matching of the polarization direction at injection, polarization preservation during acceleration and storage, measurement of the beam polarization at any orbital location, and spin manipulation at the interaction point during experimental running.

Figure 8.7 shows a schematic of the 3D spin rotator [Kondratenko2014c] for ion polarization control located at the end of the experimental straight. The rotator consists of three modules: those for control of the radial $n_x$, vertical $n_y$, and longitudinal $n_z$ components of the polarization (see Figure 8.7a). The 3D spin rotator placement in the MEIC ring is shown schematically in Figure 8.7b.

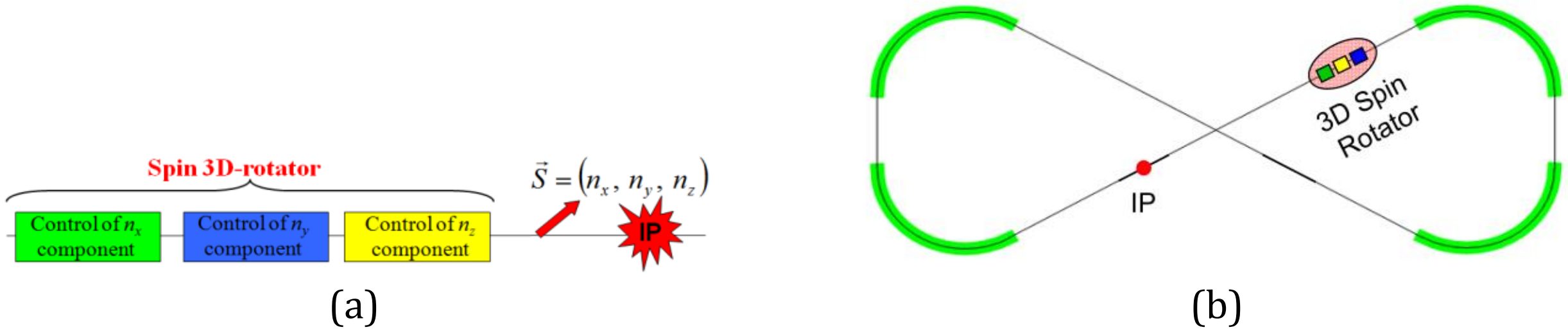

**Figure 8.7:** (a) 3D spin rotator schematic. (b) Spin rotator placement in the ion collider ring.

Figure 8.8a shows the module for control of the radial polarization component $n_x$, which consists of two pairs of opposite-field solenoids and three vertical-field dipoles producing a fixed orbit bump. The control module for the vertical polarization component $n_y$ is the same as that for the radial component except that the vertical-field dipoles are replaced with radial-field ones (Figure 8.8b). To keep the orbit bumps fixed, the fields of the vertical- and radial-field dipoles must be ramped proportionally to the beam momentum. The module for control of the longitudinal polarization component $n_z$ consists of a single weak solenoid (Figure 8.8c). There is a substantial flexibility in the placement and arrangement of these modules in the collider. For instance, to free up the space in the experimental straight, the module for control of the vertical polarization component can be installed anywhere in the arc.

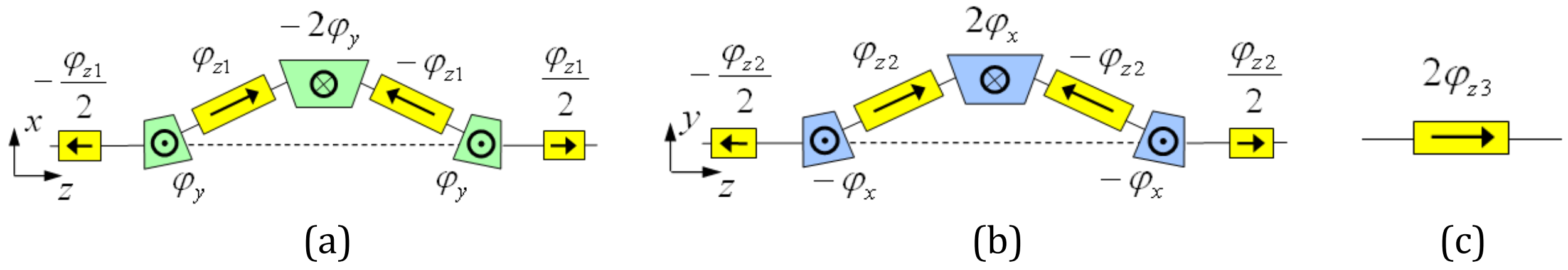

**Figure 8.8:** Modules for control of the radial (a), vertical (b), and longitudinal (c) spin components.

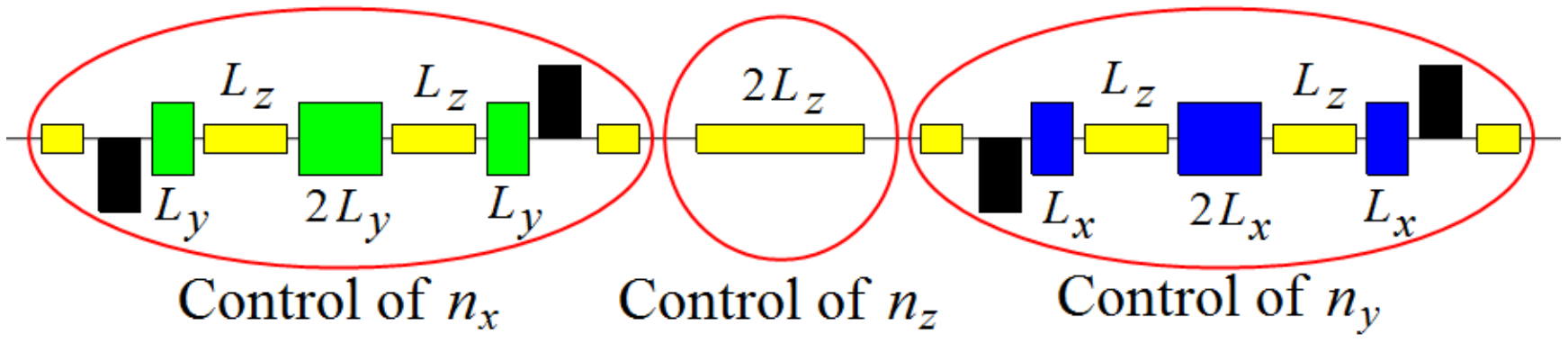

**Figure 8.9:** Schematic placement of the 3D spin rotator elements.

Schematic placement of the 3D rotator elements in the collider ring's experimental straight is shown in Figure 8.9 [Kondratenko2014c]. The structural quadrupoles are shown in black, the vertical-field dipoles are green, the radial-field dipoles are blue, and the control solenoids are yellow. With each module's length of ~6 m ($L_x = L_y = 0.6$ m, $L_z = 1.2$ m), the fixed orbit deviation in the bumps is ~12 mm in the whole momentum range of the collider. The 3D spin rotator can provide any desired polarization orientation at the interaction point. The maximum required dipole and solenoid magnetic field strengths are 3 and 3.6 T, respectively. The spin rotator shifts the proton and deuteron spin tunes from zero by sufficient amounts of 0.01 and $2.5 \cdot 10^{-4}$, respectively. The magnetic fields of the spin rotator solenoids can be changed relatively quickly on the time scale of a few seconds that allows using the 3D rotator for spin-flipping.

For polarization stability in the collider, the shift of the spin tune from zero provided by the 3D spin rotator must significantly exceed the strength of the "zero"-integer spin resonance. This resonance strength consists of coherent and incoherent parts. The coherent part is determined by lattice errors. The incoherent part is determined by the beam emittances. The coherent part significantly exceeds the incoherent one. The same type of the 3D spin rotator as described above is used for compensation of the coherent part of the "zero"-integer resonance strength, which substantially reduces the required field integrals of the 3D rotator solenoids [Kondratenko2014c]. A real collider with polarized beams then becomes equivalent to an "ideal" collider, which has its magnetic elements aligned exactly on the reference orbit so that the zero-integer resonance strength is determined only by the beam emittances. This allows for polarized beam experiments at a high precision level.

# Ion Beam Cooling

Cooling of ion beams is critical in delivering high luminosities over a broad CM energy range in MEIC. It is needed to achieve a significant reduction of the six dimensional beam emittance, and thus, for delivering a small beam spot at the interaction point. Longitudinally, a short bunch length enables a very small $\beta^*$ and crab crossing of colliding beams. In addition cooling counteracts the intra-beam scattering (IBS) induced emittance degradation, thus, extending the luminosity lifetime [Derbenev2010].

The present MEIC design utilizes conventional electron cooling [Budker1967], a well-developed technology, for cooling the ion beams. The design also adopts a scheme of multi-phase cooling [Derbenev2007, Derbenev2009] during formation of the ion beam and during collision for enhancing the cooling efficiency. This scheme is based on a simple fact that the electron cooling time is, to the first order, proportional to the six dimension beam emittance and the square of beam energy. Therefore, cooling is more efficient when the beam energy is low. At the high (collision) energy, the cooling time also sees a reduction due to a much smaller emittance as a result of the initial cooling at low energy. Combining both stages, it is expected the total cooling time should be orders of magnitude shorter than that of performing cooling only at high (collision) energy.

Table 9.1 summarizes the cooling phases in MEIC. Electron cooling is first called for in the booster ring for assisting accumulation of positive ions (from helium-3 $^{3}He^{+2}$ to partially-stripped lead ion $^{208}Pb^{32+}$) injected from the pulsed warm ion linac. For the proton or deuteron beams, MEIC utilizes a H-/D- negative ion source so no cooling is required in this accumulation state. The accumulated proton beam is boosted to 2 GeV kinetic energy, the initial cooling is performed at this energy and the transverse emittance is reduced to the design values. The proton beam is then boosted to 7.9 GeV and transferred to the collider ring. Electron cooling is used again in the collider ring during stacking of the proton beam transferred (in 8 batches), and is continued after the beam is accelerated to the collision and during *e-p* collisions. In these two stages, electron cooling is for suppressing the IBS induced emittance growth and maintaining the emittance to the design values. Cooling of ion beams will follow a similar plan according to ion beam energies in the booster and collider rings.

Two electron coolers are required to implement the MEIC cooling scheme. In the pre-booster, a DC cooler with electron energy up to 1.1 MeV is needed for the first two phases of cooling. DC cooling is a mature technology and this cooler is well within the state-of-art. The DC cooler for the Tevatron anti-proton recycler at Fermilab has an electron energy up to 4.3 MeV [Nagaitsev2006]. Recently, a 2 MeV DC electron cooler has been built and commissioned for the cooled storage ring (COSY) at FZ-Juelich, Germany [COSY2009]. This cooler can be readily adopted for the MEIC booster ring with only minor modifications.

**Table 9.1:** MEIC multi-phased electron cooling

| | Phase | Function | Proton kinetic energy (Gev/u) | Electron kinetic energy (MeV) | Cooler type |
|---|---|---|---|---|---|
| Booster | 1 | Assisting accumulation of injected positive ions | 0.11 ~ 0.19 | 0.062 ~ 0.1 | DC |
| | 2 | Emittance reduction | 2 | 1.09 | |
| Collider ring | 3 | Suppressing IBS and maintaining emittance during stacking of beams | 7.9 | 4.3 | BB (ERL) |
| | 4 | Suppressing IBS and maintaining emittance during collision | 100 | 55 | |

In the MEIC ion collider ring, an electron cooler utilizing a high-energy bunched beam will be responsible for cooling the medium energy ions (up to 100 GeV/u energy). Two accelerator technologies, an energy-recovery-linac (ERL) and a circulator ring (CR) [Martirosyan2000, Derbenev2010, Derbenev2007], will play critical roles in the success respectively of this facility [Zhang2013] and of a follow on upgrade. They provide perfect solutions to two technical challenges: the high current and high power of the cooling electron beam. As illustrated by the schematic drawing in Figure 9.1, the electron bunches from a source are accelerated in an SRF linac. In the baseline design the electrons will be merged with the ions, continuously cooling the ion bunches in a long cooling channel immersed in a strong solenoid field, and then be returned to the linac for energy recovery. Current gun and SRF capability allow the energy recovered linac beam to provide enough electron cooling for the baseline design. Table 9.2 lists the main design parameters of the ERL cooler.

For one of the higher luminosity upgrades discussed in Section 11, a full circulator ring will be deployed. In this case beam from the energy recovered linac will be deflected into a closed ring formed from the ERL beam line and a circulator connection, and will circulate in a compact ring for 25 turns. This reduces the current from the source and linac by a factor of 25. The electron bunches are then ejected out of the circulator ring and sent back to the SRF linac for energy recovery before going to the dump. As an example, a 1.5 A 55 MeV cooling beam (81 MW of power) can effectively be provided by a 60 mA, 2 MeV beam (120 kW of active beam power) from the injector/ERL if the cooling beam makes 25 turns in the circulator cooler ring. Table 9.3 lists the main design parameters for the circulator cooler.

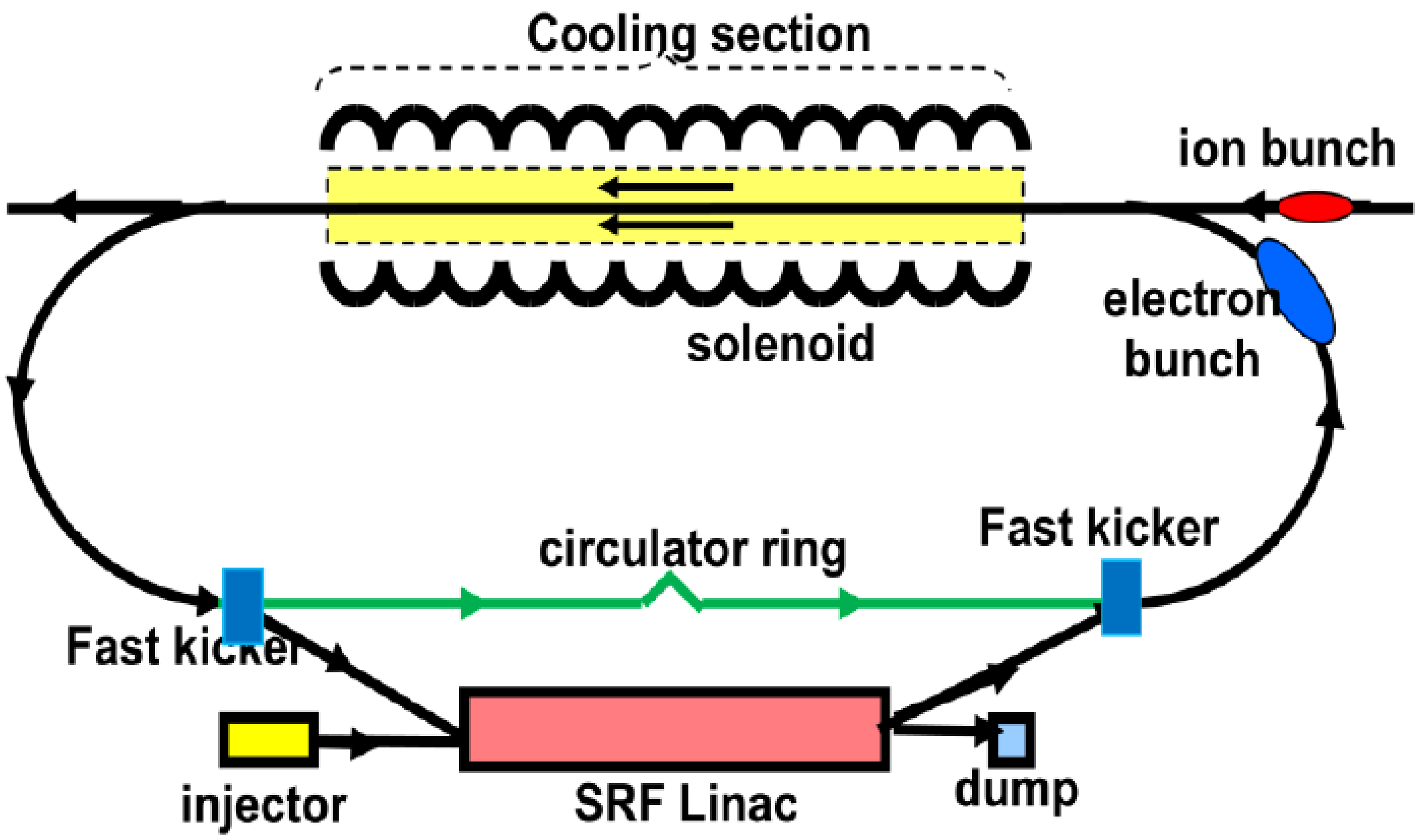

**Figure 9.1:** A schematic drawing of a bunched beam cooler based on an ERL, with upgrade circulator ring connection in green.

**Table 9.2:** MEIC bunched beam ERL cooler design parameters.

| | | |
|---|---|---|
| Proton/electron energy | GeV/ MeV | 100/55 |
| Electrons per bunch | $10^{10}$ | 0.5 |
| Current in ERL | A | 0.2 |
| Bunch repetition in ERL | MHz | 476 |
| Cooling channel length circumference | m | 60 / ~160 |
| RMS Bunch length, proton/electron | cm | 1.5 / 3 |
| Electron beam energy spread | $10^{-4}$ | 3 |
| Solenoid field in cooling sect. | T | 2 |
| Beam radius in solenoid | mm | ~1 |
| Thermal cyclotron radius | μm | ~3 |
| Beam radius at cathode | mm | 3 |
| Solenoid field at cathode | T | 0.2 |

**Table 9.3:** MEIC bunched beam cooler design parameters for the circulator cooler.

| | | |
|---|---|---|
| Proton/electron energy | GeV/ MeV | 100/55 |
| Electrons per bunch | $10^{10}$ | 2 |
| Bunch circulations in CR | | ~25 |
| Current in CR/ERL | A | 1.5 / 0.06 |
| Bunch repetition in CR/ERL | MHz | 476 / 19 |
| Cooling channel length / CR circumference | m | 60 / ~160 |
| RMS Bunch length, proton/electron | cm | 1.5 / 3 |
| Electron beam energy spread | $10^{-4}$ | 3 |
| Solenoid field in cooling sect. | T | 2 |
| Beam radius in solenoid | mm | ~1 |
| Thermal cyclotron radius | μm | ~3 |
| Beam radius at cathode | mm | 3 |
| Solenoid field at cathode | T | 0.2 |

Figure 9.2 shows a preliminary technical design [Douglas2012] of the full ERL-CC based BB Cooler in a figure-8 shape, which would be placed at the vertex of the figure-8 collider ring. The ERL loop has two ring-shaped beam lines, one is for the ERL and the other is a circulator ring. The cooling electron bunches are switched between these two rings by a fast kicker and a septum magnet. Two SRF cavities are placed in the ERL ring for chirping and de-chirping the cooling electron bunches thus enabling a transformation in the longitudinal space, namely, a long bunch length in the cooling section and a short bunch length at the SRF linac. The cooler utilizes a magnetized photo-cathode (the cathode is immersed in a 2 kG solenoid field) and the beam is transported in the specially designed beam line similar to the Fermilab DC cooler, instead of the long continuous solenoid utilized in all low energy magnetized DC coolers.

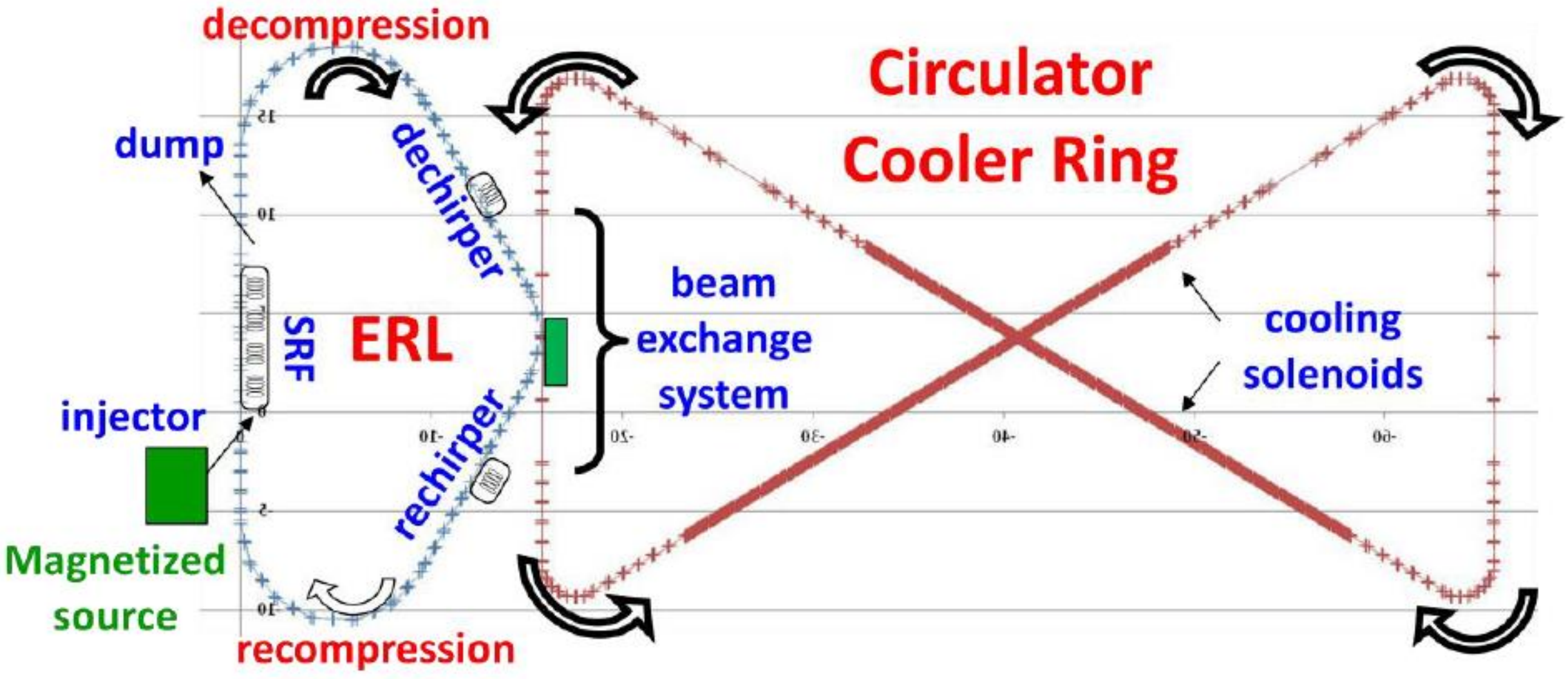

**Figure 9.2:** The beam-line layout of a technical design of the MEIC bunched beam cooler. It is located in the vertex of the MEIC figure-8 collider ring.

The MEIC bunched beam cooler requires significant R&D. The main challenges are developing an ultra-fast beam kicker and a high current magnetized electron source, as well as demonstrating suppression of collective beam effects. Demonstrating electron cooling with a bunched beam also needs to be shown. More discussions of the cooler R&D program will be presented in Section 11.

## RF/SRF for MEIC

The MEIC philosophy is to achieve high luminosity by using many bunches while keeping single bunch parameters within reasonable limits. This is the same approach adopted successfully by the B-factories more than a decade earlier, however in an electron-ion collider there are other challenges. The ion bunch rate and bunch length must match the electron beam, requiring a high frequency and high installed voltage in the ion ring as well as the electron ring. Fortunately the ion ring RF system does not have to supply as much beam power as the electron ring due to the absence of synchrotron radiation. However beam stability and collective effects are a concern in both rings and strongly HOM-damped cavities are required, with significant HOM power to be extracted. This leads to the adoption of single-cell HOM damped cavities similar to those in the B-factories.

In addition to the collider RF systems, low frequency RF systems are needed in the ion ring and booster to capture and accelerate the incoming fresh beams. These can be conventional inductively loaded, tunable, normal conducting cavities.

To preserve the high luminosity with a large crossing angle at the IP, a strong crabbing system is required with multiple cavities in both the ion ring and electron ring. A local crabbing scheme is preferred as it is compatible with multiple IP’s and eases concerns about collimation and beam losses around the rest of the machine.

Finally the high-energy ion cooler requires a high-current low-emittance electron beam, best provided by an ERL. This ERL requires a linac consisting of four 5-cell cavities and a short booster cryomodule containing four single-cell cavities. These all need HOM damping to achieve the desired high current, and would use similar components to the storage ring RF systems. Table 10.1 lists the high level RF parameters for these systems.

**Table 10.1:** Major MEIC RF parameters.

| | h | Energy (GeV) | Current (A) | Frequency (MHz) | Voltage | RF Power |
|---|---|---|---|---|---|---|
| Booster | 1 | 0.28–8 (H+) | 0.5 | 0.817 – 1.274 (457 kHz) | 32.0 kV | 8+82.4 kW (0.396 s ramp) |
| Ion storage ring | 9<br><br>6832 | 8-100 (H+) | 0.5 | 1.248 - 1.255<br><br>952.6 | 110.6 kV<br><br>18.95 MV | 27.5+165.2 kW (12s ramp)<br>(bunching) |
| Electron ring | 3416 | 3 - 10 | 3 | 476.3 | 20.6 MV | ~10 MW (SR power) |
| Cooler ERL | | 55 MeV | 0.060 | 952.6 | 56 MV | 120 kW |
| Cooler booster | | 10 MeV | 0.060 | 952.6 | 10 MV | 2000 kW |
| Future Cooler Circulator | | 55 MeV | 1.5 | 952.6 | - | |

## Ion Beam Capture and Acceleration

Multiple shots from the linac will be captured and accumulated in the booster using a low frequency RF system with a harmonic number of 1. Once sufficient beam has been accumulated the booster will ramp from 285 MeV to 8 GeV. The tunable RF system will track the revolution frequency to keep the orbit stable. The total installed voltage and RF power are determined by the ramping rate, but since this is not a rapid cycling machine the demands are modest. At 8 GeV the stored bunch will be transferred to the collider ring and captured by a similar system with harmonic number 9. Once the fill is complete the stored and cooled beam will be ramped to collision energy, again with the frequency tuning to maintain a fixed orbit. Similar cavities have been used at CSNS, Figure 10.1 [Sun2013], and JPARC, Figure 10.2 [Yoshii2007]. The parameters of this system are given in Table 10.2.

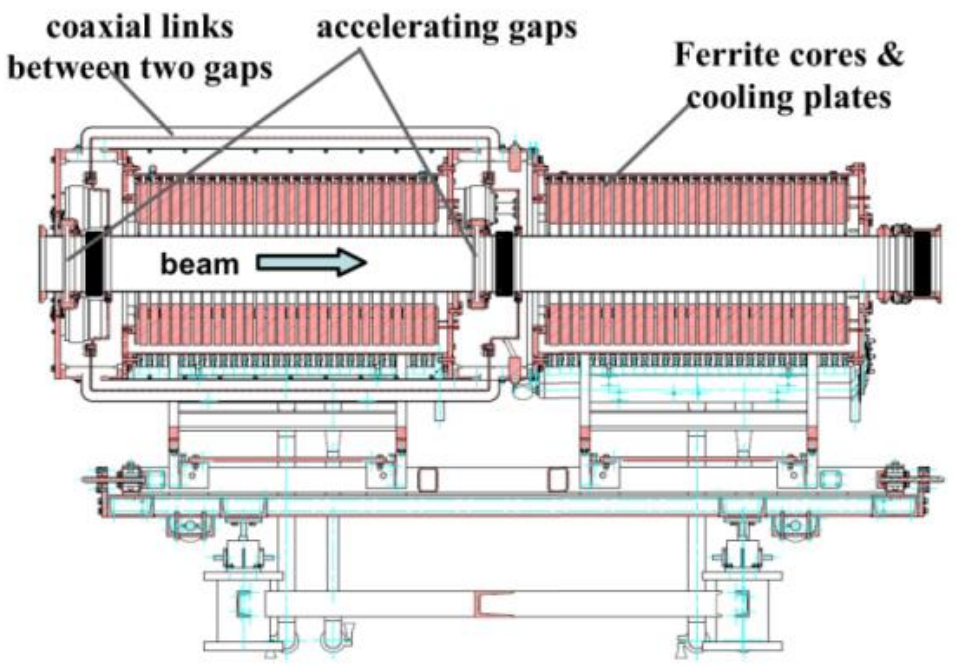

**Figure 10.1:** CSNS low frequency tunable cavity (ferrite loaded).

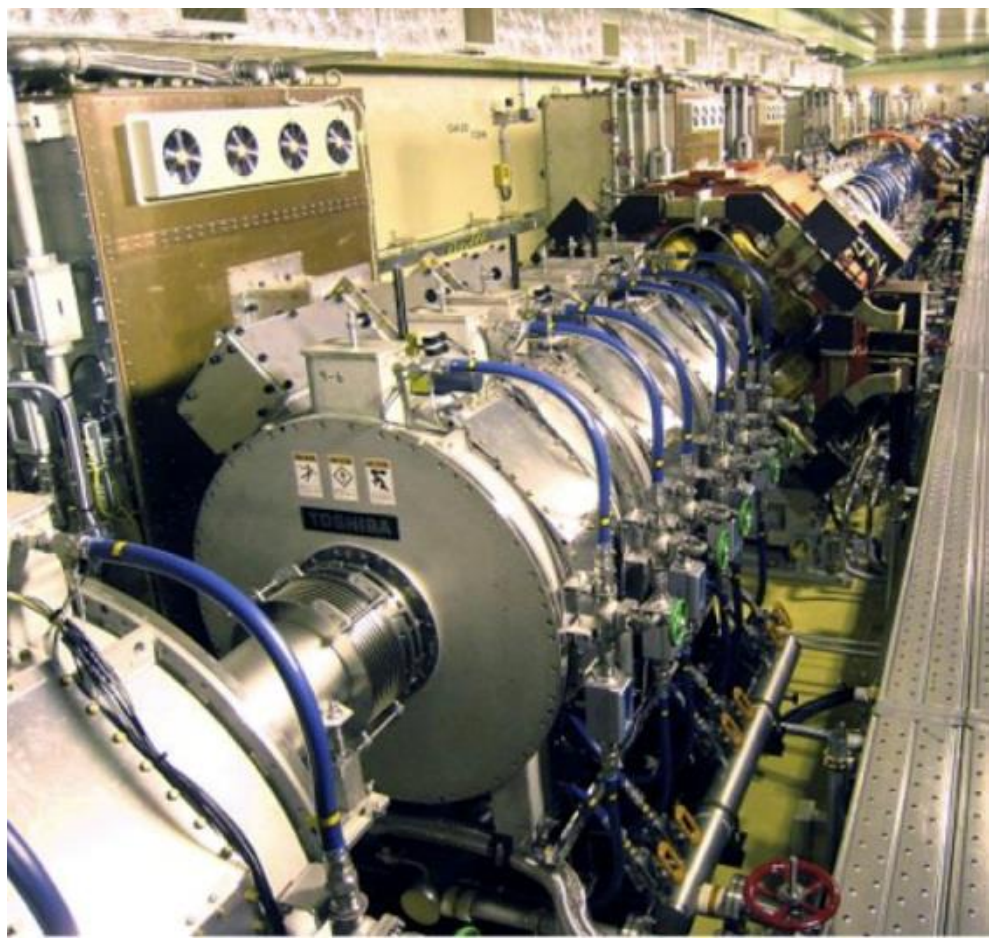

**Figure 10.2:** JPARC low frequency tunable cavity (Metglass loaded).

**Table 10.2:** Ion booster and collider ring RF parameters using low frequency cavities

| Booster | H+ | 208Pb67+ | |
|---|---|---|---|
| Energy | 0.28 ~ 8 | 0.112 ~ 3.2 | GeV |
| RF Frequency Range | 0.817 ~ 1.274 | 0.578 ~ 1.25 | MHz |
| Ramping Time | 0.396 | 0.56 | Sec |
| Vgap | 8.0 | 5.75 | kV |
| Beam Power | 8.0 | 1.85 | kW |
| Power Loss per Cavity | 41.2 | 41.2 | kW |
| Syn. Phase | 30.0° | | |
| Gaps per Cavity | 2 | | |
| Cavity Number | 2 | | |

| Collider | H+ | 208Pb82+ | |
|---|---|---|---|
| Energy | 8 ~ 100 | 3.2 ~ 40 | GeV |
| RF Frequency Range | 1.248 ~ 1.255 | 1.223 ~ 1.25 | MHz |
| Ramping Time | 12.0 | 12.2 | Sec |
| Vgap | 7.9 | 7.9 | kV |
| Beam Power | 27.5 | 10.8 | kW |
| Power Loss per Cavity | 23.6 | 23.6 | kW |
| Syn. Phase | 30.0° | | |
| Gaps per Cavity | 2 | | |
| Cavity Number | 7 | | |

## Ion Ring Bunching System

At the collision energy the beam will be re-bucketed into a high frequency bunching RF system at 952 MHz. High installed voltage will be required to achieve the desired short bunch length. The system will employ all new single-cell HOM damped SRF cavities, Figure 10.3, in a new modular cryostat based on previous JLab technology, Figure 10.4. The RF input power requirements are lower than the e-ring because there is no significant synchrotron radiation power, but the HOM extracted power is significant due to the high stored beam current and short bunches. The parameters of this system are given in Table 10.3. The bunch frequency will initially be up to 476 MHz, limited by the e-ring RF systems. In this case the ion beam will need to be pre-bunched by a single modest 476 MHz station (e.g. one spare PEP-II cavity) before the 952 MHz system is turned on. In the future if the e-ring systems are also migrated to new 952 MHz stations the bunch rate can be doubled.

**Table 10.3:** Ion ring RF parameters using 952.6 MHz SCRF cavities

| | **Proton** | | **Lead ion** | |
|---|---|---|---|---|
| **Energy** | **30** | **100** | **40** | GeV/u |
| **Current** | 0.50 | 0.50 | 0.50 | A |
| **Vpeak** | 4.94 | 18.95 | 17.93 | MV |
| **Syn. Phase** | 0.0 | 0.0 | 0.0 | degree |
| **Vgap** | 1.23 | 1.18 | 1.20 | MV |
| **Gradient** | 7.8 | 7.522 | 7.591 | MV/m |
| **Forward Power** | 51.0 | 46.94 | 109.88 | kW |
| **Cavity Power Loss** | 23.4 | 21.6 | 22.0 | W@4K |
| **Reflected Power** | 50.9 | 46.9 | 109.9 | kW |
| **Qext** | 7.07E+04 | 7.07E+04 | 3.08E+04 | |
| **Minimum Cavity Number** | 4 | 16 | 15 | |

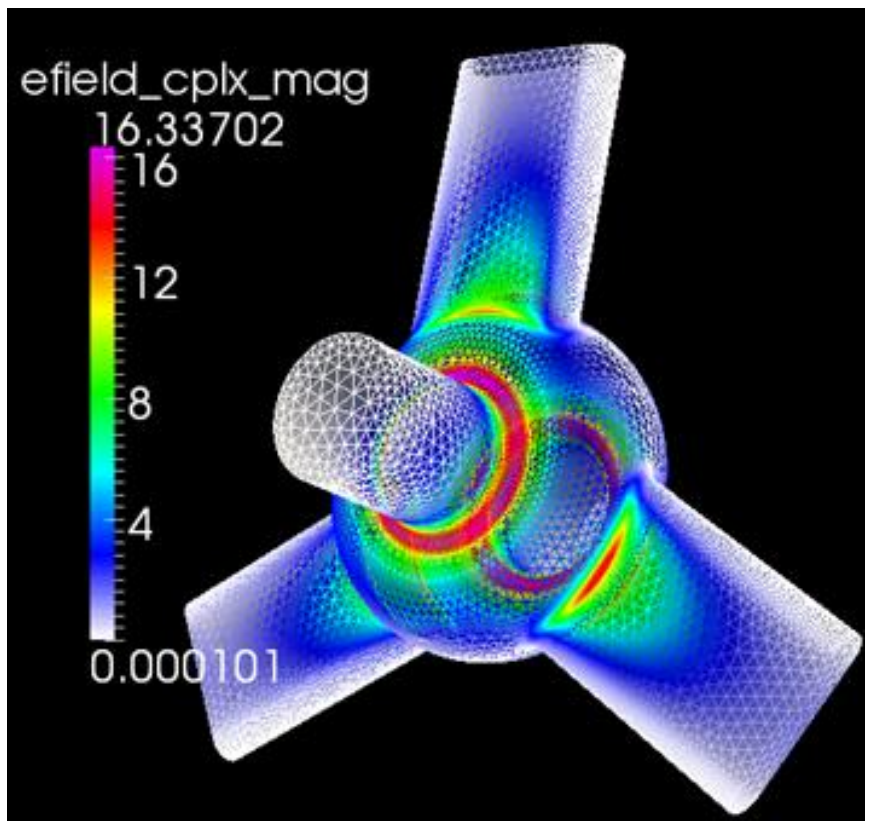

**Figure 10.3:** HOM damped cavity concept.

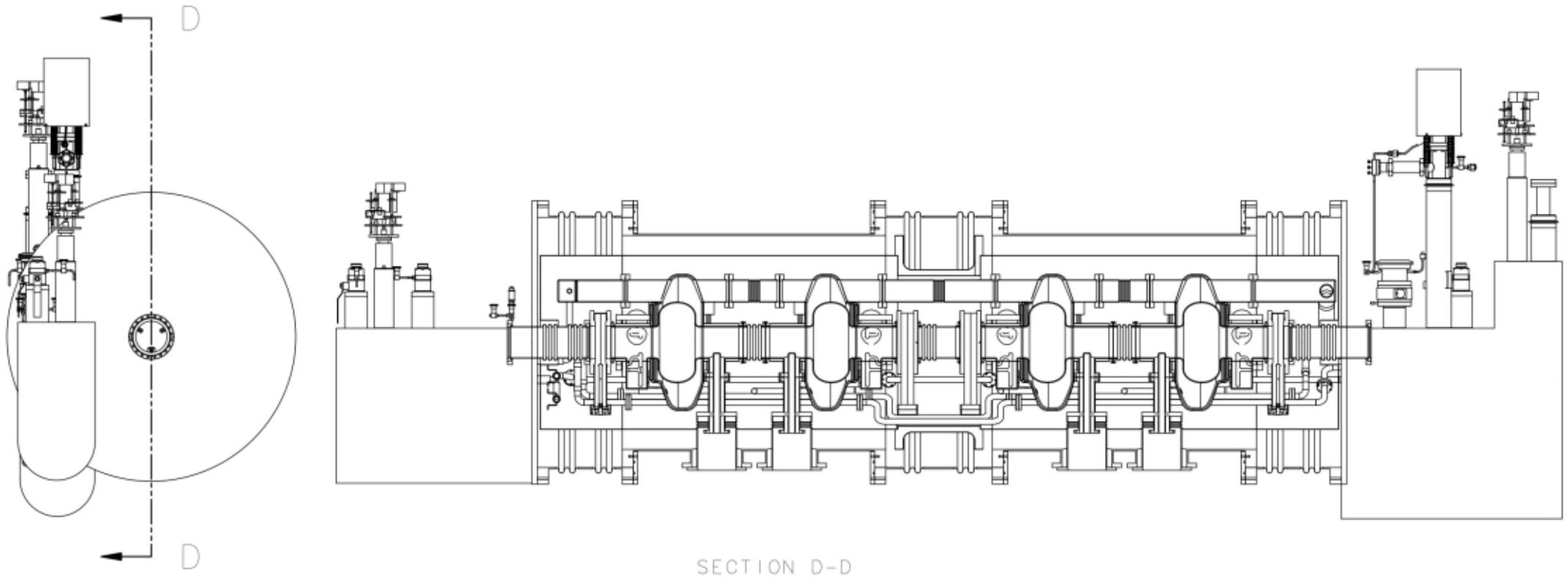

**Figure 10.4:** Concept for modular cryostat with four single-cell cavities.

## Electron Collider Ring RF System

The e-ring system may be the most challenging of the four since it has to supply up to 10 MW of synchrotron radiation power, up to 25 MV of installed voltage, present a very low impedance to the beam, safely dissipate many kW of HOM power per cavity, and operate with very high reliability. Fortunately the PEP-II 476 MHz NCRF systems [Seeman2008], Figure 10.5, were designed and operated for exactly these circumstances and are available to populate the electron ring (34 copper cavities and 13 1.2 MW klystrons and ancillary equipment are available). We will combine the 4-cavity per klystron stations of the PEP-II HER with the 2-cavity per klystron stations from the LER into one ring. With appropriate phasing the maximum beam power and installed voltage can be delivered. The parameters of this system are given in Table 10.4. For future higher energy upgrades new 952 MHz SRF cavities can be installed alongside or in place of the existing copper systems. This gives a seamless path to upgrade the energy and manage the graceful phase out of the PEP-II systems as they eventually age. The maximum beam current that can be stored in the electron ring is determined by several factors. At low energy the limit is collective instabilities driven the ring impedance, dominated by the RF cavity HOMs, hence the need for strongly damped cavities and a bunch-by-bunch feedback system. At high energies the limit is synchrotron radiation power density in the vacuum chamber or total available RF power. Thus the current has to be reduced to stay below 10 kW/m synchrotron loss in the vacuum chambers or 10 MW total in the ring. These effects are illustrated in Figure 10.6. In the central energy range between these two regimes the beam current is limited by other factors such as chamber heating, beam-beam effects or single bunch charge limits. This is the region where the luminosity is maximized and is not limited by the choice to use the PEP-II RF systems.

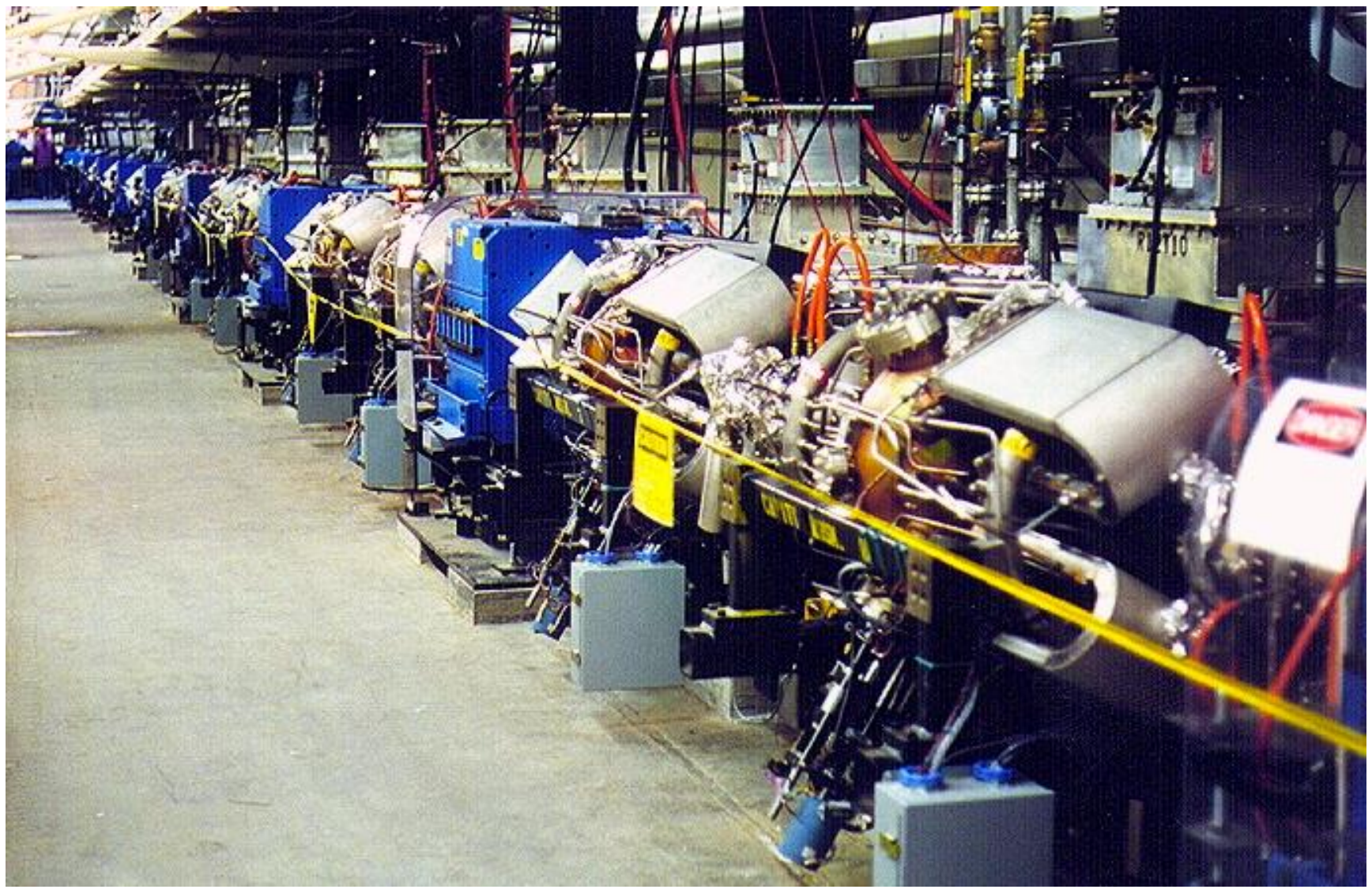

**Figure 10.5:** PEP-II NCRF systems installed in the tunnel.

**Table 10.4:** Electron ring RF parameters using 476.3 MHz NCRF cavities.

| **Energy** | **4** | **5** | **10** | GeV |
|---|---|---|---|---|
| **Current** | 3.00 | 3.00 | 0.71 | A |
| **SR power per ring** | 1.08 | 2.65 | 10.00 | MW |
| **Energy Loss per Turn** | 0.36 | 0.88 | 14.12 | MV |
| **Veff** | 0.36 | 0.88 | 14.12 | MV |
| **Vpeak** | 1.74 | 3.44 | 20.56 | MV |
| **Syn. Phase** | 12.01 | 14.89 | 43.39 | degree |
| **Vgap** | 0.43 | 0.34 | 0.79 | MV |
| **Gradient** | 1.4 | 1.1 | 2.5 | MV/m |
| **Power fed to beam per Cavity** | 271.2 | 264.8 | 384.6 | |
| **Forward Power** | 455.5 | 497.8 | 492.0 | kW |
| **Cavity Power Loss** | 27.0 | 16.9 | 89.3 | kW |
| **Reflected Power** | 157.3 | 216.1 | 18.0 | kW |
| **Qext** | 8.89E+03 | 8.89E+03 | 8.89E+03 | |
| **Minimum Cavity Number** | 4 | 10 | 26* | |

* Assumes 2 cavities per klystron. In fact some 4-cavity stations would be beneficial for more total installed voltage.

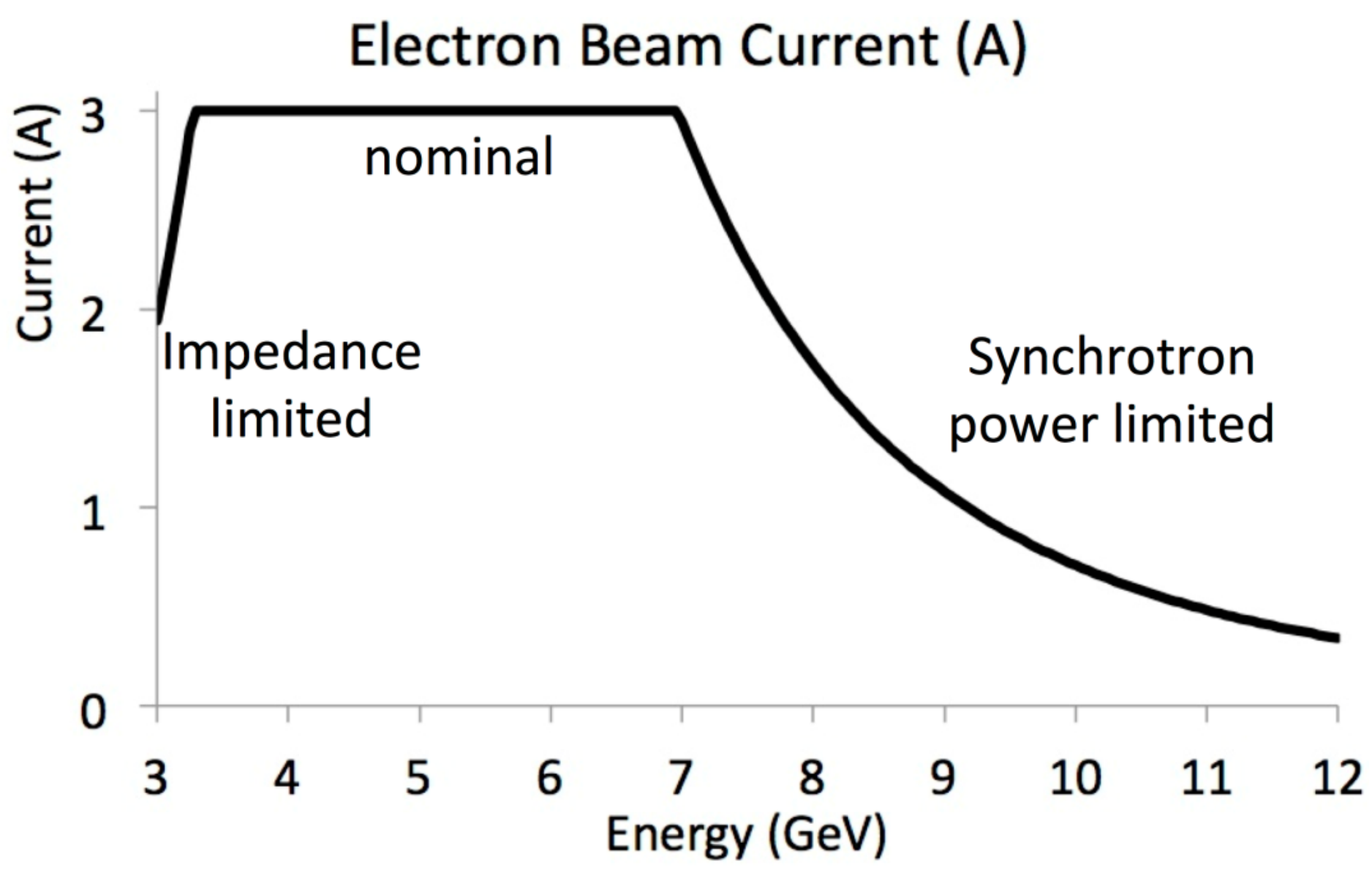

**Figure 10.6:** Beam current limits in the electron ring as a function of energy.

## Crabbing system

Because of the finite crossing angle, the luminosity would be quickly degraded without a crabbing system to compensate. A local crabbing scheme is preferred as it is compatible with multiple IP's and eases concerns about collimation and beam losses around the rest of the machine. The cavity concept is similar to those being developed for LHC [De Silva2013], Figure 10.7. These cavities also need to be HOM-damped and operate at a relatively low frequency to minimize the effects of curvature on the bunches. 952 MHz cavities will fulfill these requirements and be compatible with future upgrades. The parameters of this system are given in Table 10.5.

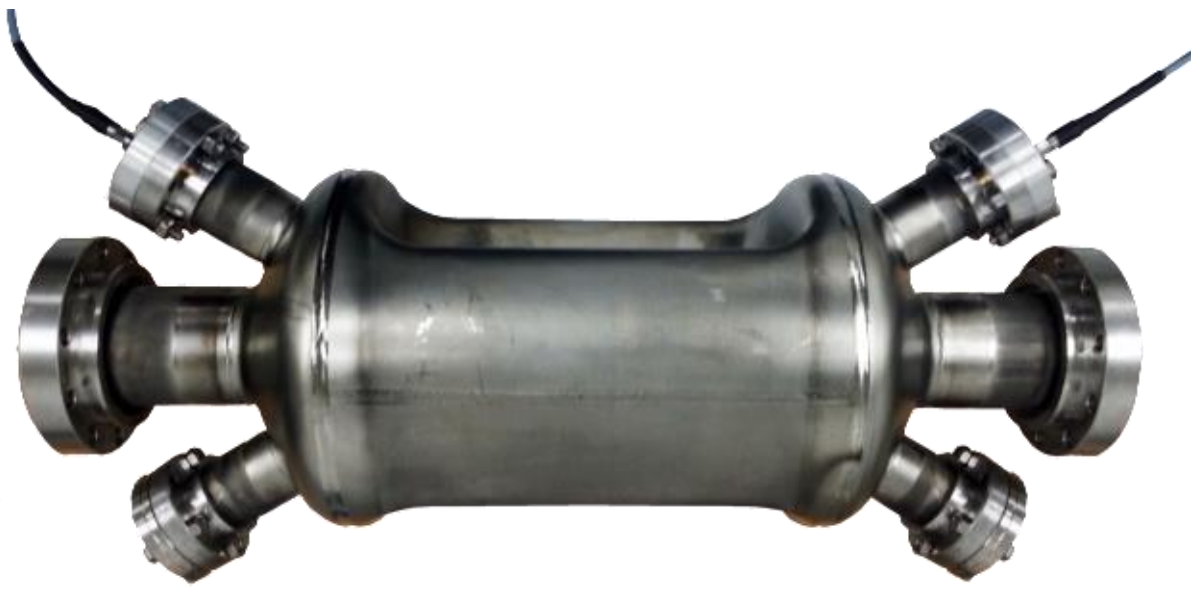

**Figure 10.7:** Prototype "RF dipole" crab cavity being developed by ODU.

**Table 10.5:** Crab cavity system parameters.

| Parameter | Units | Electron | Proton |
|---|---|---|---|
| Beam energy $E_b$ | GeV | 10 | 100 |
| Bunch frequency $n_b$ | MHz | 952.0 | |
| Crossing angle $\varphi_c$ | mrad | 50 | |
| Betatron function at the IP $\beta_x^*$ | cm | 10 | |
| Betatron fn. at the crab cavity $\beta_x^c$ | m | 200 | 750 |
| **Integrated kicking voltage $V_T$** | **MV** | **1.76** | **14.48** |
| Number of cavities (per side of IP) | -- | 2 | 6 |
| Total number of cavities (per specie) | -- | 4 | 12 |

## Cooler ERL

The high energy ion cooler requires a high-current low-emittance electron beam, best provided by an ERL, feeding a circulator ring where the beam stays for a finite number of turns before being recovered, Figure 10.8. This ERL requires an SRF linac cryomodule containing four 5-cell cavities, Figure 10.9, and a short booster cryomodule containing four single-cell cavities similar to that shown in Figure 10.4. These all need HOM damping to achieve the desired high current, and would use similar ancillary components to the ion ring RF systems. The parameters of this system are given in Table 10.6.

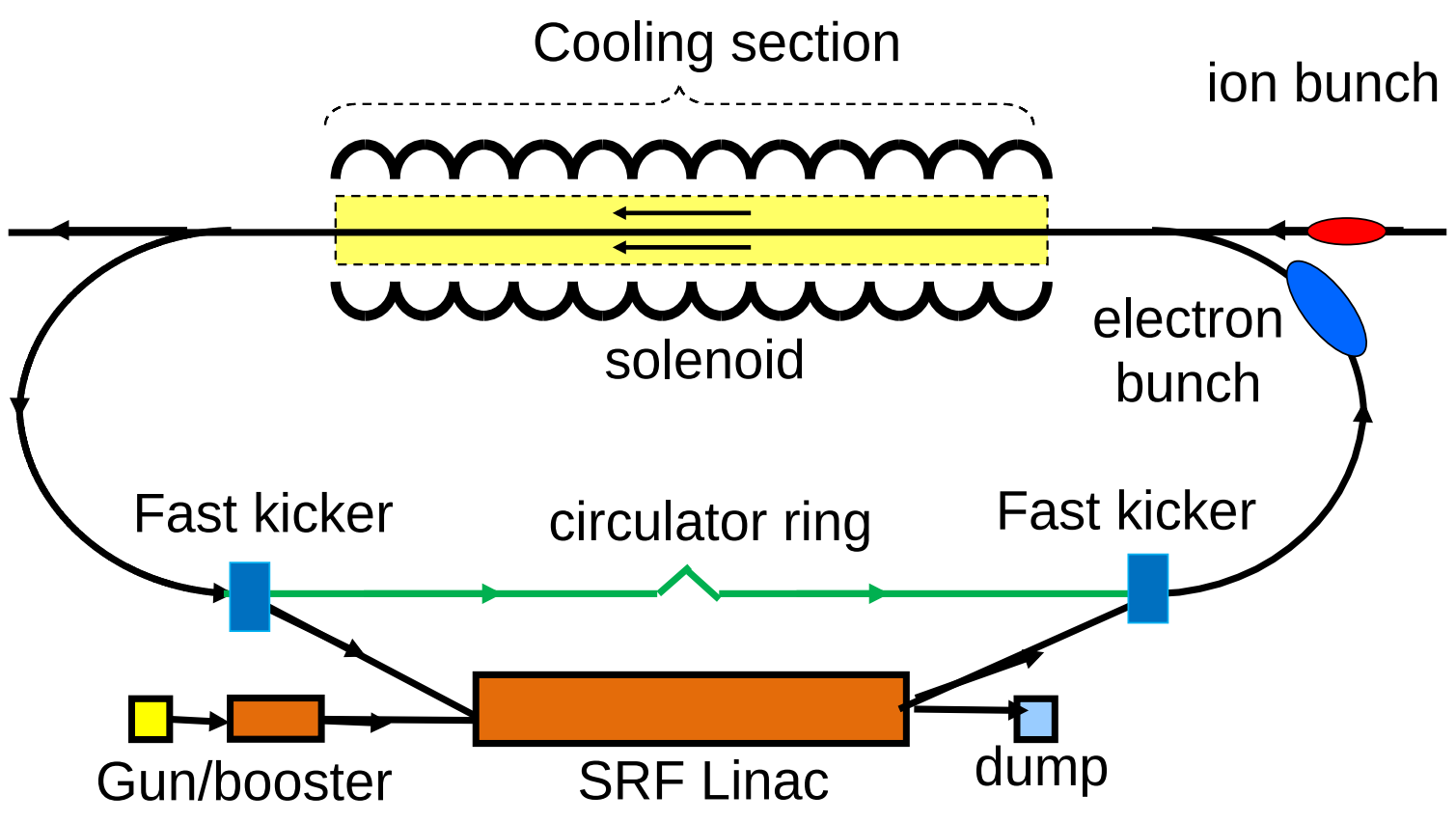

**Figure 10.8:** Schematic of the cooler ERL and ring concept.

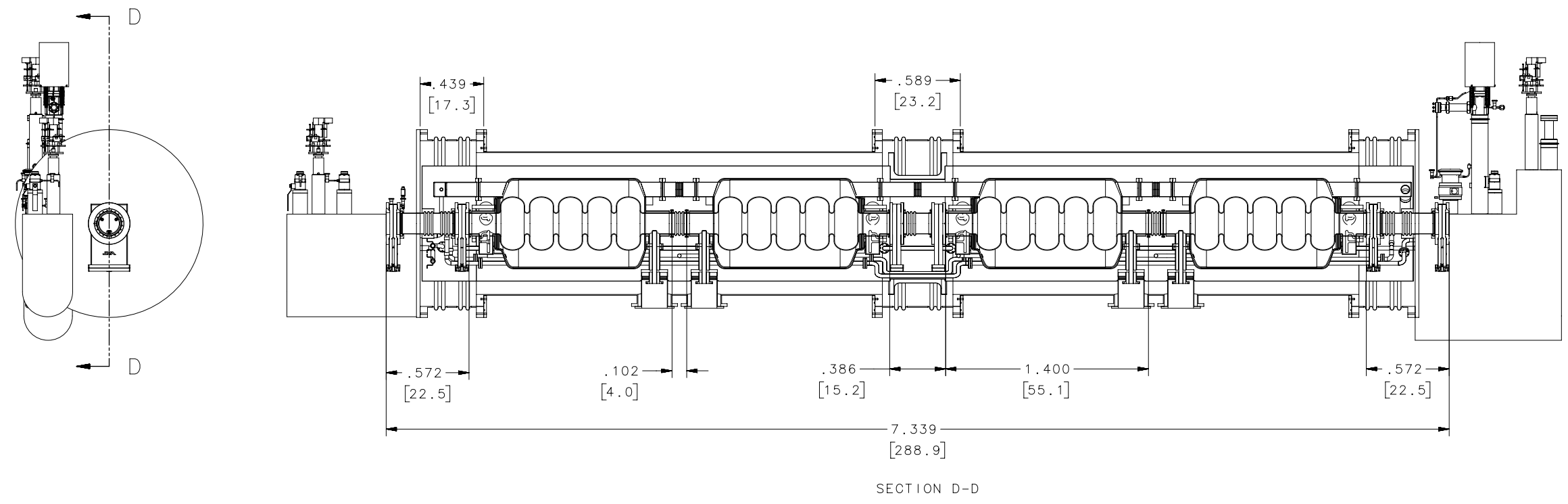

**Figure 10.9:** Concept for the multi-cell cavity cryomodule for the cooler ERL.

**Table 10.6:** RF parameters of the cooler ERL.

| | cavity type | Frequency | voltage /cav | power /cav | # cav | # CM | total volts | total power |
|---|---|---|---|---|---|---|---|---|
| | | MHz | MV | kW | | | MV | kW |
| Cooler ERL* | SRF 5 cell | 952.6 | 14 | 30 | 4 | 1 | 56 | 120 |
| Cooler booster | SRF 1 cell | 952.6 | 2.5 | 150 | 4 | 1 | 10 | 2000 |

*Assumes 200 mA gun, 10 MeV booster, and 50 MeV ERL

# Project Organization and R&D Challenges

## Project Organization

Design and R&D work on the EIC in general and specifically on the MEIC, the Jefferson Lab design, has been carried on for more than a decade. The design team has converged on the present ring-ring option in 2006 and the effort ever since has focused on advancing the design of the sub-systems, the R&D to validate the technical options and more recently the optimization of the overall design to maximize performance, reduce costs, and provide enough flexibility to allow phasing of the project and cost-effective future upgrades.

In order to better focus on the MEIC, the accelerator resources were internally organized into a project structure in 2014, with project management directly reporting to the Laboratory and Deputy Laboratory Directors, and the required personnel were explicitly matrixed to MEIC. Following the successful commissioning of the 12 GeV, accelerator resources have been internally redirected to the MEIC effort to advance design and R&D, to prepare the initial cost estimate, and to produce a Conceptual Design Report. In order to advance design and R&D, the MEIC leverages internal resources, external collaborations with national labs and universities, programmatic DoE grants, SBIR, resources from the internal LDRD program, and resources from the Commonwealth of Virginia.

## Timeline for Realization

The earliest possible realization for the EIC is shown in Figure 11.1 where the EIC timeline is overlaid with the 12 GeV Project and FRIB construction schedules. Substantial construction costs for the EIC are assumed to be available after the end of the FRIB construction. The timeline assumes endorsement for an EIC at the end of the ongoing NSAC Long Range Plan process. It also assumes that the relevant accelerator R&D for the down-select process are resolved by the end of 2016 and a complete Conceptual Design Report (CDR) is produced prior to CD1. We are in the process of establishing a schedule for the CDR and a realistic completion date would be in 2017. Regular 12 GeV CEBAF operations may continue during almost all of the MEIC construction project period.

## R&D for Enabling Technologies and Risk Mitigation Strategies

The overall development strategy for the MEIC has been to optimize machine performance while keeping the number of technical risks at a minimum. Whenever possible, conservative design choices have been selected. An overview of R&D challenges for the MEIC design, an assessment of the risk level and mitigating strategies is summarized in Table 11.1.

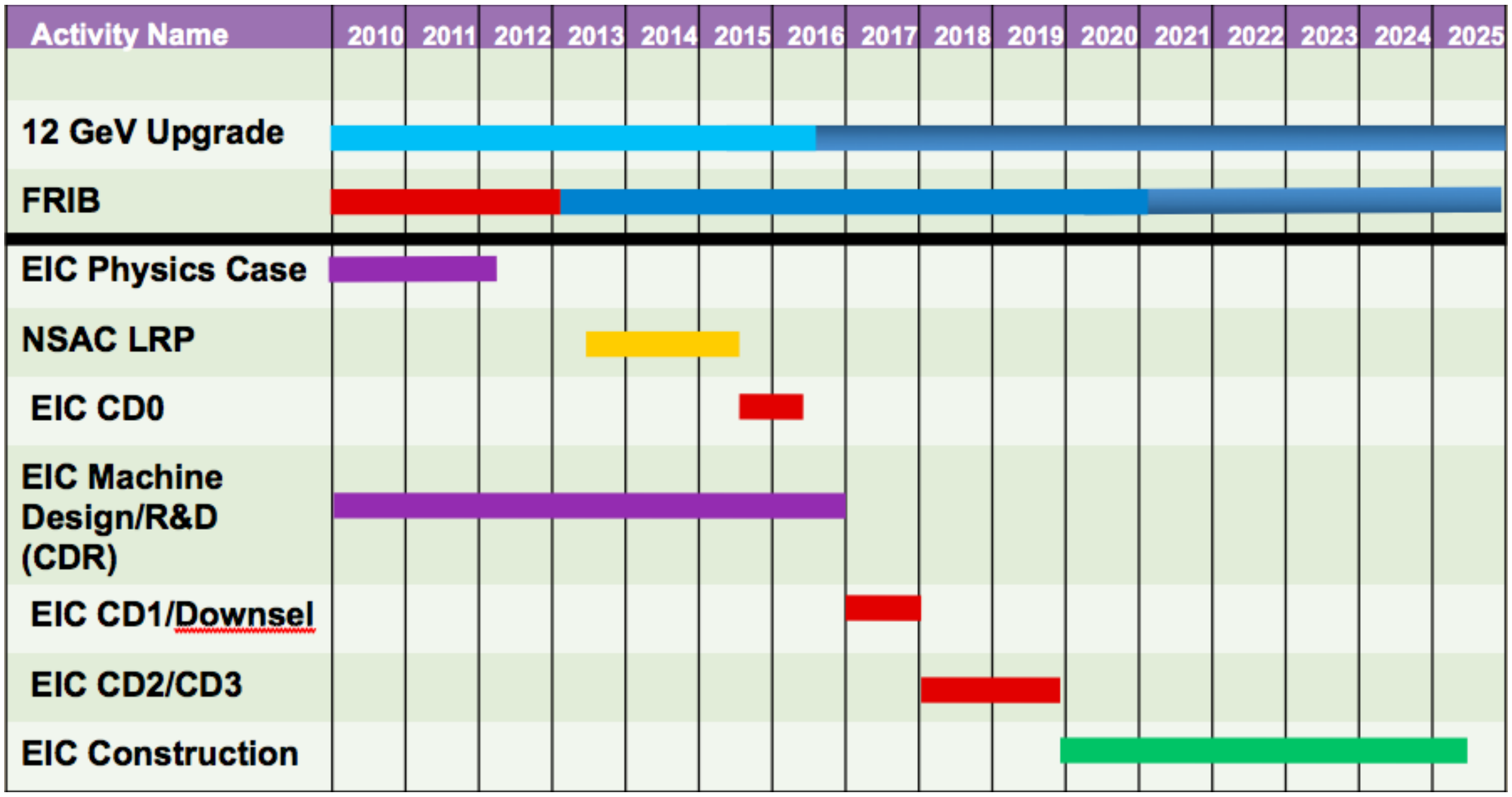

**Figure 11.1:** Time line showing earliest realization of the MEIC

The main R&D challenge in the MEIC design is the high energy bunched beam cooling, necessary to reduce emittance and counteract IBS growth in the ion ring. DC cooling is by now a very well established technology for ion beams up to 8 GeV, but bunched beam cooling has not been attempted yet at high energy. The baseline luminosity in Table 3.1 can be achieved with "weak" cooling, with an ERL cooler without the circulator cooler; therefore there is no unmitigated high risk in the baseline design. The circulator cooler is an option for performance enhancement (See Table 11.2). We have an R&D program that aims to demonstrate bunched beam cooling with an ERL and circulator cooler, based on simulation and a series of tests validating critical aspects of the design. The design and test of the ultra-fast kicker needed to transfer bunches between the ERL and the circulator cooler is the goal of a funded LDRD. An experimental test is planned to measure CSR in CEBAF, thus validating the CSR models. The concepts of ERL and circulator cooling can be tested at the JLAB FEL, where the UV and IR lines can be reconfigured into a recirculator cooler with minimum hardware modification. A first validation of bunched beam cooling can be achieved at the IMP DC cooler in Lanzhou by adding an RF drive to the existing DC cooler, and a collaboration is in place to realize this test. In addition, the present program for low energy bunched beam cooling at RHIC, if successful, will be a proof-of-principle validation for the technology.

**Table 11.1:** Overview of the MEIC R&D challenges and mitigating strategies.

| R&D challenge | Risk level | Mitigating strategies | Mitigated Risk Level |
|---|---|---|---|
| Magnetized bunched beam electron cooling **Circulator Cooler** | HIGH | • Simulation<br>• Test of ultra-fast kicker (LDRD)<br>• Test of ERL+ circulator cooler at FEL | MEDIUM |
| Magnetized bunched beam electron cooling **ERL only** | MEDIUM | • CSR experiment at CEBAF(LDRD)<br>• Test of bunched beam cooling at IMP(LDRD)<br>• Experience from RHIC low energy e-cooling<br>• 100-200 mA unpolarized e- source | LOW |
| Low $\beta$* in the ion ring | MEDIUM | • Chromatic correction scheme<br>• IR non-linear correction scheme<br>• Dynamic Aperture tracking with errors and beam-beam forces<br>• Operational experience at hadron colliders (Tevatron, RHIC, LHC) | LOW |
| Space charge dominated beams | MEDIUM | • Simulation<br>• DC cooling in booster<br>• Operational experience at space-charge dominated rings UMER and IOTA | LOW |
| Figure-8 layout | MEDIUM | • Spin-tracking simulations | LOW |
| Super-ferric magnet designs | MEDIUM | • Prototypes (SSC, GSI)<br>• Early MEIC prototype (FY15-16)<br>• Operational experience at GSI<br>• Alternative cos θ designs | LOW |
| Crab cavities | MEDIUM | • Simulations<br>• Prototypes<br>• Operational experience at KEK-B and LHC<br>• Test of crab cavity in LERF (FEL) | LOW |

The MEIC design has low $\beta^*$s of 10 cm horizontal and 2 cm vertical (and respectively 4 cm and 0.8 cm in the high luminosity IR) in the ion ring. This value is made possible by the very short bunch lengths in the MEIC rings, but is beyond operational experience at hadron colliders. On the other hand the same operational experience at RHIC and LHC have validated the main mitigation strategy, based on chromatic correction, non-linear local IR error corrections at the IR final focus triplets, and performance prediction by dynamic aperture simulation including field and alignment errors and the beam-beam effect.

Space charge is a factor in the Booster dynamics; we plan to quantify the effects through simulation with 2-stage DC cooling, initially for conditions at Booster injection and at 3 GeV. A vibrant program of space-charge studies is in progress at the UMER ring at the University of Maryland and is planned at the IOTA ring at FNAL, and we can benefit by increased experimental experience with space charge dominated beams and possibly execute beam experiments that simulate the space charge parameters at the MEIC.

The main strategy for spin preservation is a figure 8 layout for the booster and the ion ring. The main validation strategy is spin-tracking; preliminary results confirm the spin preservation capability of the figure 8 layout.

Super-ferric magnets are simpler, cheaper to build, and cheaper to operate than cos θ superconducting magnets. Three full length 38 m prototype 3 T superferric magnets designed for the SSC were built, tested, and met design performance. Two Tesla super-ferric magnets are the chosen technology for the NICA design in Dubna and for the SIS100 rapid-cycling ring at the FAIR complex at GSI. For that project 113 curved-geometry superferric dipoles were recently built and tested, and provide a useful cost comparable. A full prototype of the 3 T superferric dipole for MEIC is planned to be built at Texas A&M University and tested in FY15-16. Furthermore an ongoing R&D on cable-in-conduit based superferric magnets and new conductors holds the promise of higher fields and better-performing magnets for the MEIC.

## Phasing and Collider Upgrades

The present MEIC baseline, the results of performance and cost optimization, offers the possibility of phasing-in project components in order to defer costs if desired and offers an early start to physics. Similarly, the design offers the potential of very cost effective future upgrades

Phasing options include:

- Starting with 1 interaction region only and phasing in the second IR (installation of IR special magnets and second detector).
- Starting the physics program with DC cooling in the Booster only, while phasing in the bunched beam cooling systems based on the ERL, and later on the circulator cooler.
- Building up SRF incrementally in the ion ring starting the physics program at the medium center of mass energy and phasing in physics at higher energies.
- Starting with one ion source, and phasing in the second.
- Phasing in ion operations.

Possible upgrades include:

- Raising the e-ring energy to 12 GeV by adding more RF to the e-ring.
- Adding damping wigglers to the e-ring to reduce emittance (a factor 2 reduction was demonstrated in LEP operations).
- Further reducing the e-ring emittance with a TME lattice (We have a lattice solution already and it requires swapping a fraction of the PEP-II dipoles with new ones).
- Modifying the ion ring lattice to avoid transition crossing for all ion species through replacement of present FODO arcs with a newly developed negative momentum compaction optics, which yields imaginary transition gamma for the ion ring.
- Raising the ion-ring energy to 120 GeV (That requires an increase of the super-ferric magnet field strength by 20%).
- Raising the ion-ring energy to >250 GeV (a 5.5 T design for cos-theta RHIC-like magnets has been already considered and appears feasible, and we can capitalize on R&D on high field superconducting magnets for the LHC upgrade and beyond).

Table 11.2 offers an example of possible performance enhancements from the upgrades. It assumes a smaller electron emittance (about a factor 2, as could be achieved by adding damping wigglers and new optics in the electron ring), a larger beam-beam parameter (allowed by running only 1 IP), and doubles the repetition rate.

**Table 11.2**: Upgraded MEIC main design parameters for a full-acceptance detector, with reduced emittance and doubled repetition rate

| CM energy | GeV | 21.9 (low) | | 44.7 (medium) | | 63.3 (high) | |
|---|---|---|---|---|---|---|---|
| | | *p* | *e* | *p* | *E* | *p* | *e* |
| Beam energy | GeV | 30 | 4 | 100 | 5 | 100 | 10 |
| Collision frequency | MHz | 952 | | 952 | | 159 | |
| Particles per bunch | $10^{10}$ | 0.66 | 3 | 0.66 | 2.6 | 2.0 | 2.8 |
| Beam current | A | 1 | 4.5 | 1 | 4 | 0.5 | 0.72 |
| Polarization | % | >70 | >70 | >70 | >70 | >70 | >70 |
| Bunch length, RMS | cm | 2.5 | 1.2 | 0.7 | 1 | 2.75 | 0.7 |
| Norm. emittance, hor/ver | µm | 0.45/0.45 | 30/30 | 0.35/0.07 | 40/20 | 0.35/0.07 | 440/220 |
| Horizontal and vertical $\beta^*$ | cm | 1/1 | 3.7/3.7 | 5/1 | 4/0.8 | 7.5/1.5 | 4/0.8 |
| Vert. beam-beam parameter | | 0.008 | 0.05 | 0.015 | 0.061 | 0.06 | 0.061 |
| Laslett tune-shift | | 0.061 | small | 0.055 | small | 0.013 | small |
| Detector space | m | 7/3.6 | 3.2/3 | 7/3.6 | 3.2/3 | 7/3.6 | 3.2/3 |
| Hour-glass (HG) reduction factor | | 0.65 | | 0.86 | | 0.65 | |
| Lumi./IP, w/HG correction, $10^{33}$ | $cm^{-2}s^{-1}$ | 6.7 | | 32.6 | | 4 | |

We conclude that the MEIC baseline design fulfills the performance requirements for the EIC with few manageable technical risks and a series of performance enhancements are possible to increase the luminosity by about a factor 10.